\documentclass[12pt,leqno]{amsart}
 
\usepackage{amsmath}
\usepackage{graphicx}
\input amssym.def
\input amssym
\usepackage{pdfsync}

\long\def\symbolfootnote[#1]#2{\begingroup%
\def\thefootnote{\fnsymbol{footnote}}\footnote[#1]{#2}\endgroup}

\newtheorem{theorem}{\sc Theorem}[section] 
\newtheorem{lemma}[theorem]{\noindent {\sc Lemma}} 
\newtheorem{corollary}[theorem]{\sc Corollary}
\newtheorem{proposition}[theorem]{\sc Proposition}

\newtheorem{definition}{Definition}[theorem]
\newtheorem{conjecture}{Conjecture}[theorem]

\theoremstyle{plain}

\renewcommand{\a}{\alpha}
\renewcommand{\b}{\beta}
\renewcommand{\d}{\delta}
\newcommand{\D}{\Delta}
\newcommand{\e}{\varepsilon}
\renewcommand{\th}{\theta}
\newcommand{\g}{\gamma}
\newcommand{\G}{\Gamma}
\renewcommand{\l}{\lambda}
\renewcommand{\k}{\kappa}

\newcommand{\s}{\sigma}

\renewcommand{\t}{\tau}
\newcommand{\cal}{\mathcal}
\newcommand{\Z}{{\Bbb Z}}
\newcommand{\R}{{\Bbb R}}
\newcommand{\C}{{\Bbb C}}

\renewcommand{\o}{\omega}

\renewcommand{\i}{\infty}
\newcommand{\p}{\partial}

\newtheorem{example}[theorem]{\sf Example}
\renewcommand{\thefootnote}{\fnsymbol{footnote}}
\newcommand{\nat}{\natural}
\renewcommand{\thefootnote}{\fnsymbol{footnote}}
\addtocontents{toc}{\hfill Page\endgraf}
\addtocontents{lof}{\ Figure\hfill Page}

\makeatletter
\renewcommand\tableofcontents{%
  \newfont{\scaledfont}{cmr12 scaled 7000}
  \parindent\z@\raggedright
  \scaledfont\contentsname\par\normalsize%
  \rule{\textwidth}{1pt}
  \nobreak
  \vskip 40\p@
  \@starttoc{toc}%
}
\makeatother
\begin{document}
 \author[J. Harrison  Department of Mathematics  U.C. Berkeley]{Jenny Harrison
\\Department of Mathematics
\\University of California, Berkeley}
\title[Chainlet Geometry]{Lectures on Chainlet Geometry --  New Topological Methods in Geometric Measure Theory}
 \begin{abstract}  These draft notes  are from a graduate course given by the author in Berkeley  during the spring semester of 2005.  They cover the basic ideas of a new, geometric approach to geometric measure theory.  They begin with a new theory of exterior calculus at a single point.  This infinitesimal theory extends, by linearity, to a discrete exterior theory, based at finitely many points. A general theory of calculus culminates by taking limits in Banach spaces, and is valid   for domains called ``chainlets'' which are defined to be elements of the Banach spaces.  Chainlets include manifolds,  rough domains (e.g., fractals), soap films, foliations, and Euclidean space.  Most of the work is at the level of the infinitesimal calculus, at a single point. The number of limits needed to get to the full theory is minimal.  Tangent spaces are not used in these notes, although they can be defined within the theory.  This new approach is made possible by giving the Grassmann algebra more geometric structure.  As a result, much of geometric measure theory is simplified.  Geometry is restored and significant results from the classical theory are expanded.  Applications include existence of solutions to a  problem of Plateau,  an optimal Gauss-Green theorem and new models for Maxwell's equations.   \end{abstract}
\maketitle

\section{Polyhedral chains}\label{polyhedral}  Chainlet geometry is first developed for domains in Euclidean space (e.g., submanifolds), and later expanded to abstract manifolds\symbolfootnote[1]{The author thanks Morris Hirsch and James Yorke for their support and encouragement over the years, and for their helpful editorial comments. She also thanks the students in her graduate class and those in Ravello for their patience and interest, all of whom helped make this draft possible. These notes will appear, in a similar draft form, in the Proceedings of Ravello Summer School for Mathematical Physics, 2005, and will be further developed into several research papers, as well as a full monograph.
 }. 

 Let $v \in \R^n$ and $X \subset \R^n$.  We let $$T_v X:= \{ p + v: p \in X\}$$ denote the {\itshape \bfseries translation} of $X$ through the vector $v$.  Generally, we will let $n$ denote the dimension of the ambient space we are using and $k$ will denote the dimension of a subspace.  
 A $k$-dimensional {\itshape \bfseries affine subspace} $E$ of $\R^n$ is  the translation of a $k$-dimensional  linear subspace $L$ through a fixed vector $v$, that is, $E := T_v L.$   Let $E$ be a $k$-dimensional affine subspace of $\R^n$ and $F$  a $(k-1)$-dimensional affine subspace of $\R^n$ with  $F \subset E \subset \R^n.$       A $k$-dimensional {\itshape \bfseries half-space}  of $\R^n$ is the closure of one of the two components   of $E -F$.  Denote such a half-space by $E_F.$
 
 To define an $n$-cell, we set $E = \R^n$. Let $F_i, i = 1, \dots, j$, be $(n-1)$-dimensional affine spaces.    An {\itshape \bfseries $n$-cell} $\s$  in $E$ is any finite intersection of $n$-dimensional half-spaces $\s = \cap_{i = 1}^j E_{F_i}$ satisfying:
\begin{enumerate}
\item $\s$ has nonempty interior in $E$.
\item   $\s$ is compact.  \\ We can furthermore assume
\item  Each $ F_i \cap {\s}, 1 \le i \le j,$ has nonempty interior in $F_i$
\end{enumerate}  since only those  $F_i$ that satisfy (3) affect the set $\s$.


An $n$-cell will always be convex since it is the intersection of convex sets\footnote{A $2$-cell is  often called a {\itshape \bfseries polygon,} and a $3$-cell a {\itshape \bfseries polyhedron}.     Every time  a block of cheese is cut into small pieces with a knife, all the pieces will be   $3$-cells.  A set $X \subset \R^n$ is {\itshape \bfseries convex} if whenever $p,q \in X$ then $tp +(1-t)q \in X, 0 \le t \le 1.$    An $n$-cell is sometimes called a {\itshape \bfseries convex polytope}.}  

Each $\t_i = F_i \cap {\s}$   is called a {\itshape \bfseries facet}  of $\s.$  The {\itshape \bfseries frontier} of a subset of $E$ is defined to be its closure less its interior.    Clearly, the frontier of an $n$-cell $\s$ in $\R^n$ is the union of its facets.  We will shortly define the boundary of $\s$ which is somewhat different from its frontier.     
 
  Given a cell $\s$, we may easily identify the half-spaces which generate it.  The $F_i$ are the unique $(n-1)$-dimensional affine spaces that contain the facets of $\s$.   The half-space $E_{F_i}$  coincides with the half  of $E - F_i$ that intersects $\s$.   The collection of such half-spaces $\{E_{F_i}\}$ is said to  {\itshape \bfseries generate} $\s$.

  For $k < n$, a {\itshape \bfseries $k$-cell} $\s$ in $\R^n$ is the intersection of an $n$-cell with a $k$-dimensional affine subspace $E$ of $\R^n$ that meets the interior of $\s$, called the {\itshape \bfseries subspace} of $\s$.  The {\itshape \bfseries subspace} of $\s$ is uniquely determined as the minimal affine subspace containing $\s$.   The {\itshape \bfseries dimension} of a cell $\s$ is   the dimension of its subspace $E$.     The following lemma is immediate.   
\begin{lemma} \label{facet}
Each facet of a $k$-cell is a $(k-1)$-cell.  Two $k$-cells  intersect in a  cell with dimension $\le k$.  
\end{lemma}

Our $k$-cells may be given an orientation as described below.

Assume $k >0$.  An {\itshape \bfseries orientation} of a $k$-dimensional linear subspace $L$ of $\R^n$   is an equivalence class of ordered bases of $L$ where two bases are {\itshape \bfseries equivalent} if and only if their transformation matrix has positive determinant.  An {\itshape \bfseries orientation} of a $k$-dimensional affine subspace $E$  is defined to be an orientation of the linear subspace $L$ {\itshape \bfseries parallel} to $E$.  That is, there exists a vector $v$ such that $L = T_v E.$     An {\itshape \bfseries orientation} of a $k$-cell $\s$ is defined to be an orientation of its affine subspace $E$.   A $k$-cell and its affine subspace $E$ may have different orientations chosen for them, since $E$ has two possible orientations.    We say a $k$-cell $\s$ is {\itshape \bfseries oriented} if we have assigned an orientation to it.  That is, we have specified a class of ordered bases within the subspace of $\s$.    In the future, we assume each $k$-cell  $\s$ is oriented.  If the orientation of a $k$-cell $\s$ is the same as that chosen for its affine subspace $E$,  we say that $\s$ is {\itshape \bfseries postively oriented} with respect to the orientation of $E$.  Otherwise, it is {\itshape \bfseries negatively oriented}  with respect to the orientation of $E$.     
No orientation need be assigned to 
$0$-cells which turn out to be single points $\{x\}$ in $\R^n$.    It might be  tempting to try and orient all affine subspaces $E$ of $\R^n$, once and for all, given an orientation on $\R^n.$   A little thought will convince the reader that this may not be done naturally. However, we may orient finitely many of them.

The {\itshape \bfseries $k$-direction}\footnote{This is called a $k$-direction because of the analogy with the directions of a compass for $k = 1$.}  of $\s$ is defined as the linear subspace $L$ parallel to the affine subspace $E$ of $\s$, oriented to match the orientation on $\s$.

We assign a translated cell $T_v \s$ the same orientation as $\s$.
 
         A {\itshape \bfseries list} $U^j := (u_1, \dots, u_j)$ is an ordered set of $j$-vectors $u_i$ in $\R^n,$ with duplications permitted.   
 
\subsection*{Parallelepipeds}.  If $X \subset \R^n$ is some set, the {\itshape \bfseries convex hull} of $X$ can be described as the set of points of the form $\sum_{i=1}^N t_i x_i$,   where $N$  is an arbitrary natural number, the numbers $t_i$ are non-negative and sum to $1$, and the points $x_i$ are in $X$.
Let $V^k:= (v_1, \dots, v_k)$ denote a list of linearly independent vectors.   A {\itshape \bfseries $k$-parallelepiped} $\s(V^k)$ is the convex hull of the vectors $V^k,$ oriented according to the orientation given by the $V^k$. We also call the translation of any $\s(V^k)$ through a vector $u \in \R^n$ a $k$-parallelepiped.  Clearly, $k$-parallelepipeds are $k$-cells.  The $k$-direction of $\s(V^k)$ is the subspace spanned by $V^k$.   {\itshape \bfseries Degenerate} $k$-parallelepipeds arise if the list $V^k$ is not linearly independent.

\subsection*{Algebraic $k$-chains}    Form the vector space of formal sums of $k$-cells over a field $\mathbb{F}$.  Call an element of this space a {\itshape \bfseries $k$-chain.} Let $\s'$ denote the $k$-cell obtained from $\s$ by reversing its orientation.     The $k$-chain $\s + \s '$ is not the zero $k$-chain as a ``formal sum''.     We next equate $\s + \s'$ with $0$.
  
 Formally, the vector space 
of {\itshape \bfseries algebraic $k$-chains} $S$ with coefficients in $\mathbb{F}$ is the quotient of the subspace generated by chains of the form $\s + \s'$. 
    In other words, to create an algebraic $k$-chain, we simply take a linear combination of oriented $k$-cells, making sure to   ``collect like terms''  to obtain a unique and canonical sum. It only remains to treat two cells $\s$ and $\s'$ with the same support, but opposite orientation.  By equating (replacing) $\s'$ with $-\s$ we are able to add two such cells together.   Thus, if two $k$-cells of the $k$-chain have the same support, their coefficients are added, taking into account their signs which are determined by orientation.    But it their supports differ even slightly, they are represented as distinct terms in the sum.    
 An important example of an algebraic $k$-chain is the boundary of a $(k+1)$-cell.  

 An algebraic $k$-chain $S$ is said to be {\itshape \bfseries non-overlapping} if   any two cells of $S$ only meet in facets, if at all.  We call the unoriented pointset underlying a $k$-cell its {\itshape \bfseries support}, denoted $supp(\sigma).$    The {\itshape \bfseries support} of a non-overlapping algebraic $k$-chain $S = \sum a_i \s_i$ is defined as $$supp(S) := \cup supp(\s_i).$$ We say that an algebraic $k$-chain $S$ is {\itshape \bfseries supported in} $X \subset \R^n$ if $supp(S) \subset X.$     

Remarks:  
\begin{itemize}
\item When we write $2 \s$ we are not doubling the size of $\s$, but considering two superimposed copies of $\s$.   Scalar multiplication by a positive scalar changes {\itshape \bfseries multiplicity}, or {\itshape \bfseries density}, but not support.   
\item Roughly speaking, an algebraic $k$-chain with integer coefficients is merely a choice of a finite number of oriented $k$-cells, each with integer multiplicity, placed in $\R^n$.  This is quite different from taking unions of point sets, although the figures may look the same since the support of a non-overlapping chain is the union of the supports of its cells.    Analysis based on  unions and intersections of sets is substantially different. The vector space structure of algebraic $k$-chains   brings algebra of multiplicity and orientation into the mathematics at an early stage.
  
\end{itemize}
\subsection*{Induced orientation on the boundary}  
\begin{figure}[t]
\begin{center}
\resizebox{3.5in}{!}{\includegraphics*{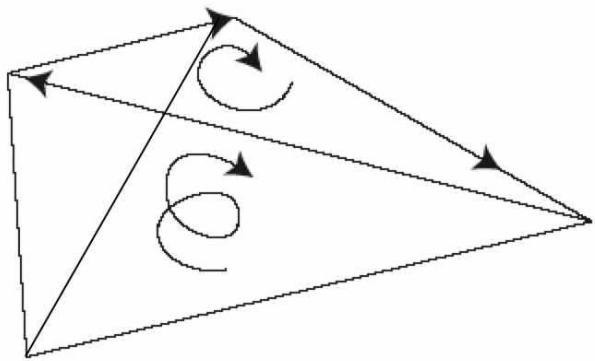}}\label{figure1}
\caption{Induced orientation on the boundary}
\end{center}
\end{figure}

Let $E$ be a $k$-dimensional affine subspace of $\R^n$, $k > 1$.   Let $E_F$ be one of the half-spaces determined by  $F \subset E$ where $F$ is a $(k-1)$-dimensional affine subspace.    We will show that an orientation of $E$ and a choice of half-space $E_F$ together naturally induce an orientation on $F$, the frontier of $E_F$.  We may  assume that $0 \in F$.    Choose a basis $B$   of   $F$.  Choose a vector $v$ in $E_F$, not in $F$ and append it to     $B$    to obtain a basis $B_1$ of $E$.   
If the orientation given by $B_1$ is positive (with respect to our original orientation on $E$) then we say that   $B$ is a positively oriented basis for $F$.  For $k = 1$ $F$ will be zero dimensional. The argument is slightly different since there is no meaning of orientation of a point.    Choose a vector $v$ in $E_F$.  This vector is a basis of $E$.  If it is positive, we assign the coefficient $+1$ to the point $F$, otherwise, we assign $-1$ to $F$.   
  In practice,  the induced orientation   can be captured by a simple figure.  (See Figure \ref{figure1}.)  
   

  This leads to a natural definition of the {\itshape \bfseries boundary} of a $k$-cell $\s$.  Suppose $k > 1$. Each facet $\t_i$ of $\s$ has an orientation induced from that assigned to $\s$.  Suppose $k = 1.$  Observe that if $A, B \in \R^n$ then $A- B$ denotes both a zero chain and a vector.  The boundary of a  $1$-cell $\s$ with endpoints $A, B$ is  the zero chain $A -B$  if the orientation  of $\s$ agrees with the vector $A-B.$   Define $$\p \s := \sum \t_i.$$  Note that $\p \s$ is non-overlapping and its support is the frontier of $ \s.$      This extends linearly to algebraic $k$-chains $$\p \sum a_i \s_i : = \sum a_i \p \s_i.$$
  
 For example,    consider  a triangle $\s$ with vertices $A, B, C$.  Then $\p( \p\s) = 0.$   This can be seen by adding up the vectors of the boundary facets $(A-B) + (B-C) + (C-A) = 0.$  We next give a general proof to this assertion.      

\begin{lemma}\label{bbl}
Let $\t_i$ and $\t_j$ be two $k$-cells of the boundary of a $(k+1)$-cell $\s$ that intersect in a common facet $\k$.  Their induced orientations on $\k$ are opposing.
\end{lemma}

\begin{proof}     Let $F_i$, $F_j$, $G$, $E$ be the affine subspaces of $\t_i$, $\t_j$, $\k$ and $\s$, respectively. 
 Let $C$ be a $(k-1)$-dimensional affine subspace containing  $G$ so that $\t_i$ and $\t_j$ are in opposing half-spaces of $E - C.$ (There are lots of these.)
   Then $F_i$ and $F_j$ are in opposite half spaces of $E-C.$  Choose a basis of $G$, extend it to $C$ and then to $E$ in two ways:  one by choosing a basis vector in $F_i$, the other by choosing a basis vector in $F_j.$  These bases have the opposite orientation.  The result follows since the induced orientations on $\k$   are opposing.      \end{proof}

\begin{theorem}\label{bb}
$$(\p \circ \p) \s = 0$$ for all $k$-cells $\s$.
\end{theorem}

\begin{proof}   This follows directly from   Lemma \ref{bbl}
. The boundary of a $k$-cell is the sum of its facets:   $\p \s = \sum \t_i.$  Two facets  $\t_i$ and $ \t_j$ are either disjoint or meet in a common facet $\k$.   By the preceding lemma, the orientations induced on $\k$ are opposing.  Thus  $\k$ contributes nothing to the sum when we add $\p \t_i + \p \t_j$ to find $\p(\p \s).$
\end{proof}

We wish to further simplify our basic chains  since the   definition of an algebraic $k$-chain can be  inefficient, recording information we do not care about.  
   Choose an orientation of $\R^n$.  Let $S= \sum_{i = 1}^j a_i \sigma_i$ be an algebraic $n$-chain.    We orient each $\sigma_i$ positively and adjust the sign of $a_i$, accordingly.    Define the function $S(x) := \sum a_i$ where the sum is taken over all $i \in \{1, \ldots, j\}$ satisfying $x \in \sigma_i.$   Set  $S(x) := 0$ if $x$ is not in  any $\sigma_i$.We say that algebraic $n$-chains $S$ and $S'$ are {\itshape \bfseries equivalent} and write $S \sim S'$ iff the functions $S(x)$ and $S'(x)$ are equal except in a finite set of cells of dimension $< n.$   
   
   Suppose $S$ is an algebraic $k$-chain with $k < n.$ Orient the subspaces of  the $k$-cells of $S$.   The choice of orientations uniquely determine the expression  $S = \sum a_i \s_i$, as above.  We define a similar coefficient function $S(x)$, depending on  these orientations.   Given two algebraic $k$-chains $S$ and $S'$, make certain to choose the same orientations for any subspaces in common.       Algebraic $k$-chains $S$ and $S'$ are {\itshape \bfseries equivalent} $S \sim S'$ iff the functions $S(x)$ and $S'(x)$ are equal except in a finite set of cells of dimension $< k.$   (See Figure 2.)

   \begin{figure}[t]
\begin{center}
\resizebox{3.5in}{!}{\includegraphics*{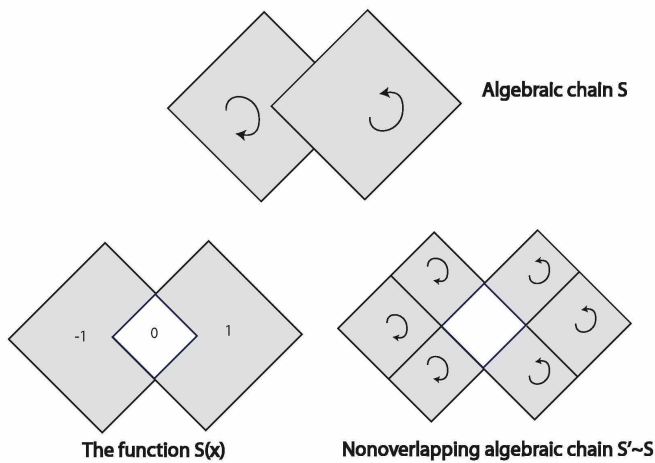}}\label{polyhedralfig}
\caption{Representing a chain as a sum of cells with non-overlapping interiors}
\end{center}

\end{figure}

The {\itshape \bfseries Cartesian product} of  two subsets $X  \subset \R^n$, $Y  \subset \R^m$ is defined as a subset of $\R^{n+m}$:  $$X \times Y := \{(p,q) \in \R^{n+m}:p \in X, q \in Y\}.$$    
     Suppose $S$ and $S'$ are algebraic $k$-chains.  We say that $S'$ is a {\itshape \bfseries subdivision} of $S$ if $S \sim S',$ and  each cell of $S$ is a sum of non-overlapping cells of $S'.$   In the previous two examples, $S'$ is a subdivision of $S$. 
    
     A  {\itshape \bfseries polyhedral $k$-chain $P$} is defined as an equivalence class of 
algebraic $k$-chains.  If $S$ is an algebraic chain, $P = [S]$ denotes the polyhedral chain of $S$.  
 This clever definition of Whitney \cite{whitney} implies  that if $S'$ is a subdivision of the algebraic chain $S$, as in the preceding example, then $S$ and 
$S'$ determine the same polyhedral chain $P$ which behaves nicely for integrating 
forms. In particular, algebraic $k$-chains $S$ and $S'$ are equivalent if and only if integrals of smooth differential $k$-forms agree over them.   This property is sometimes taken as the definition of a polyhedral chain, but we wish to define it without reference to differential forms since we have not even defined them yet.  The {\itshape \bfseries support} of a polyhedral $k$-chain $P = [S]$ is defined to be the support of the function $S(x)$, namely the closure of the set of points $x$ such that $S(x) \ne 0.$   This set is independent of the choice of representative algebraic chain $S$.     Denote the linear space of polyhedral $k$-chains by $\cal{P}_k.$

\begin{proposition}If $S \sim S'$ are algebraic $k$-chains, then $S$ and $S'$ have a common subdivision by an algebraic $k$-chain.

 \end{proposition}

\begin{proof}  The proof 
reduces to showing that the formal sum   $a_1 \s_1 + a_2 \s_2$  is equivalent to a non-overlapping chain.   A natural choice is found by subdividing $\s_1$ with the affine subspaces determining $\s_2$ and vice versa. (See Figure 2.)
   Then  $\s_1$ is subdivided into non-overlapping cells, each with constant multiplicity  and so is $\s_2$. If any two of these cells overlap, they have the same support and thus are collected together in the algebraic sum.  Hence the representative is non-overlapping.     The proof follows by induction.
\end{proof}

It follows that every polyhedral $k$-chain $P$ has a  representative by a non-overlapping algebraic $k$-chain. Observe that the {\itshape \bfseries support} of $P$ is independent of any non-overlapping representative.   The {\itshape \bfseries translation} of $P$ through a vector $v$ is denoted $T_vP$ and is well defined by translating any algebraic representative of $P$.

   \subsection*{Boundary of a polyhedral chain}  The boundary operator on $k$-cells extends to algebraic $k$-chains $S = \sum a_i \s_i$ by linearity:     $$\p S := \sum a_i  \p \s_i.$$

\begin{proposition}
Suppose $S$ is an algebraic $k$-chain.  
If $S'$ is a subdivision of $S$, then $\p S \sim \p S'.$
\end{proposition} 
   
   The proof is similar to that of  
  Theorem \ref{bb} and we omit it. 

\begin{lemma}
If $S \sim S'$ are algebraic $k$-chains, then $\p S \sim \p S'.$
\end{lemma}

\begin{proof}
There exists a common subdivision $S''$ of $S$ and $S'.$  By the previous exercise, $\p S\sim \p S'' \sim \p S'.$
\end{proof}

 It follows from this lemma that the boundary operator is well defined on polyhedral $k$-chains for $k > 0$.  If $P = [S]$, then 
 $$\p P : =  [\p S].$$
  For $k = 0 $ we set $\p P := 0$.  We deduce
$$\cal{P}_n \buildrel \p \over \to \cal{P}_{n-1}
 \buildrel \p \over \to   \cdots   \buildrel \p \over \to \cal{P}_1 \buildrel \p \over \to \cal{P}_0$$
is a {\itshape \bfseries chain complex} (meaning that $\p\circ \p = 0).$ 

  \newpage
  \section{Banach spaces of chainlets}\label{Banach}    
In this section we will define a family of norms on $\cal{P}_k.$
 The norms are initially defined for polyhedral $k$-chains in Euclidean space $\R^n$ and it is shown later how to extend the results to  singular chains $\sum a_i f_i \s_i$ in manifolds $M^n$.   Our norms are defined using an inner product, but a different choice leads to an equivalent norm, and thus the same topology, the same space of chainlets. 
 
\subsection*{Mass of polyhedral chains}  Choose an inner product $<\cdot, \cdot>$ on $\R^n.$     The {\itshape \bfseries mass} of a $k$-parallelepiped, $k \ge 1$, determined by vectors  $(v_1, \dots, v_k),$ is defined by $$m(\s(v_1, \dots, v_k) ) := \sqrt{det(<v_i, v_j>)},$$  that is, the determinant of the $k \times k$ matrix of inner products.
Observe that $m(\s(\l v_1, \dots, \l v_k) ) =  \l^k m(\s(v_1, \dots, v_k) ).$     Let $\s$ be any $k$-cell in $\R^n$ and cover it with finitely many $k$-parallelepipeds.    Add these terms and infimize over all such sums to define $m(\s).$\symbolfootnote[2]{We may also begin with any volume form, or   any translation invariant, finitely additive measure $m$ of $\R^n$.}  Every 0-cell $\s^0$ takes the form $\s^0 = \{x\}$ and we set $m(\s^0) = 1$.

In order to prepare ourselves for the norms to come, we give three equivalent definitions of the mass of a polyhedral $k$-chain $P$. Definition 1 uses integration over differential forms which we include at this stage for expository purposes only.  Definition 2 is an implicit definition, not usually seen.  We will use Definition 3 for developing chainlet geometry, and will then define differential forms and their integrals over polyhedral chains without resorting to the standard definitions from the classical theory.     All three definitions will be shown to be equivalent in Proposition \ref{massequiv}. 
\subsection*{Definition 1}[Differential forms definition of mass] If $\o$ is a differential $k$-form  define $$\|\o\|_0 := \sup_{\s} \frac{\int_\s \o}{m(\s)}$$ where $\s$ denotes a $k$-parallelepiped. 

\subsection*{Integration over polyhedral chains}  It is well understood that integration over  cell-like domains $\int_{\s} \o$ is a linear function of $\o$. For a polyhedral chain $P = \sum a_i \s_i$ we define $$\int_P \o := \sum a_i \int_{\s_i} \o.$$  Thus, the integral is linear in both its integrand $\o$ and its domain $P$.

 If $P$ is a polyhedral $k$-chain, define its ``sup'' norm as follows:
$$M_1(P) := \sup_{\o} \frac{\int_P \o}{\|\o\|_0}$$
where $\o$ denotes a $k$-form.  The following result is immediate and says the integral is continuous using the norms $\|\quad\|_0$ and $M_1.$

\begin{proposition}
$$\left|\int_P \o \right| \le \|\o\|_0 M_1(P).$$
\end{proposition}

 \begin{proposition}\label{equalmass}
$M_1(\s) = m(\s).$
\end{proposition}

\begin{proof}
Let $\o_0$ be the $k$-volume form in the $k$-direction of $\s$.  Then  $ \int_{\s} \o_0 = m(\s).$ Hence $M_1(\s) \ge m(\s).$   But $M_1(\s) \le m(\s)$ by the definition of $\|\o\|_0.$
\end{proof}

\subsection*{Definition 2}[Implicit definition of mass]    $M_2$ is the largest seminorm in the space of seminorms $|\cdot|_s$ of polyhedral $k$-chains satisfying $$|\s|_s \le m(\s)$$ for all $k$-parallelepipeds $\s$. 

\subsection*{Definition 3}[Explicit definition of mass]
$$M_3(P) := \inf\left\{\sum |a_i| m(\s_i): P = \sum a_i \s_i\right\}.$$

\begin{proposition}\label{M3}   
$$M_3(P) = \sum_{i=1}^m
|a_i|m(\sigma_i)$$ 
where  $\sum_{i=1}^m a_i\sigma_i$ is  non-overlapping.
\end{proposition}
\begin{proof}   Clearly, $M_3(P)$ is bounded above by the RHS.   If $P = \sum a_i \s_i$ has overlapping cells, any non-overlapping subdivision will have smaller mass.   Finally, observe that any two non-overlapping subdivisions will have the same mass. The result follows. 
\end{proof}

\begin{proposition}\label{massequiv}
All three definitions of mass are equivalent.
\end{proposition}

\begin{proof}
 
$M_1 \le M_2$:  By definition of $\|\o\|_0$ we know
$\int_{\s} \o \le m(\s)\|\o\|_0.$  Hence  $M_1(\s) \le m(\s)$ and therefore $M_1(P) \le M_2(P).$

$M_3 \le M_2$:   This follows immediately from the definition of $M_2$ and Proposition \ref{equalmass} which tells us  $M_3(\s) = m(\s).$ 
 
       $M_3 \le M_1$:   By Proposition \ref{equalmass} 
       $M_3(\s) = m(\s)  = \int_{\s} \o_0 \le M_1(\s).$ By linearity of $M_1$ and $M_3$, $M_3(P) \le M_1(P).$
       
       $M_2 \le M_3$:  Suppose $|\cdot|$ satisfies $|\s| \le m(\s).$   Suppose $P = \sum a_i \s_i$ is non-overlapping.  Then $|P| \le \sum |a_i| |\s_i| \le \sum |a_i| m(\s_i) = M_3(P).$  The result follows since $M_2(P)$ is defined to be the supremum of such norms.  
\end{proof}

We therefore may use the notation $M(P) = M_i(P), i = 1, 2, 3,$ and $M(\s) = m(\s)$.    

\begin{lemma}
Mass is a norm on the space of polyhedral $k$-chains $\cal{P}_k.$
\end{lemma}

\begin{proof} If $\s \ne 0$,  then $M(\s) = m(\s) > 0.$  
Suppose $P \ne 0.$  Write $P = \sum_{i = 0}^j a_i \s_i$ as a non-overlapping decomposition into $k$-cells where  $a_0 \ne 0$ and $\s_0 \ne 0,$  say.  By Proposition \ref{M3} $M(P) = \sum |a_i| M(\s_i) \ge |a_0|M(\s_0) > 0.$ 
\end{proof}

 Let $\cal{N}_k^0$ denote the Banach space obtained upon completion of $\cal{P}_k$ with the mass norm.   Elements of $\cal{N}_k^0$ are called {\itshape \bfseries $k$-chainlets of class $N^0$}.

    \begin{figure}[b]
\begin{center}
\resizebox{3.5in}{!}{\includegraphics*{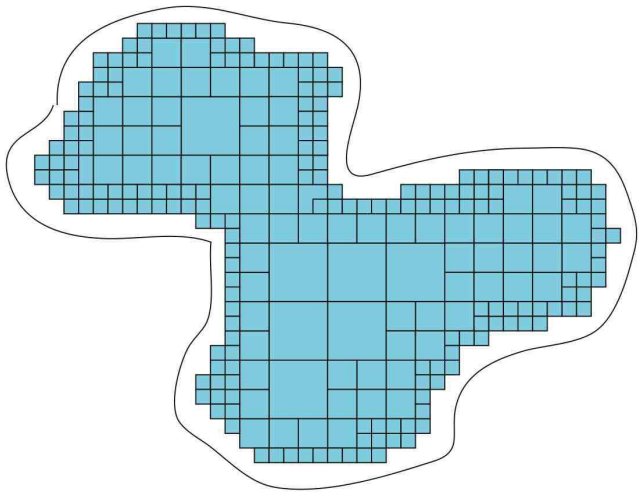}}\label{whitneylabel}
\caption{Whitney decomposition}
\end{center}
\end{figure}

\subsection*{Example --The Whitney decomposition}  Let  $U \subset \R^n$ be bounded and open,   e.g.,  the interior of the Van Koch snowflake.    (See Figure 4.)   Using the Euclidean inner product, Whitney found a  useful way to write $U$ as a union of binary cubes.  Let $k \ge 0.$  Define $\Z/2^k := \{z/2^k: z \in \Z\}.$  A {\itshape \bfseries $k$-binary  cube} in $\R^n$ is a closed cube with edge length $2^{-k}$ and with vertices   from the $k$-binary lattice defined to be $\{\Z/2^{k} \times \cdots \times  \Z/2^{k}\}$ of $\R^n$, where $ k \ge 0$.  We say that a cube $Q$ is {\itshape \bfseries acceptable} to $U$ if $Q$ and all binary cubes meeting $Q$ are contained in $U$.  Let $k_0$ be the smallest integer such that there exists a  $k_0$-binary cube acceptable to $U$.  Let $F_0$ be the union of all such $k_0$-binary cubes.  Let $k_s = s + k_0, s \ge 1, s \in \Z.$  For each $s \ge 1$, let $F_s$ be the union of $k_s$-binary cubes acceptable to $U - (F_1 \cup \cdots \cup F_{s-1}).$  Then $U = \cup F_s.$  Let $P_s$ denote the polyhedral chain obtained from adding the oriented cubes of $F_s$.  It is not hard to prove that $R_j := \sum_{s=0}^j P_s$ forms a Cauchy sequence in the mass norm.   Thus $R_j \to J$ in $ \cal{N}_k^0.$    (See Figure 3.)

   Although mass is a norm on polyhedral $k$-chains, it is limited in scope.    For example, suppose $v \ne 0$ is not in the affine subspace of $\s$.  Then the sequence of cells $T_{2^{-i}v} \s$ does not converge to $\s$ in the mass norm since $M(  T_{2^{-i}v} \s - \s) = 2M(\s)$ for all $i \ge 1.$

 \subsection*{Natural seminorms on polyhedral chains} In these notes, we give three equivalent definitions of the {\itshape \bfseries $r$-natural seminorm} $|\cdot|^{\natural_{r}},$ for all  $r \ge 0,$ on polyhedral $k$-chains that are parallel to the three definitions just given for the mass norm.   The first definition uses differential forms and is developed in  \S \ref{isosection}.  As before, we include it in this section for expository purposes.  
 
 The norm $|\o|^{\natural_{r}}$ of a   form $\o$ is defined to be the supremum of differentials of $\o$ up to order $r.$  (See  \S \ref{discretecalc} for more details.) We say a form is of class $B^{r}$ if $|\o|^{\natural_{r}} < \i.$ 
 
 \subsection*{Definition 1}[Differential forms definition of the natural norms]  For $r \ge 0$, define $$|P|_1^{\natural_{r}}  := \sup\frac{\int_P \o}{|\o|^{\natural_{r}}}.$$     This approach parallels that taken in geometric measure theory (GMT), but has quite different results.  

 The initial difference is that the forms are different at the outset.  The differential forms used in GMT   \cite{derham} to define currents  have compact support.   Our forms have no restriction on their supports and, thus, form a larger space.  It follows that our spaces of chainlets form proper subspaces of currents.   
  It has been our experience that this difference in the choice of forms leads to a significant simplification of geometric measure theory, bringing a geometrical foundation to the theory.

 In these notes we introduce a number of linear operators on chainlets, all of which are bounded in the chainlet spaces.  The Banach spaces of chainlets are not reflexive since they are separable and differential forms are not separable \cite{continuity}.  Thus $\cal{N}_k^{\natural_{r}}$  is a proper subspace of $(\cal{N}_k^{\natural_{r}})^{**}$.  Of course, any space can be viewed as a subset of its second dual.  In particular, each $k$-chainlet can be viewed as a linear operator
 $$\o \mapsto \int_P \o$$ on the space of $k$-forms.  By working in the space of chainlets rather than its double dual, the theory is simplified and expanded in unexpected ways. 

\subsection*{Definition 2}[Implicit definition of the natural norms]   The implicit definition  has the advantage of brevity over the explicit definition and we take this as our starting definition.  The definition is recursive.    Define $|P|_2^{\natural_{0}} := M(P)$ for all polyhedral $k$-chains, $0 \le k \le  n.$  Assume the seminorm $|P|_2^{\natural_{r-1}}$ has been defined.

The {\itshape \bfseries $r$-natural seminorm} $|\cdot|_2^{\natural_{r}}$ on  polyhedral $k$-chains, $0 \le k \le n$, is the largest  seminorm in the space of translation invariant seminorms $| \cdot|_s$ satisfying 
\begin{enumerate}
\item  $|P|_s \le M(P)$  
\item $|P -T_v P|_s \le |v||P|_2^{\natural_{r-1}}$ for all polyhedral  $k$-chains $P$ and all vectors $v \in \R^n$.
\item $|\p Q|_s  \le |Q|_2^{\natural_{r-1}}$ for all polyhedral $(k+1)$-chains $Q$ in $\R^n$. 

\end{enumerate}
 
 Of course,if $k = n$, the last term is vacuous.  By ``largest'' we mean that if $|\cdot|_s$ satisfies the above conditions, then $|P|_s \le |P|_2^{\natural_{r}}$ for all polyhedral $k$-chains $P$.
The trivial seminorm $|P|_s = 0$  satisfies the conditions, so we know this collection of seminorms is not empty.  It is not clear that there is a nonzero seminorm   satisfying the two conditions, nor that the supremum of such seminorms turns out to be a norm\footnote{The $1$-natural norm is equivalent to Whitney's sharp norm \cite{whitney}.  His sharp theory was severely limited as it had no continuous boundary operator and therefore had no possibility of a  Stokes' theorem. Whitney's flat norm had no possibility of a Divergence theorem. Whitney tried hard to reconcile the two norms for over twenty years, but was not successful. The author acknowledges his great classic \cite{whitney} with its numerous, timeless results.}.  

\begin{proposition}\label{decreasing}
$|P|_2^{\natural_{r+1}} \le |P|_2^{\natural_{r}}.$
\end{proposition}
\begin{proof} First observe that $|P|_2^{\natural_{1}} \le |P|_2^{\natural_{0}}.$  Assume $|P|_2^{\natural_{r}} \le |P|_2^{\natural_{r-1}}.$  It follows that $|P|_2^{\natural_{r+1}}$ satisfies the three conditions in the definition of the norm $|\cdot|_2^{\natural_{r}}.$  The result follows since $ |\cdot|_2^{\natural_{r}}$ is the largest seminorm satisfying the three conditions.
 
\end{proof}
The explicit definition\symbolfootnote[1]{  Banach spaces of polyhedral chains supported in a bounded, open set $U \subset \R^n$ can also be defined. This is straightforward, and the interested reader may consult \cite{whitney} where there is a chapter devoted to necessary changes and examples in the flat and sharp norms.  The definitions for the natural norms are parallel, and we postpone this for the time being.}   of the $r$-natural norm is given in the strong topology and is postponed until \S \ref{naturalnorms}.  
The Banach space  obtained upon completion is denoted $\cal{N}_k^{r}$.    Elements of $\cal{N}_k^{r}$ are called {\itshape \bfseries $k$-chainlets of class $N^{r}$}.   
 
\subsection*{Remarks}  
\begin{itemize}
\item  $\cal{N}_k^{r}$ is the smallest Banach space containing the Banach space $\cal{N}_k^0$ with norm satisfying inequalities (2) and (3) above.
\item Norms are different from measures.  For example, measures are scale invariant, whereas norms may not be.    The Banach spaces are scale invariant, though.  Later, we will see there are ``hidden units'' and how the terms scale in the extrinsic definition of the norm.  (See Theorems \ref{scale} and \ref{rho}.)
\item  We rarely need to know the exact value of the natural norm of a chainlet. Comparisons are important to us,  rather than exact values.  Inequalities (1)-(3) above are usually sufficient to provide upper bounds on sequences, letting us establish the Cauchy criterion.
  Lower bounds are found by using the weak version of the natural norm. Usually, a reasonably intelligent choice of polyhedral approximators  will provide suitable upper bounds, and a reasonably intelligent choice of a differential form will provide a decent lower bound. It is recommended to try a few examples such as those given below. 
\end{itemize}

   In   section \S \ref{naturalnorms}, we  construct an explicit seminorm satisfying the above inequalities and show it coincides with the implicit definition given above.      In section \S \ref{integralsofchainlets}, we use the explicit definition to prove that the natural seminorm is a norm. In \S \ref{isosection}, we show the   definition using differential forms is equivalent to the implicit and explicit definitions.

Although we will eventually prove that the natural seminorms are norms, we must treat them as seminorms until then.  A seminorm leads to a normed space as follows.  First complete the space of polyhedral $k$-chains with the seminorms $|\cdot|^{\natural_r}.$    Equate $P \sim P'$ in the completion if $|P-P'|^{\natural_r}=0$.  The equivalence classes, denoted $\widetilde{P},$ form a Banach space $\widetilde{\cal{N}}_k^{\natural_{r}}$ with norm  $|\widetilde{P}|^{\natural_r} := |P|^{\natural_r}$.   Later, we show that $P \sim P' \iff P = P'$  in   ${\cal{N}}_k^{\natural_{r}}$. Hence $P = \widetilde{P} $ and thus ${\cal{N}}_k^{\natural_{r}} = \widetilde{\cal{N}}_k^{\natural_{r}}.$  Members of the Banach space obtained upon completion are called {\itshape \bfseries $k$-chainlets} of class $\widetilde{N}^{\natural_{r}}.$  For simplicity of notation, we omit the tilde in the following discussion.    
  
  \subsection*{Examples}  
\begin{enumerate}
\item The boundary of a bounded, open set $U \subset \R^n.$    We have seen that  the Whitney decomposition of $U$ sums to an $n$-chainlet   $J$ of class $N^0$.      It follows that $\p  J$   is an $(n-1)$-chainlet of class $N^1$.  For example, the Van Koch snowflake curve supports a $1$-chainlet of class $N^1$.  (See Figure 4.) 

 \begin{figure}[b]
\begin{center}
\resizebox{3.5in}{!}{\includegraphics*{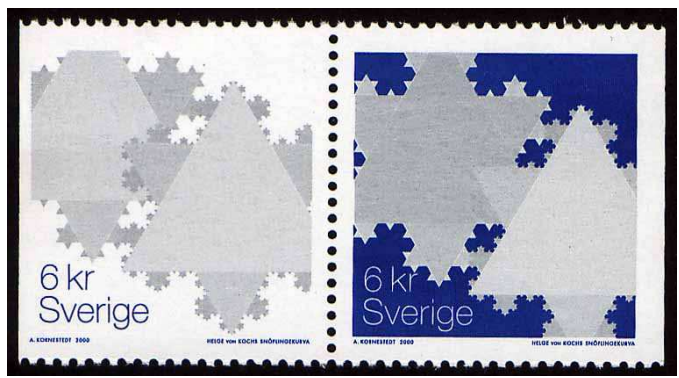}}\label{figsnow}
\caption{The Von Koch snowflake on a stamp}
\end{center}
\end{figure}

\item  In the plane, define a sequence of polyhedral chains $P_i$ as follows:   Let $\s_i$ denote the positively oriented square centered at the origin with edge $2^{-i}.$  Let $P_i = 2^{2i} \s_i.$   Then the $P_i$ form a Cauchy sequence in the $1$-natural norm.  Hence the boundaries $\p P_i$ also form a Cauchy sequence in the $2$-natural norm.

\item  Let $f:[a,b] \to \R$ be a nonnegative  continuous function.    The {\itshape \bfseries subgraph} of $f$ is the region of the plane between the graph of $f$ and the $x$-axis.    Using the Whitney decomposition and subtracting off the $x$-axis chainlet and intervals connecting $a$ to $f(a)$ and $b$ to $f(b)$ (suitably oriented),   we can easily see that the graph of $f$ supports\footnote{In the next chapter, we will define the {\itshape \bfseries support} of a chainlet.  For now, we use the term only for examples and loosely describe it as the set ``where the chainlet lives''.  } a chainlet of class $N^1$.  

\item We can use this to define a {\itshape \bfseries $k$-submanifold} of Euclidean space $\R^n.$    A $k$-submanifold is the sum of $k$-chainlet graphs of functions defined over bounded, open domains contained in $k$-dimensional affine subspaces.  Later, we will see that basic results of calculus are valid for all $k$-chainlets and thus for all $k$-submanifolds, even those built from such rough functions as the Weierstrass nowhere differentiable function.  (See Figure 5.)


 \begin{figure}[t]
\begin{center}
\resizebox{3.5in}{!}{\includegraphics*{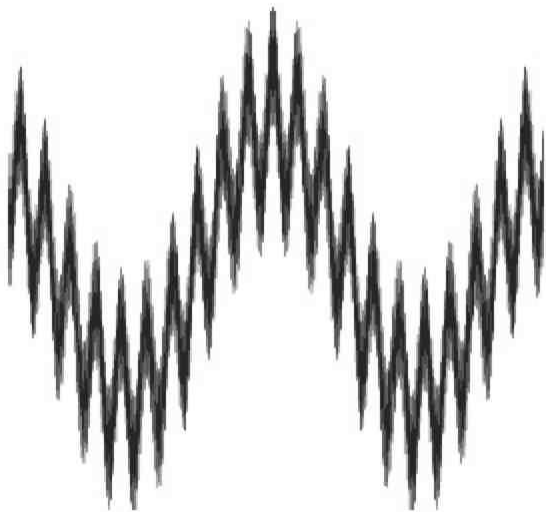}}\label{fignonsmooth}
\caption{The Weierstrass nowhere differentiable function}
\end{center}
\end{figure}

\item Suppose $f:[a,b] \to \R$ is Lebesgue integrable and  nonnegative.   This is equivalent to saying that $f$ is the limit of an increasing sequence of simple, step functions $g_i \to f$ in the $L^1$ norm. $|f|_{L^1} = |\int_a^b f|.$   Each step function $g_i$ naturally detemines a polyhedral chain $P_i$ supported in its graph.  It is a straightforward exercise to show that the $P_i$ form a Cauchy sequence in the $1$-natural norm.   Thus $P_i \to J$ in $\cal{N}_1^{1}$ and the graph of $f$ supports $J$.  For smooth functions, this process does not yield the same chainlet as the previous method, but a chainlet we can think of as  the x-component of $J$.\footnote{Components of chainlets are described in \cite{continuity}.}

\item We may   extend the idea of a submanifold considerably by using graphs of $L^1$ functions for local stuctures, rather than graphs of continuous functions.  We may even define abstract versions of these by   requiring that overlap maps be diffeomorphisms where the local charts are chainlets obtained from $L^1$ functions. 
\end{enumerate}

\newpage 
 \section{Explicit definition of the natural norms}\label{naturalnorms}

 \subsection*{Difference chains}       We introduce a notion of the  derivative of a geometric object such as a $k$-cell, by first considering geometric divided differences.

 Let $U^j := (u_1, \dots, u_j)$ denote a list of vectors in  $\R^n, j \ge 1.$   Set $U^0 := \emptyset.$ 

   Let $\s$ be a $k$-cell in $\R^n$.  
 
   For $j=1$, define the {\itshape \bfseries $1$-difference $k$-cell} 
   $$ \s^1 = \Delta_u \s  :=   T_{u}\s - \s.$$   
   
 \begin{figure}[b]
\begin{center}
\resizebox{3.5in}{!}{\includegraphics*{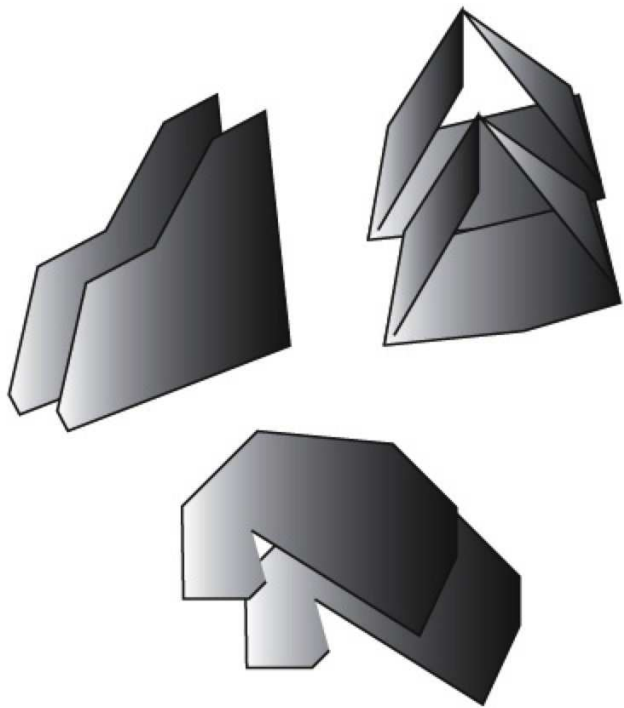}}\label{figdiff}
\caption{A $1$-difference $2$-chain}
\end{center}
\end{figure}

This is a chain consisting of two cells, oppositely oriented.  A simple example is the sum of the opposite faces of a cube, oppositely oriented.  The chain is supported in the two faces.  
  For an algebraic chain $S = \sum a_i \s_i$, extend linearly to define 
   $$ \Delta_u S  :=  \sum a_i \D_u \s_i.$$

Now suppose  $j > 1$.  Let $U^j = (u_1, \dots, u_j)$ denote a list of $j$-vectors and let $U^{j-1} = (u_1, \dots, u_{j-1}).$  Define the {\itshape \bfseries $j$-difference $k$-cell}\footnote{This terminology is distinguished from the $j$-differential $k$-cells  that we will define later which are geometric directional derivatives, and thus come from infinitesimal translations.} inductively by
$$\s^j = \D_{U^j} \s := \D_{u_j} (\D_{U^{j-1}} \s).$$
 Typically, $\D_{U^j} \s$ consists of $2^{j}$ cells, although some overlapping cancellation is possible. The integer $k$ is the {\itshape \bfseries dimension} of the difference cell, $j$ is its {\itshape \bfseries order}.

 A {\itshape \bfseries $j$-difference $k$-chain} $D^j$  in $\R^n$ is defined to be a polyhedral chain of $j$-difference $k$-cells,
 
 $$D^j = \sum_{i=1}^m a_i \s_i^j.$$ 
 
with coefficients $a_i \in \mathbb{F}.$   Observe that the  vectors  $(U^j)_i$ determining $\s_i^j$ may vary with $i$.  (See Figure 6.)  
The vector space of    $j$-difference $k$-chains is denoted $\cal{D}_k^j$.  It is a subspace of polyhedral $k$-chains $\cal{P}_k.$ 

  Given a polyhedral $k$-chain $P$ and a vector $u \in \R^n$ define 
the  {\itshape \bfseries $1$-difference $k$-chain} as
$$ \D_u P :=    T_u P - P .$$  

Given a list $U^j = (u_1, \dots , u_j)$ let $ U^{j-1} = (u_1, \dots, u_{j-1})$ and define  
$$ \D_{U^j} P :=  \D_{u_j}(\D_{U^{j-1}} P).$$  

\begin{lemma} The difference chain $\D_{U^j}P$ is independent of the order of vectors in the list $U^j$.
\end{lemma}

\begin{proof}
This follows since translation operators commute:  $T_u \circ T_w = T_w \circ T_u.$
\end{proof}

\subsection*{Difference norms}  
The {\itshape \bfseries $j$-difference norm} of $\s^j = \D_{U^j} \s$ is defined by $$\|\s^j\|_j := M(\s) |u_1| \cdots |u_j|.$$

Define the {\itshape \bfseries $j$-difference norm} of a $j$-difference $k$-chain $D^j $  as $$\|D^j\|_j :=  \inf\left\{ \sum_{i=1}^m |a_i|\|\s^j_i\|_j: D^j =  \sum_{i=1}^m a_i  \s_i^j, a_i \in \mathbb{F}  \right\}.$$

    \begin{lemma}
The $j$-difference norm $\|\cdot\|_j$ is a norm on $\cal{D}_k^j$.
\end{lemma}

The proof that this is a seminorm is left as an exercise.    We show it is a norm   in $ \ref{normj}.$   
 \subsection*{Explicit definition of the  natural norms}     Let $P \in \cal{P}_n$ be a polyhedral $n$-chain.  Define $$|P|_3^{\natural_{r}} := \inf\left\{\sum_{j=0}^{r} \|D^j\|_j: P = \sum_{j=0}^{r} D^j, D^j \in \cal{D}_n^j\right\}.$$  For $0 \le k < n$, the definition is recursive.  Assume  $|P|_3^{\natural_{r}}$ has been defined for polyhedral $(k+1)$-chains $P$.
 
 Let $P \in \cal{P}_k$ be a polyhedral $k$-chain.
 Define $$|P|_3^{\natural_{r}} := \inf \left\{\sum_{j=0}^{r} \|D^j\|_j + |B|_3^{\natural_{r-1}}: P = \sum_{j=0}^{r} D^j + \p B, D^j \in \cal{D}_k^j, B  \in \cal{P}_{k+1} \right\}.$$    There is at least one decomposition, a trivial one, where $D^0 = P$ and where $B$ and the other $D^j$ are zero chains.

There is a straightforward proof available in \cite{hodge} that $|\cdot|_3^{\natural_{r}}$ is a norm that makes use of  the classical   Stokes' theorem for polyhedral chains. (This is essentially the proof in Theorem \ref{oldintegral}.)  However, in these notes, we wish to define forms and develop the calculus from first principles and cannot make use of the exterior calculus until we rebuild it.

\begin{lemma}\label{translatelemma}
$$|P- T_vP|_3^{\natural_{r+1}} \le |v||P|_3^{\natural_{r}}.$$
\end{lemma}

\begin{proof}  Let $\e > 0.$  There exists $r \ge 0$ and a decomposition $P = \sum_{j=0}^r D^j + \p B$ such that 
$$|P|_3^{\natural_{r}} > \sum_{j=0}^r \|D^j\|_j + |B|_3^{\natural_{r-1}} - \e.$$  Then $P-T_vP = \sum_{j=0}^r E^{j+1} + \p (B- T_v  B)$   where $E^{j+1} = D^j - T_v D^j .$  Observe that $\|E^{j+1}\|_{j+1} \le |v|\|D^j\|_j$ by definition of the difference norm.
The proof proceeds by induction.  Assume $P$ is $n$-dimensional.  Then $B = 0.$  Hence 
$$|P -T_v P|_3^{\natural_{r+1}} \le \sum_{j=0}^r \|E^{j+1}\|_{j+1}  \le |v| \sum_{j=0}^r\|D^j\|_j < |v|(|P|_3^{\natural_{r}} + \e) .$$   Since the result holds for all $\e > 0$ we have
$$|P -T_v P|_3^{\natural_{r+1}} \le |v||P|_3^{\natural_{r}}$$ for all polyhedral $n$-chains.

Now assume the result holds for $(k+1)$-chains.  Then if $P$ is a $k$-chain we have
$$\begin{aligned} |P -T_v P|_3^{\natural_{r+1}} &\le \sum_{j=0}^r \|E^{j+1}\|_{j+1} + |B-T_vB|_3^{\natural_{r}}  \\&\le 
 |v|\sum_{j=0}^r\|D^j\|_j + |B|_3^{\natural_{r-1}} \\&  <|v|( |P|_3^{\natural_{r}}+ \e) .\end{aligned}$$  The lemma follows since this result holds for all $\e > 0.$  
\end{proof}

\begin{proposition}\label{natequal}
$|P|_3^{\natural_{r}} = |P|_2^{\natural_{r}}.$
\end{proposition}

\begin{proof}
It suffices to show that
 $|\cdot|_3^{\natural_{r}}$ is the largest seminorm in the class of translation invariant seminorms $|\cdot|_s$ satisfying 
\begin{enumerate}
\item $|P|_s  \le M(P)$.  
\item $|P-T_vP|_s \le |v||P|_3^{\natural_{r-1}}$ for all polyhedral $k$-chains $P$ and $v \in \R^n$.
 \item $|\p Q|_s  \le |Q|_3^{\natural_{r-1}} $ for all polyhedral $(k+1)$-chains $Q$, if $k < n.$
\end{enumerate}

Clearly, $|\cdot|_3^{\natural_{r}}$ is translation invariant.  Part (2) follows from Lemma \ref{translatelemma}.  Parts (1) and (3) are immediate from the definition.  Hence $|\cdot|_3^{\natural_{r}} \le |\cdot|_2^{\natural_{r}}.$

Properties (1) and (2) imply that $|D^j| \le \|D^j\|_j$ for all difference chains $D^j$.
 Let $P$ be a polyhedral chain.   
Given $\e > 0$ there exists $r \ge 0$ and $P = \sum_{j=0}^r D^j + \p B $ such that 
$$|P|_3^{\natural_{r}} > \sum_{j=0}^r \|D^j\|_j + |B|^{{r-1}} - \e.$$  If $k = n$, then the same inequality holds, except   $B = 0$.    In this case we have $$|P|_s \le \sum_{j=0}^r |D^j| \le \sum_{j=0}^r \|D^j\|_j < |P|_3^{\natural_{r}} + \e.$$  Since the result holds for all $\e > 0$ we conclude $|P|_s \le |P|_3^{\natural_{r}}.$
 Now assume the result holds for polyhedral $k$-chains $B$, we prove it holds for polyhedral $(k-1)$-chains $P$.  

 Thus by induction and (2)
$$
\begin{aligned} |P|_s &\le \sum_{j=0}^r |D^j|  + |\p B|  \le  \sum_{j=0}^r \|D^j\|_j + | B|  \\&\le 
\sum_{j=0}^r \|D^j\|_j + |B|_3^{\natural_{r-1}}  < |P|_3^{\natural_{r}} +\e.
\end{aligned}
$$

It follows that  $|\cdot|_3^{\natural_{r}} \ge |\cdot|_2^{\natural_{r}}.$

\end{proof}

 Henceforth, we use the  notation $|\cdot|^{\natural_{r}}$ instead of $|\cdot|_i^{\natural_{r}}, i = 2, 3.$    \\

   The next result follows from the definition and by taking limits in the Banach space $\cal{N}_k^r.$
\begin{lemma}\label{basicinequalities}  If $J$ is a   $k$-chainlet of class $N^r$, $r \ge 0$,  then 
\begin{enumerate} 
\item $\p: \cal{N}_k^r \to \cal{N}_{k-1}^{r+1}$ with $|\p J|^{\natural_{r+1}} \le |J|^{\natural_{r}}$
\item $|J -T_v J|^{\natural_{r+1}} \le |v||J|^{\natural_{r}}.$
\end{enumerate}
\end{lemma}

By Proposition \ref{decreasing}   the norms $|P|^{\natural_{r}}$ are decreasing as $r \to \i.$  For completion, we consider the limit norm.   The norm $|P|^{\natural_{\i}}$ is defined as the largest seminorm in the space of seminorms $|\quad|_s$
satisfying 
\begin{enumerate}
\item $|P|_s \le M(P)$
\item $|T_vP -P|_s \le |P|$ for all polyhedral $k$-chains $P$ and $v \in \R^n$
\item $|\p Q|_s \le |Q|_s$ for all polyhedral $(k+1)$-chains $Q$ in $\R^n$.
\end{enumerate}

\begin{proposition} Let $P$ be a polyhedral  $k$-chain.  Then $$|P|^{\natural_{\i}} = \lim_{r \to \i} |P|^{\natural_{r}}.$$
\end{proposition}    

The proof follows readily  from the definitions.  We see later that $|\cdot|^{\natural_{\i}}$ is a norm.   The Banach space obtained is denoted $\cal{N}_k^{\i}$.  This norm is of less interest to us as it is not  scale invariant.   
  
If $J^0 \in \cal{N}_k^r$ and $U^j = (u_1, \dots, u_j)$ let  $J^j := \D_{U^j} J^0$ denote the {\itshape \bfseries difference chainlet} (defined in the same way as difference cells.)  If $J^0$ has finite mass, define 
$$\|J^j\|_j :=   |J^0|^{\natural_{0}}|u_1| \cdots |u|_j .$$   It follows from Lemma \ref{basicinequalities} that $|J^j|^{\natural_{j}} \le \|J^j\|_j.$

The next corollary permits decompositions of chainlets into difference chainlets plus a boundary to estimate the natural norms.  
  \begin{corollary}\label{natnorms}  $$|J|^{\natural_r} = \inf\left\{\sum_{j=0}^r \|J^j\|_j + |K|^{{\natural_{r-1}}}: J = \sum_{j=0}^r J^j + \p K, J^j  \in \cal{N}_k^r, K  \in \cal{N}_{k+1}^{r-1}\right \}.$$  
\end{corollary}
\begin{proof}
Denote the RHS by $| J|_{<r>}$.  This is clearly a seminorm.   By the implicit definition of the $r$-natural norm, we know that $  |J|_{<r>} \le |J|^{\natural_r} .$  

On the other hand, $|J^j|^{\natural_r} \le \|J^j\|_j$ for difference chainlets $J^j$.   Let $\e > 0$.  There exists $J = \sum J^j + \p K,  J^j  \in \cal{N}_k^r, K \in   \cal{N}_{k+1}^{r-1}$ such that $$|J|_{<r>} > 
 \sum_{j=0}^r \|J^j\|_j + |K|^{{\natural_{r-1}}} - \e.$$  
 Then $$\begin{aligned} |J|^{\natural_r} &\le \sum_{j=0}^r |J^j|^{\natural_r} + |K|^{{\natural_{r-1}}} 
   \le \sum_{j=0}^r \|J^j\|_j + |K|^{{\natural_{r-1}}}   < |J|_{<r>} + \e. \end{aligned}$$  It follows that $  |J|_{<r>} \ge |J|^{\natural_r} .$  
 
 \end{proof}
 
     \subsection*{Directional derivative $\nabla_u$ of a chainlet }   
    Given a chainlet $J = \lim_{i \to \i} P_i$ and a vector $u \in \R^n$, define the {\itshape \bfseries translation} of $J$ through $u$ by $T_u J:= \lim_{i \to \i}  T_u P_i$.  Since $|T_uP|^{\natural_{r}} = |P|^{\natural_{r}}$ it follows that  $T_uJ$ is well defined and satisfies $$|T_uJ|^{\natural_{r}} = |J|^{\natural_{r}}.$$  Although translation is not continuous in the mass norm, it is continuous in the $r$-natural norm. It follows from Lemma \ref{basicinequalities}(2) that, for $r > 0$, $$|T_{hu}J - J|^{\natural_{r}} \to 0 \mbox{ as } h \to 0.$$

         If $J$ is a $k$-chainlet $J$ of class $N^r$ and  $u \in \R^n,$ define the $1$-difference $k$-chainlet $ \D_u J =    T_u J - J .$  If $U^j = (u_1, \dots, u_j)$ and $U^{j-1} = (u_1, \dots, u_{j-1})$, recursively define a $j$-difference $k$-chainlet $$\D_{U^j} J := \D_{u_j} \D_{U^{j-1}}J.$$ Observe that $\D_{U^j} J$ is independent of the order of vectors in the list $U^j$.
         \begin{theorem} \label{difference}  If $J \in \cal{N}_k^r$, then
$$\left| \D_{U^j} J\right|^{\natural_{r+j}} \le   |u_1|\cdots |u_j||J|^{\natural_{r}}.$$
\end{theorem}

\begin{proof}   By Lemma \ref{basicinequalities}(2) we have
$$\left| \D_{U^1} J\right|^{\natural_{r+1}}
= \left| T_{u_{1}}J -J \right|^{\natural_{r+1}} \le   |u_1||J|^{\natural_{r}}.$$  The general result follows by induction.  

 \end{proof}

          Define
   the {\itshape \bfseries $1$-differential chainlet}   $$\nabla_uJ := \lim_{h \to 0^+} \frac{\Delta_{hu} J}{ h}.$$  The next proposition shows this limit exists.
 The operator $\nabla_u$ is the {\itshape \bfseries directional derivative} in the direction $u$.  
 \begin{proposition}
 If $J$ is a $k$-chainlet of class $N^{r}$, then $\nabla_uJ$ is uniquely defined as a $k$-chainlet of class $N^{r+1}.$  Furthermore,
 $$|\nabla_u J|^{\natural_{r+1}} \le |u||J|^{\natural_{r}}.$$
 \end{proposition}

 \begin{proof}  We show that if $P$ is a polyhedral $k$-chain, then $ \frac{\Delta_{hu} P}{ h}$ forms a Cauchy sequence.
 Let $v_{\ell} = 2^{\ell} 2^{-i-j}u, 0 \le \ell \le 2^j.$  
Using the triangle inequality, we have
$$\begin{aligned}\left|\frac{T_{2^{-i-j}u} P -P}{2^{-i-j}} - \frac{T_{2^{-i}u}P -P}{2^{-i}}\right|^{\natural_{r+1}}  \\  &\hspace{-1in}\le 
 \sum_{\ell = 1}^{2^j -1}\left|\frac{(T_{2^{-i-j}u} P - P)}{2^{-i}} - T_{v_{\ell}} \frac{(T_{2^{-i-j}u} P - P)}{2^{-i}}\right|^{\natural_{r+1}}   \\&\hspace{-1in}\le
2^{{i+j}} 2^{-i}|u| |T_{2^{-i-j}u} P - P|^{\natural_{r}} \\&\hspace{-1in}\le
 2^{-i}|u|^2|P|^{\natural_{r-1}}.
\end{aligned}$$ 
It follows that $\frac{T_{2^{-i}u}P -P}{2^{-i}}$ forms a Cauchy sequence  in the $(r+1)$-natural norm.    Thus its limit $\nabla_u P$ exists as a chainlet of class $N^{r+1}.$

From the definition of the $r$-natural norm
$$|\nabla_u P|^{\natural_{r+1}} = \lim \frac{|\Delta_{hu} P|^{\natural_{r+1}}}{h}  = \lim \frac{|T_{hu} P -P|^{\natural_{r+1}}}{h}  \le |u||P|^{\natural_{r}}.$$   Now suppose $J$ is a $k$-chainlet of class $N^r.$  Let $P_i \buildrel \natural_r \over \to J$.  Since $|P_i|^{\natural_{r}}$ forms a Cauchy sequence, so does  $|\nabla_u P|^{\natural_{r+1}}.$  Then $$\nabla_u J = \lim_{i \to \i} \nabla_u P_i.$$ and $$|\nabla_u J|^{\natural_{r+1}} \le |u||J|^{\natural_{r}}.$$

  \end{proof}

We recursively define {\itshape \bfseries $j$-differential   $k$-chainlets} by taking geometric differentials of $(j-1)$-differential $k$-chainlets. Let $$U^j = (u_1, \dots, u_j),   j \ge 1,$$ denote a list of vectors.  Setting $ {U}^{j-1} := (u_1, \dots, u_{j-1})$ define
$$\nabla_{U^j} J := \nabla_{u_j} \nabla_{U^{j-1}} J.$$
 
\begin{proposition} If $J$ is a $k$-chainlet of class $N^r$, then $\nabla_{U^j} J$ is a $k$-chainlet of class $N^{r+j}$ with
$$|\nabla_{U^j} J|^{\natural_{r+j}} \le |u_1| \cdots |u_j||J|^{\natural_{r}}.$$
\end{proposition}

  \newpage
 \section{Differential elements} 
 \begin{verse}
{\em And what are these fluxions? The velocities of evanescent increments?  They are neither finite quantities, nor quantities infinitely small, nor yet nothing. May we not call them ghosts of departed quantities}
\end{verse}
\begin{flushright} -- Bishop George Berkeley \end{flushright}

      A {\itshape \bfseries homothetic replica} of $X \subset \R^n$ is defined to be a linear contraction or expansion of $X$ followed by a translation. By taking smaller and smaller homothetic replicas, we shrink a parallelogram to a point and find the limit in our Banach space of chainlets.
 
 Recall that $\s(V^k)$ denotes the parallelepiped determined by the vectors in the list $V^k = (v_1, \dots, v_k).$
  \begin{lemma} The 
sequence $2^{ki} \s(2^{-i}V^k) $ is Cauchy in the $1$-natural seminorm.  
\end{lemma}
\begin{proof}   First observe that $  M(2^{ki} \s(2^{-i}V^k)) = M(\s(V^k))$ for all $i \ge 0.$   Let $Q_i = \s(2^{-i}V^k).$ 
We use $1$-difference cells to estimate the difference between  the $i $ and $i+\ell$ terms $$|2^{ki}Q_i - 2^{k(i+\ell)}Q_{i + \ell}|^{\natural_{1}}.$$
 
    Notice that $Q_i$ can be subdivided into $2^{k\ell}$ cubes $Q_i = \sum R_{i,j}$  each a homothetic replica of $Q_{i+ \ell}.$    When we also consider the multiplicities we can form pairs  $$2^{ki}Q_i - 2^{k(i+\ell)}Q_{i + \ell} = \sum_{j=1}^{2^{k(i+\ell)}} R_{i,j} - Q_{i+\ell}.$$  Therefore
$$|2^{ki}Q_i - 2^{k(i+\ell)}Q_{i + \ell}|^{\natural_{1}} \le  \sum_{j=1}^{2^{k(i+\ell)}} \|R_{i,j} - Q_{i+\ell}\|_1.$$  The distance of translation between each pair is bounded by the diameter of $Q_i$ which is bounded by a  constant $C$ times $2^{-i}.$  Therefore the RHS is bounded by 
$$C 2^{k(i+\ell)} 2^{-i} M( Q_{i+\ell}) \le C   2^{-i}.$$  This tends to $0$ as $i \to \i.$  
\end{proof}

We call the limit a {\itshape \bfseries simple $k$-element}\footnote{This is essentially the same as a {\itshape \bfseries geometric monopole} of Dirac, a {\itshape \bfseries point measure} and   an {\itshape \bfseries atom} of Whitney's sharp norm.} and denotes it by $$\a(V^k).$$  Its {\itshape \bfseries support}   $$supp(\a(V^k)):= 0.$$  Its {\itshape \bfseries mass} is defined by $$M(\a(V^k)) := M(\s(V^k)).$$      Later, we will define mass and support   for all chainlets and prove that these definitions coincide.

Its {\itshape \bfseries $k$-direction}  is defined to be the oriented linear subspace containing $\s(V^k)$ with the same orientation as that determined by the  list $V^k.$  

A chain of simple $k$-elements is called a {\itshape \bfseries $k$-element}.  We denote a $k$-element by $\a$, as well. This terminology is analogous, of course, to   simple $k$-vectors and $k$-vectors.  But the analogy only goes so far.  We will be able to define operators on $k$-elements, which is not possible for $k$-vectors.  For example, we may translate a $k$-element to any point in $\R^n.$

Let $p \in \R^n$ and $\a$ be a $k$-element.  A {\itshape \bfseries $k$-element supported at $p$}  is defined by  $$\a_p  := T_p \a .$$    
We form chains of   $k$-elements in the Banach space of chainlets as below, $$A = \sum a_i \a_{p_i}.$$
A subset $X$ of a Banach space $Y$ is said to be {\itshape \bfseries dense} in $Y$ if every open set $U$ in $Y$ contains some point of $X$.  
\begin{theorem}\label{approx}
Chains of simple $k$-elements  are dense in the Banach spaces $\cal{N}_k^r, r \ge 1$.  
\end{theorem}  
\begin{proof}  
We first show that any $k$-cell $\s$  may be approximated by chains of simple $k$-elements.  First subdivide $\s$ into its Whitney decomposition and write $\s = \lim W_j$ in the mass norm, where the $W_j$ are chains of cubes of diameter $\ge 2^{-j}$.  (See Figure 3.) For each $W_j$ subdivide all its cubes  into binary cubes of the same edge length of order $2^{-j}.$  Denote the sum of these smaller cubes by $P_j = \sum_i Q_{j,i}.$  Since $P_j$ is a subdivision of $W_j$ it follows that $P_j \to \s$ in the mass norm.   Fix $j$ and choose a point $p_{j,i} \in Q_{j,i}$.  Consider the simple $k$-element $  \a_{ p_{j,i}}(Q_{j,i})$.  It has mass the same as the mass of $Q_{j,i}$, the same $k$-direction as $\s$, and is supported in $p_{j,i}.$   Write $E_j = \sum \a_{ p_{j,i}}(Q_{j,i}).$   It follows from using the difference chain portion of the $1$-natural norm that  $|P_j - E_j|^{\natural_{1}} \le  M(P_j) C 2^{-j}$.   By the triangle inequality  $$|\s -  E_j|^{\natural_{r}} \le |\s - P_j|^{\natural_{r}} + |P_j - E_j|^{\natural_{r}}.$$  But the last two terms are small if $j$ is large since $$|\s  - P_j|^{\natural_{r}}  \le |\s  - P_j|^{\natural_{0}}$$ and 
$$|P_j - E_j|^{\natural_{r}} \le |P_j - E_j|^{\natural_{1}}.$$
\end{proof}

 We now show that we may use any $k$-cell $\s$  to define a simple $k$-element, not just a $k$-cube.  Let $\phi_j:\R^n \to \R^n$ denote the linear contraction by a factor of  $2^{-j}.$   Let $\s_j = 2^{-kj}\phi_{j*} \s.$  We show the sequence $  \s_j$ is Cauchy in the $1$-natural norm.   Approximate $\s$ by a simple $k$-element chain $E = \sum \a_{\ell}.$  Then $E_j = \sum 2^{-kj}\phi_{j*} E$ approximates  $\s_j$.  Therefore, by the triangle inequality, 
 
     $$|\s_j - \s_{j +\ell}|^{\natural_{1}} \le |\s_j - E_{j}|^{\natural_{1}} + |E_{j} - E_{j+\ell}|^{\natural_{1}} + |E_{j+\ell} - \s_{j+\ell}|^{\natural_{1}},$$ each term of which is small.    Define $$\a(\s)  := \lim_{j \to \i} \s_j$$  
and $$\a_p(\s) = T_p(\a(\s)).$$

\begin{corollary}
$\a(\s) = \a(\s')$ if and only if the $k$-directions and masses  of $k$-cells $\s$ and $\s'$ are the same.  
\end{corollary}

\begin{proof}
Suppose $\s$ and $\s'$ are two $k$-cells with the same $k$-direction and mass.   
Approximate $\a(\s)$ and $\a(\s')$ with   simple $k$-element chains $A$ and $A'$, respectively, with   $supp(A) \cup supp(A') \subset B_{\e/2}(0).$    Since $\a(\s)$ and $\a(\s')$ have the same mass we may assume $M(A) = M(A').$   The difference $$|A -A'|^{\natural_{1}} \le \sum \|\a_i - \a_i'\|_1 \le M(A)\e.$$
Therefore $$|\a(\s) - \a(\s')|^{\natural_{1}} \le |\a(\s) - A|^{\natural_{1}} + |A-A'|^{\natural_{1}} + |A' - \a(\s')|^{\natural_{1}}.$$   It follows that $\a(\s) = \a(\s').$  
  
Conversely, assume $\a(\s) = \a(\s').$   Let $\s(V^k)$ and $\s((V^k)')$  be cubes with the same mass and direction as $\s$ and $\s'.$   By the first part we know $\a(\s) = \a(\s(V^k))$ and $\a(\s') = \a(\s((V^k)')).$  Thus
$$\begin{aligned} M(\s) = M(\s(V^k)) &= M(\a(\s(V^k))) = M\a(\s((V^k)')) \\&= M(\s((V^k)')) = M(\s').\end{aligned}$$
 
 The proof is similar for the $k$-direction and we omit it.
 
 \end{proof}
 
 Define $${  Vec(\s)}:= \a(\s).$$  By the Corollary above, $Vec(\s)$ depends only on the $k$-direction and mass of $\s$.

\begin{lemma}\label{sums}
If $\s = \sum \s_i$ is a non-overlapping subdivision, then $$Vec(\s) = \sum Vec(\s_i).$$
\end{lemma}

\begin{proof}
This follows since the $k$-directions of the cells $\s_i$ are the same as the $k$-direction of $\s$ and $M(\s) = \sum M(\s_i).$  
\end{proof}

   If   $S = \sum a_i \s_i$ is an algebraic $k$-chain, define  $${ Vec(S)} := \sum a_i Vec(\s_i).$$   It follows that $Vec(S) \in \cal{N}_k^1.$
   
   The operator $Vec$ and its boundary $\p Vec$ will form the foundation for differentiation, integration and measure.

Example:  Let $\s = \sum_{i=1}^{\i} \s_i$ be the Whitney decomposition of  a $k$-cell $\s$.  Then $Vec(\s) = \sum_{i=1}^{\i} Vec(\s_i) \in \cal{N}_k^1.$

\begin{lemma}
If $S \sim S'$ are algebraic $k$-chains, then $Vec(S) = Vec(S').$
\end{lemma}
 
\begin{proof}
The proof reduces to assuming that $S'$ is a subdivision of $S$.  It further reduces to assuming that $S$ is a cell $\s$ and $S'$ subdivides $\s$.   The result follows from Lemma \ref{sums}. \end{proof}

 As  a corollary, we know that $Vec(P)$ is well defined for  polyhedral chains $P$.  
 
 \begin{proposition}\label{Veczero}
If $P$ is a polyhedral $k$-chain, then $Vec(\p P) = 0.$
\end{proposition}

\begin{proof}
This follows since $Vec(\p \s) = 0$ for every $k$-cell $\s$.
\end{proof}

Later in these notes,  we will prove that $Vec(J)$, $M(J)$ and $supp(J)$  are well defined for all chainlets $J$ of class $N^1$  and extend the previous definitions for polyhedral $k$-chains and $k$-elements. Moreover, if $Vec(J)$ is a $k$-element $\a$ with $M(J) = M(\a)$ and $supp(J) = supp(\a)$, then $J = \a.$  (See Theorem \ref{point}.)
 
 Recall $\phi_j$ denotes linear contraction by a factor of  $2^{-j}.$   Let $P$ be a polyhedral $k$-chain with compact support.  Let $P_j = 2^{kj}\phi_j(P)$.

\begin{theorem}  $$\lim P_j = Vec(P).$$ \end{theorem}

This implies that the shrinking annulus, for example, limits to the same $2$-element as the shrinking square.  Their boundaries limit to the same boundary in the chainlet norm.   So it makes no sense to speak of a particular shape of space at the smallest scales.  The standard model  for smooth $k$-manifolds, using a $k$-tangent space is correct, but misleading, as it is not the only way to view a local structure.   As we will see, it is also more limited than our $k$-element models.  
 
 \subsection*{Vector spaces of $k$-elements}
  
 \begin{proposition} \quad
\begin{enumerate}
\item $\a(v_1, \dots, tv_i, \dots, v_k) = t \a(v_1, \dots, v_i, \dots, v_k)$;
\item $\a(v_1, \dots,   v_i, v_{i+1},   \dots, v_k) = - \a(v_1, \dots,    v_{i+1},v_i,  \dots, v_k);$
\item $\a(v_1, \dots, v_i +v_i', \dots, v_k)=\a(v_1, \dots, v_i, \dots, v_k)\\+\a(v_1, \dots, v_i', \dots, v_k)$. 
\end{enumerate}
\end{proposition}

\begin{proof} For the first part, observe that   the $k$-parallelepiped   formed by  vectors $(v_1, \dots, tv_i, \dots, v_k)$ has the same orientation and direction, and also has $t$ times the mass of the $k$-parallelepiped formed by $(v_1, \cdots, v_k).$    The second part is found by considering the determinant of the linear transformation which changes the order of the bases.

For the last part,  assume $k = 1$ and $i = 1.$  Recall that $\s(v_1, \dots, v_k)$ denotes the $k$-cell that is the parallelepiped formed by the vectors $(v_1, \dots, v_k).$ By the triangle inequality
$$\begin{aligned} |\s(v_1 + v_1') - \s(v_1) - \s(v_1')|^{\natural_{1}} &\le |\s(v_1 + v_1') - T_{v_1} \s(v_1') - \s(v_1)|^{\natural_{1}} \\&+ |T_{v_1} \s(v_1') - \s(v_1') |^{\natural_{1}}.\end{aligned}$$        

There exists a $2$-cell $\t$ such that $\s(v_1 + v_1') - T_{v_1} \s(v_1') - \s(v_1) = \p \t.$ Therefore, the definition of the $1$-natural norm yields  $$|\s(v_1 + v_1') - T_{v_1} \s(v_1') - \s(v_1)|^{\natural_{1}}  \le  M(\t).$$
The second term $ T_{v_1} \s(v_1') - \s(v_1') $ is a $1$-difference cell and thus 
$$|T_{v_1} \s(v_1') - \s(v_1') |^{\natural_{1}} \le \|T_{v_1} \s(v_1') - \s(v_1')\|_1.$$

We are now ready to shrink the chains.  We take mass renormalized homothetic replicas of each $1$-cell, with diameters reduced by $2^{-k}$  and obtain similar inequalities for each $k$.     Let $\t_k$ denote the masss renormalized version of $\t$.  Then $2^kM(\t_k)  \le 2^kM(\t) 2^{-2k} = M(\t)2^{-k}.$
Let $w_k = 2^{-k} v_1, w_k' = 2^{-k} v_1'.$  Then
   $$\begin{aligned} \|T_{w_k} 2^k\s(w_k') - 2^k\s(w_k')\|_1   \le M( 2^k\s(w_k'))|w_k| &= 2^kM(\s(w_k'))  2^{-k}|v_1| \\&= 2^{-k} |v_1||v_1'|.\end{aligned}$$
It follows that as $k \to \i$, then
$$|2^k\s(w_k + w_k') - 2^k\s(w_k) - 2^k\s(w_k')|^{\natural_{1}}   \to 0.$$  We do not yet know $\a(v_1 + v_1') = \a(v_1) +\a(v_1')$    in the seminormed space but they are  equal in the reduced normed space.  
 
  \end{proof}

 One can form a vector space $\cal{A}^0_k = \cal{A}^0_k(0)$ of $k$-element chains 
 with coefficients in $\mathbb{F},$  supported at the origin.   We do this by taking linear combinations of $k$-element cells supported at $0$ and  using the three relations above to combine ``like terms'', much as with polyhedral chains.    The vector space of $k$-element chains supported at $p$ is denoted $\cal{A}_k^0(p).$   The next lemma follows directly from the definition of a $1$-element as $\a = \a(v)$ where $v \in \R^n.$
 
\begin{lemma}
$\cal{A}_1^0(p) \cong \R^n$.
\end{lemma}
 Soon   we will see that $\cal{A}_1^0(p)$ has much more structure associated to it than $\R^n$.   
 
 \begin{theorem}\label{vec}  $Vec$ is a   linear operator
$$Vec: \cal{P}_k \to \cal{A}_k^0$$ with $$M(Vec(P)) \le M(P)$$ for all polyhedral chains $P \in \cal{P}_k.$
\end{theorem}

\begin{proof}
This follows since $M(Vec(\s)) = M(\s)$ for every $k$-cell $\s$.
\end{proof}

     \subsection*{Inner products on $\cal{A}$}  An inner product for $\R^n$ induces one on $\cal{A}_k^0$ as follows.      Define an inner product for simple $k$-elements   $\a, \a' \in \cal{A}_k^0$ as follows:
$$<\a,\a'> := det(<v_i, v_j'>)$$ where $\a = \a(v_1, \dots, v_k)$, $\a' = 
\a(v_1', \dots, v_k').$  Since determinant is an alternating, $k$-multilinear function of the $v_i$'s and $v_j'$'s we obtain a scalar valued function, linear in each variable
$$<\cdot, \cdot>: \cal{A}_k^0 \times \cal{A}_k^0 \to \R.$$ Note that $<\a,\a'> = <\a',\a>$ since the determinant of the transpose of a matrix is the same as the determinant of the matrix.  

It is left to show that the inner product is well defined. (This is not completely trivial.)   We use inner product as a kind of "scaffolding".  A different choice leads to an equivalent theory.

\subsection*{Orthonormal basis of $\cal{A}_k^0$}  Choose an orthonormal basis $(e_1, \cdots e_n)$ of $\R^n$.  For a finite indexing set $H = \{h_1, h_2, \cdots, h_k\}$, let $$e_H :=  \a(e_{h_1},   e_{h_2}, \cdots ,e_{h_k} ).$$ Then the collection 
of simple $k$-elements  
$$(e_H: H = \{h_1 < h_2 < \cdots < h_k\})$$ forms a basis of $\cal{A}_k^0.$  This follows since  $H \ne L \implies $the matrix has a row of zeroes.  $H = L \implies$ all but the diagonal elements vanish.   Thus $<e_H, e_L> = \pm 1$ if $H$ and $L$ contain the same indices with no repeats, and is zero otherwise.    

\subsection*{Mass of $\a$} For a simple $k$-element $\a$, it follows directly from the definitions that $M(\a) = \sqrt{<\a,\a>}.$ 

\subsection*{ Volume element}   Let $(e_1, \dots, e_n)$ be a positively oriented orthonormal basis of $\R^n$.
The {\itshape \bfseries volume element} $vol$ is well defined (with respect to the chosen orientation).  It is the unique $n$-vector in $\cal{A}_k^0$  with $$vol = \a(e_1, \dots, e_n).$$  Observe that $M(vol) = 1.$  Furthermore, $vol$ is an orthonormal basis of $\cal{A}_n^0$ and $<vol, vol> = 1$.

 \subsection*{Differential elements}    
 \begin{verse} {\em He who can digest a second or third fluxion need not, methinks, be squeamish about any point in divinity.}
\end{verse}
\begin{flushright} -- Bishop George Berkeley \end{flushright}
   Differential elements of higher order lead us to define the boundary of a $k$-element.   
  
  Apply the directional derivative operator $\nabla_u$   to $k$-elements $\a.$  For example,  in $\R^1$, $\nabla_{e_1}\a(e_1)$ is much like the {\itshape \bfseries Dirac dipole}, or the derivative of the Dirac delta function.  
  \begin{proposition} \quad \\
\begin{enumerate}
\item $t \nabla_u \a = \nabla_{tu} \a =\nabla_u t\a.$
\item $\nabla_{u+u'} \a =  \nabla_u \a + \nabla_{u'} \a.$
\item $\nabla_u \a(v_1, \dots, v_i +v_i', \dots, v_k) = \nabla_u \a(v_1, \dots, v_1, \dots, v_k) \\\hspace{.5in}+ \nabla_u \a(v_1, \dots,  v_i', \dots, v_k).$
\item $\nabla_u \a(v_1, \dots, v_i, v_{i+1}, \dots, v_k) = -\nabla_u \a(v_1, \dots, v_{i+1}, v_i, \dots, v_k).$
\end{enumerate}
\end{proposition}

Form a vector space $\cal{A}_k^1 \subset  \cal{N}_k^2$ of chains of differential $k$-elements $$ \sum a_i  \nabla_{u_i} \a_i,$$ subject to the above relations.  Since a $k$-element $\a$ is a $k$-chainlet of class $N^1$, its  boundary $\p \a$ is a  well defined  $(k-1)$-chainlet of class $N^2.$  Furthermore,
  
\begin{proposition}
$\p: \cal{A}_k^0 \to \cal{A}_k^1$ is a homomorphism.
\end{proposition}

\begin{proof}  Let $\a \in \cal{A}_k^0.$  We show that $\p \a \in \cal{A}_k^1.$  Let $Q$ be a parallelepiped such that $\a = \a(Q).$  Then $\a(Q) = \lim_{j \to \i} Q_j$ where $Q_j$ is a homothetic replica of $Q$, that is, a copy of $Q$ scaled by a factor of $2^{-j},$  weighted to have the same mass as $Q$.  Since the boundary operator is continuous we conclude
$$\p \a = \lim_{j \to \i} \p Q_j. $$   Now the boundary of $Q$ can be written as a sum of $k$ pairs of oppositely oriented and parallel $(k-1)$-parallelepipeds, each pair limiting to a  $(k-1)$-element of order $1$.  
 Linearity is clear.  
\end{proof}
 
 The vector space $\cal{A}_k^j \subset \cal{N}_{k}^{j+1}$ of $j$-differential $k$-element chains 
 $$ \sum a_i  \nabla_{U^j} \a_i$$ is defined in a similar fashion to $\cal{A}_k^1$, but also subject to linearity relations of the translation vectors.   The boundary operator $\p:\cal{A}_k^j \to \cal{A}_{k-1}^{j+1}$ satisfies $\p \circ \p = 0.$
We remark that the translation vectors of difference elements do not satisfy linear relations, but the translation vectors of differential elements do.   This implies that a fixed basis $W^n = (w_{1}, \dots, w_{n})$ of $\R^{n}$ naturally generates a finite basis of each $\cal{A}_k^j.$  We will see how this   leads to simpler computations and matrix methods in a later section.  
 
\begin{lemma}  Let $j, k, \ell \ge 0.$
$\cal{A}_k^{j+\ell} \subset \overline{\cal{A}_k^{j} }$ with completion taken in the $j$-natural norm.  
\end{lemma}
The proof is similar to the proof for $j = 0$ and we omit it.  

\subsection*{$Vec^j$}  (Under construction, not necessary for the rest of the notes)  The operator 
 $$Vec: \cal{A}_k \to \cal{A}_k^0$$ has higher order analogues.  We know that it extends to chainlets $J$.  However, $Vec(\a^j) = 0$ for any $j > 0.$  (This can easily seen by taking limits of difference cells.)   We sometimes wish to identify ``higher order parts'' of a chainlet.   Let $Vec^0 := Vec,$ for consistency of notation.  
 
We define a projection operator $$Vec^j: \cal{A}_k \to \cal{A}_k^j.$$   This will extend to an operator $$Vec^j: \cal{N}_k^{\natural_{}} \to  \cal{A}_k^j.$$     This reduces to defining $Vec^j(\a^i).$  It can be demonstrated that $\a^j$ depends only on a $k$-element $\a^0$ and the directions of translation.   We have seen that $\a^0$ depends only on its $k$-direction and mass.     It follows that $Vec^j(\a^i) = 0$ if $i \ne j,$ and $Vec^j(\a^j) = \a^j.$

As a corollary, we are able to define the inverse to the directional derivative operator $\nabla_u,$  up to a constant.

\subsection*{Norms on $\cal{A}_k^j$}
For $U^j = (u_1, \dots, u_j)$ and $\a \in \cal{A}_k^0$, define $$\|\nabla_{U^j} \a\|_j := |u_1| \cdots |u_j|M(\a).$$
Extend by linearity to   define a norm on each $\cal{A}_k^j$ as follows:
$$\left\| \sum a_i \nabla_{(U^j)_i}\a_i \right\|_j : =  \sum 
|a_i| \|\nabla_{(U^j)_i} \a_i\|_j.$$

As before, it is straightforward to prove this is a seminorm and we will be able to deduce it is a norm, once we know that the $j$-natural seminorm is a  norm.      $\cal{A}_k^j$ is a subspace of $\cal{N}_k^j$, the space of $k$-chainlets of class $N^j$.    We often simplify the notation and denote $$\a^j = \a_k^j = \nabla_{U^j} \a(V^k), $$ if the context is sufficiently clear.  When we want to consider several $k$-elements at once, we might use symbols $\a, \b, \g \in \cal{A}.$

  \section{Algebras of $k$-elements and $k$-coelements}\label{algebras}
 The Banach algebra defined in this section significantly extends the Grassmann algebra.  Define
$$\cal{A}_k :=   \cal{A}_k^0 \oplus \cal{A}_k^1 \oplus \cdots$$ 
and
   $$\cal{A} := \R \oplus \cal{A}_0 \oplus \cal{A}_1 \oplus \cdots \oplus \cal{A}_n.$$ 
   
   We define operators and products on $\cal{A}$.   
   \subsection*{Boundary operator on $\cal{A}$}     Since $$\p: \cal{A}_k^j \to \cal{A}_{k-1}^{j+1}$$ is naturally defined and satisfies $\p \circ \p = 0$,  the boundary operator is well defined on $\cal{A}.$  $$\p:\cal{A} \to \cal{A}.$$
  
 \subsection*{Exterior product on $\cal{A} \times \cal{A}$}  The exterior product $\wedge$ will turn $\cal{A}$    into a   {\itshape \bfseries differential graded  Banach algebra}.  
   
The 
    {\itshape \bfseries exterior product} 
    $$\wedge:\cal{A}_k^j \times \cal{A}_{k'}^{j'} \to \cal{A}_{k+k'}^{j+j'}$$     is defined by  
   $$\nabla_{U^j} \a(V^k) \wedge \nabla_{\tilde{U}^{j'}} \a(\tilde{V}^{k'}) :=  \nabla_{(U^j, \tilde{U}^{j'})} \a(V^k, \tilde{V}^{k'}).$$
\begin{lemma}   The exterior product is associative, bilinear and graded commutative.  In particular, suppose $\a, \b, \g \in \cal{A}.$   Then
\begin{enumerate}
\item $(\a \wedge \b) \wedge \g = \a \wedge (\b \wedge \g).$
\item $(t\a) \wedge \b = t(\a \wedge \b) = \a \wedge (t\b), \, \forall t \in \mathbb{F};$
\item $(\a + \b) \wedge \g = (\a \wedge \g) + (\b \wedge \g)$;
\item $\a \wedge \b = (-1)^{dim(\a)dim(\b)}\b \wedge \a$ if $\a$ and $\b$ are simple.  
\end{enumerate}
\end{lemma}

\begin{proof}  The vectors of translation do not affect the sign as they are unordered.    However, $\a(V^k, V^{k'}) = (-1)^{kk'} \a(V^{k'}, V^k).$  
\end{proof}

It follows that $\a$ and $\b$ are linearly dependent simple $k$-elements iff $\a \wedge \b = 0.$   All $k$-elements in $\R^3$ are simple.  However, if $(v_1, v_2, v_3, v_4)$ is a basis of $\R^4$, then $\a(v_1, v_2) + \a(v_3, v_4)$ is not simple.  

\begin{lemma} If $\a, \b \in \cal{A}$, then  $$\p(\a \wedge \b) = (\p \a) \wedge  \b + (-1)^{dim( \a)} \a \wedge (\p \b).$$
\end{lemma}

The next lemma follows directly from the definition of the norm $\|\cdot \|_j.$      
\begin{lemma}
$$\|\nabla_{U^j} \a(V^k) \wedge \nabla_{\tilde{U}^{j'}} \a(\tilde{V}^{k'}) \|_{j+j'}\le \|\nabla_{U^j} \a(V^k) \|_j \| \nabla_{\tilde{U}^{j'}}  \a(\tilde{V}^{k'}) \|_{j'}.$$
\end{lemma}

Suppose $A = \sum_i a_i (\a_k^j)_i \in \cal{A}_k^j$ and 
$B = \sum_{\ell} b_{\ell} (\a_{k'}^{j'})_{\ell} \in \cal{A}_{k'}^{j'}$.
    Define $$A \wedge B := \sum_i \sum_{\ell} a_ib_{\ell} (\a_k^j)_i \wedge  (\a_{k'}^{j'})_{\ell}  \in \cal{A}_{k +k'}^{j+j'}.$$    

\begin{proposition}  Assume $A  \in \cal{A}_k^j$ and $B \in \cal{A}_{k'}^{j'}.$  Then 
$$\|A \wedge B\|_{j+j'} \le \|A\|_j\|B\|_{j'}.$$
\end{proposition}

Finally, let $A = A_{k_1}^{j_1} + \cdots + A_{k_p}^{j_p} \in \cal{A}$ and $B = B_{\ell_1}^{n_1} + \cdots + B_{\ell_m}^{n_m}  \in \cal{A}$  define $$A \wedge B := \sum A_{k_i}^{j_i}\wedge B_{\ell_s}^{n_s}.$$  Define a norm 
$$\|A\|:= \sum \|A_{k_i}^{j_i}\|_{j_i}$$ by taking sums of the individual norms.  
\begin{theorem} $\|A \wedge B\| \le \|A\|\|B\|.$
\end{theorem}
 
Thus $\cal{A}$ is a Banach algebra.  It is unital  since $\mathbb{F}$ has a unit $1$ and $1 \wedge B = B = B \wedge 1.$ Each element of $\cal{A}$ is nilpotent for $k > 0$ since its square is zero.      We will later find subalgebras that are division algebras and will show that $\cal{A}$ has an inner product  structure, making it into a Hilbert space.   It is not finite dimensional since $j$ can be arbitrarily large.  However, it is finitely generated.  That is, a basis of $\R^n$ leads to a basis for each $\cal{A}_k^j$.  

 Each operator and product we define on $\cal{A}$ has extensions to certain pairs of chainlets.   All extend to pairs of element chains supported in the same finite set of points.  However,  wedge product is not continuous in the chainlet topology and poses a general obstruction to extensions of products and operators that make use of $\wedge$ to all chainlets.  We introduce operators and products on $\cal{A}$ in this section and later discuss limits of extensions in the chainlet spaces. 

\subsection*{Exterior product on $\cal{A}$}  We may think of exterior product as an operator on $k$-elements.  Fix   $\b \in \cal{A}$ and define $e_{\b}: \cal{A} \to \cal{A}$ by  $$e_{\b} \a := \a \wedge \b. $$    \subsection*{Linear pushforward} Suppose $L:\R^{n} \to \R^{n}$ is linear and $U^{j} = (u_{1}, \dots, u_{j})$ is a list of vectors.   The {\itshape \bfseries pushforward}  $L_{*}(U^{j})$ is defined by applying $L$ to each of the vectors in the list $U^{j},$  $$L_{*}(U^{j}) :=  (L(u_{1}), \dots, L(u_{j})).$$  Define $$L: \cal{A} \to \cal{A}$$ by
$$L_{*}\a(V^{k}):= \a(L_{*}(V^{k})).$$
\begin{proposition}\label{boundary} If $L: \R^n \to \R^n$ is a linear transformation and $\a \in \cal{A}$, then 
  $$L_*(\p \a) = \p( L_*(\a)).$$
\end{proposition}

\subsection*{Hodge dual $\perp$}
      Choose an inner product  $<\cdot, \cdot>$  on $\R^n$.   Suppose $V^k$ is linearly independent, $0 \le k < n.$ Choose a basis  $W^{n-k}$ for the subspace orthogonal to the subspace spanned by $V^k.$     Order the vectors of $W^{n-k}$ so that the basis $(V^k, W^{n-k})$ is positively oriented and further adjust the norms of the vectors so that $M( \a(V^k)) = M(\a(W^{n-k})$.   Define
 $$\perp (\a(V^k)) := \a(W^{n-k}).$$   The choice of vectors in the list $W^{n-k}$ is not in any way unique, but the subspace and mass determining them are unique. Thus $\perp$ is a well defined operator on $k$-elements.  It represents multiplication by $i$ in the complex plane.  $$\perp \a(v) = \a(iv).$$   Define the volume element $$vol := \a(U^n)$$ where $U^n$ is a positively oriented basis of $\R^n$ such that $M(\s(U^n)) = 1.$   Define $\perp(\a(V^n)$ to be the signed mass of $\a(V^n)$, with the sign taken according to the orientation of $V^n$.
It follows easily that  $$\a \wedge \perp \a = M(\a)^2 vol$$ and  
  $$\perp\perp\a = (-1)^{k(n-k) }\a.$$

Then if $(V^k)$ comes from an orthonormal basis,  $(\a(V^k))$ forms an orthonormal basis of $\cal{A}_k^0.$  

If $V^k$ is linearly dependent, then $\a(V^k)$ is a {\itshape \bfseries degenerate} $k$-element.  Its mass is zero, but we still consider its dimension to be $k$.  It is the zero $k$-element.  The zero $k$-element has no $k$-direction, only   mass, which is zero. It is therefore unique.   Thus $\a \wedge \a$ is the zero $2k$-element if $\a$ is any simple $k$-element. The zero $0$-element is the origin in $\R^n$.  If $n = 0$ we get the zero of the reals, i.e., the real zero!  
If $\a$ is degenerate,  define $\perp \a$ to be the zero $(n-k)$-element.

\subsection*{Products combining $\perp$ and $\wedge$}  The following products may be extended to certain chainlets since $\perp$ is a continuous operator on $\cal{A}$.   Later, we will see extensions where at least one member of the pair is a chainlet $J$.  The other member is permitted to be the chainlet associated to a differential form $\o$ through Poincare duality homomorphism.     

We consider ``interior products'', ``exterior products'' and ``intersection products'' of $J$ and $\o$, of arbitrary dimension and order.   These are products mapping $$ \cal{A} \times \cal{A} \to \cal{A}$$ with limited continuity in the chainlet topology.     
  
  When dimensions are allowed to vary, we obtain division product which has proven quite important in chainlet geometry.   
  
  \subsection*{Cross product}   
  
  Define $$\times: \cal{A}_{k_1}^{j_1} \times \cal{A}_{k_2}^{j_2}  \to \cal{A}_{n-k_1-k_2}^{j_1+j_2}$$ by
  $$\a \times \b := \perp(\a \wedge \b).$$
  For $k_1 = k_2 = n/3$ this product combines pairs of $k$ elements and produces a $k$-element.   Of course, when $n= 3$ this corresponds to the standard cross product of vectors. 
   
In the degenerate case with $v = w$, then $Vec(v) \wedge Vec(w)$ is the zero $2$-element.  Thus  $\perp$ of it is the zero $1$-element.  On the other hand, if $v$ is orthogonal to $w$, then the wedge product is a $2$-element with mass the same as the product $|v||w|$, so its perp corresponds to a vector with norm $|v||w|$ that is orthogonal to this $2$-element.

\subsection*{Intersection product}     Let $k_1 + k_2 \ge n.$  Let $\a \in \cal{A}_{k_1}^{j_1}$,  $\b \in \cal{A}_{k_1}^{j_2}$.  Define
$$\a \cap \b := \perp(\perp \a \wedge \perp \b).$$
 
For $j_1 = j_2 = 0$ this identifies the {\itshape \bfseries intersection product} of $\a$ and $\b$.

It is of interest to study sums of the intersection product with division and exterior products.  
 
  Is $\a \wedge \b /\a = \b?$
  
  Need $(\perp \a) \cap (\a \wedge \b) = \b$

\subsection*{Projection}  If $\a \in \cal{A}_{k+j}$, $\b \in \cal{A}_k$ define $$\pi_{\a} \b :=  \perp(\perp \b \cap \a) \cap \a.$$

\subsection*{Interior product}   Suppose $\a, \b \in \cal{A}$ with  $k$ the dimension of $\a$   and  $j$ the dimension of $\b$.    Suppose $j \le k$.      
Define the {\itshape \bfseries interior product} $$\a/\b := \perp(\b \wedge \perp \a).$$

\begin{proposition} If $j = k$, then interior product determines an inner product of $\a$ and $\b$:  $$<\a, \b> := \a/\b.$$  
\end{proposition}

\begin{proof}   Interior product is linear since the operators $\perp$ and $\wedge$ are linear.  Since
 $ \b \wedge \perp \a   =  (-1)^{k(n-k)} \perp \a \wedge \b = \a \wedge \perp \b$ we know $<\a, \b> = <\b, \a>.$
 $\a \wedge \perp\a = M(\a)^2$.  Hence interior product is nonnegative. 
 Finally, it is nondegenerate since $M(\a) = 0$ implies $|\a|^{\natural_{1 }} = 0$ implies $\a = 0.$  We rely on the fact that the natural seminorm is a norm.
\end{proof}

If $j < k$, then interior product becomes {\itshape \bfseries division product}, sometimes called {\itshape \bfseries slant product}. 

 \symbolfootnote[1]{We will see later that    $\int_A dV = \perp A$ for $A \in \cal{A}_n^0.$  Hence  $<\a, \b>  :=   \int_{\a \wedge \perp\b} dV.$ }     That is, if $\a = \b$ and $k = 1$, then $\a \wedge \perp \b = M(\a)M(\b)vol$.  Hence $<\a,b>  = M(\a)M(\b).$    If they are orthogonal, $\a \wedge \perp \b$ is the zero 3-element.  Thus its perp is   $0 \in \mathbb{F}.$

  This makes each algebra $\cal{A}_k^j$ into a Hilbert space.\symbolfootnote[2]{   The entire space of chainlets is not a Hilbert space since it contains a subspace isomorphic to $L^1$ functions.}

\subsection*{Orthogonal projection}   Suppose $\b$ has smaller dimension than $\a$.    $\perp_L(\perp_L \b)$    appears to be the  orthogonal projection into $L$.  
  
 \section{Algebra of coelements}   A {\itshape \bfseries $k$-coelement of order $j$} is an alternating member $\g^j$   of the dual space $(\cal{A}_k^j)'$ of linear functionals.  The space of all $k$-coelements of order $j$ is denoted $\cal{C}_k^j.$  Observe that a $k$-coelement is an {\itshape \bfseries alternating tensor} since reversing the order of any two vectors in a list of vectors generating a $k$-element $\a$ reverses the sign of $\a.$

\begin{lemma}
$$\cal{C}_k^j \subset \overline{\cal{C}_k^{j-1}}.$$
\end{lemma}
Since each $\cal{A}_k^j$ is finite dimensional,   so is its dual space $\cal{C}_k^j.$   

Let $((\a_k^j)_{1}, \dots, (\a_k^{j})_{d})$ be a basis of $\cal{A}_k^j$.  A basis of $\cal{C}_k^j$ is given by 
$\g_{i}^j((\a_k^{j})_{\ell}) := \d_{i,\ell}$ where $\d_{i,\ell} = 0$ whenever $i \ne \ell$ and $\d_{i,\ell} = 1$ whenever $i = \ell.$

 The dual norm is defined by
 $$\|\g^{j}\|_j := \sup\frac{\g^{j}(A^{j})}{\|A^{j}\|_j}, A^{j} \in \cal{A}_k^j.$$

Define $$\cal{C}_k := \cal{C}_k^0 \oplus \cal{C}_k^1 \oplus \cdots$$   
  and $$\cal{C} := \R \oplus \cal{C}_0 \oplus \cdots \cal{C}_n$$ 
 with norm defined linearly. 
 
The {\itshape \bfseries exterior derivative} $d:\cal{C}_k^{j} \to \cal{C}_{k+1}^{j-1}$ is given by
$$d C_{k}^{j} (\a_{k+1}^{j-1}) : = C_{k}^{j} (\p \a_{k+1}^{j-1}),$$  making $\cal{C}$ into a differential graded algebra.

{\itshape \bfseries Wedge product}  The spaces $\cal{A}_k^j$ and $\cal{C}_k^j$ are canonically isomorphic via the inner product on $\R^n$.   Wedge product of chainlet elements carries over to wedge product of chainlet coelements.  Wedge product may also be defined axiomatically, without reference to an inner product, as in \cite{grassmann}. 

\subsection*{Differentials} The {\itshape \bfseries differential $dx^i$} of $\R^n$  is defined to be the linear functional on $\cal{A}_n^0$ given by $dx^i(\a(a_1 e_1, \dots, a_n e_n))  := a_i.$   Wedge products of lists of $k$ differentials $\{dx^1, dx^2, \cdots, dx^n\}$ form a basis of  $\cal{C}_k^0$ if repetitions are omitted.    We follow the common practice of omitting the wedge sign for $k$-vectors e.g., $dx^i \wedge dx^j = dx^i dx^j.$

 \begin{theorem}  The mapping $a^H dx^H \mapsto a^H \a(e^H)$ induces an isomorphism   $$\cal{C}_k^0 \cong  \cal{A}_k^0.$$
 \end{theorem}
 
 A similar result holds for higher order elements by replacing $dx^H$ with $\nabla_{e^K} dx^H.$

\subsection*{Linear pullback} Suppose $L:\R^{n} \to \R^{n}$ is linear and $U^{j} = (u_{1}, \dots, u_{j})$ is a list of vectors.      Define
$$L_{*} (\nabla_{U^j}\a(V^k)):= \nabla_{(L_{*}U^j)}\a(L_{*}V^k).$$

 Define the {\itshape \bfseries linear pullback} $$L^{*}\g(A)  : = \g(L_{*}A).$$
 
 \begin{lemma}
 \begin{enumerate}
\item $L^*(\g + \eta) = L^* \g + L^* \eta;$
\item $L^*(\g \wedge \eta) = L^* \g \wedge L^* \eta;$
\item $dL^* = L^* d$
\item $(L \circ K)^* = K^* \circ L^*$
\end{enumerate}
\end{lemma}
Examples.  
1.  Let $f:\R^2 \to \R$ be defined by $f(x,y) = x-y$  and $\o = dt$ which is a 1-form on $\R.$   Then 
$$\begin{aligned}
f^*\o_{(x,y)} (v)   &= \o_{f(x,y)}( Df_{x,y}(v)) \\&= dt_{x-y}(v_1 -v_2) \\&= v_1 -v_2 = (dx -dy)_{(x,y)}(v).
\end{aligned}$$  Therefore $f^* dt = dx -dy.$  

Remark.  The standard approach (commonly unjustified) is 
$$t = x-y \implies dt = dx - dy.$$

2. Let $f: \R^1 \to \R^2$ be defined by   $f(t) = (t^2, t^3).$  Let $\o$ be the $1$-form in $\R^2$ given by $\o = x dy$.   Then 
$$\begin{aligned}
f^*\o_t( 1) &= \o_{f(t)}( Df_t(1)) \\&= \o_{(t^2,t^3)} (2, 3) \\&= xdy_{(t^2,t^3)} (2, 3) \\&=   3t^2 \\&= 3t^2dt_t(1).
\end{aligned}
$$  Therefore $f^*(xdy) = 3t^2 dt.$

\subsection*{Hodge star}  For $\g \in \cal{C}_k^0$ define $$\star \g(\a) := \g(\perp \a).$$

 \subsection*{Interior product}
 For $\b \in \cal{A}$, define
       $$i_{\b} \g(\a) :=  \g(e_{\b}(\a)).$$

\subsection*{Directional derivative}    For $v \in \R^n$, define the directional derivative of a coelement $\g$ by 
$$\nabla_v \g(\a):= \g(\nabla_v(\a)).$$
\subsection*{Products of coelements}  
 Generally, the products for $\cal{A}$ have dual versions for $\cal{C}.$  

The inner product of coelements takes a familiar form:  
  $$<\g,\eta>:= \star(\g \wedge \star \eta).$$

Intersection product can be defined for appropriate pairs of dimensions.     $$\g \cap \eta := \star(\star \g \wedge \star \eta).$$ 
  
Cross product: $$\g \times \eta := \star(\g \wedge \eta).$$

Projection: $$\pi_{\g} \eta:= \star(\star \g \cap \eta) \cap \eta.$$

The Riesz Representation Theorem gives us a canonical isomorphism between elements and coelements via the inner product.  We use $\flat \a = \g$ and $\sharp \g = \a$ to refer to this isomorphism.  
Suppose $\g$ has dimension $k_1$ and $\a$ has dimension $k_2$.  If $k_1 = k_2$ we have $$\g(\a) = <\sharp \g, \a>.$$
If $k_1 < k_2$, then $$\g(\a) = \a/\sharp \g.$$
If $k_1  > k_2$, then $$\g(\a) = \flat(\sharp \g/\a).$$

  The algebra of elements $\cal{A}$ may be used to quantize matter (particles) and the algebra of coelements $\cal{C}$ may be used to quantize fields.   These are analogous to the bras and kets of Dirac \cite{dirac}, giving them new geometric meaning and the power of chainlet operators.     

\subsection*{Operator algebras} Under development:  Naturally arising are the operator algebras of $\cal{A}$ and $\cal{C}.$  Denote these by $B(\cal{A}) $ and $B(\cal{C}).$
  One can construct element bundles and coelement bundles whose sections are element fields and coelement fields, respectively.  Of especial interest are sphere bundles and perp of these.  Study Whitney and Steenrod.

   Define the algebra  $$\cal{T} = \cal{A} \oplus \cal{C}.$$ The product will be an extension of wedge product.    Here we find new results similar to Stokes' theorem but for $k$-element  and $k$-coelement valued forms.  

For $\g \in \cal{C}_{k+j}$ and $\a \in \cal{A}_k$, define 
$$\g \wedge \a \in \cal{C}_j$$ by 
$$\g \wedge \a(\b) := \g(\a \wedge \b)$$ for all $\b \in \cal{A}_j.$
For $\g \in \cal{C}_{k}$ and $\a \in \cal{A}_{k+j}$, define 
$$\g \wedge \a \in \cal{A}_j$$ by 
$$\eta(\g \wedge \a) := \eta \wedge \g(\a)$$  for all $\eta \in \cal{C}_j$.

\begin{lemma}
$$\|\g \wedge \a\| \le \|\g\|\|\a\|.$$
\end{lemma}  

\subsection*{Infinitesimal calculus}  We now have the ingredients for calculus at a single point\symbolfootnote[2]{This list is not complete, for the sake of brevity, and does not include all products and operators discussed in this paper.  For example,  Lie derivative is defined later, but also has meaning at a point. } .  Extend $\g^j$ to $\cal{A}$ by setting $\g^j(\a^i) = 0 $ if $i \ne j$.  Let $\a = \sum \a^j \in \cal{A}$  and $\g \in \cal{C}$ 
\begin{itemize}
\item Integral:  $\int_{\a} \g := \sum_j \g^j(\a^j).$
\item Exterior derivative:  
$d \g(\a) := \sum_j  \g^{j+1}(\p \a^{j}).$
\item  Change of variables: $L^* \g(\a) := \sum_j  \g^j(L_* \a^j).$
\item Hodge star: $ *\g(\a) := \sum_j   \g^j(\perp \a^j).$
\item Directional derivative: $\nabla_v \g(\a) := \sum_j \g^j(\nabla_v \a^j).$
\item Exterior product: $e_{\b}( \a) := \sum_j \b \wedge \a^j.$
\item Interior product: $i_{\b}\g(\a)  := \sum _j \g^j(e_{\b}( \a^j)).$
\end{itemize}

Infinitesimal versions of Stokes', divergence and curl theorems are given by $$\int_{\a^{j}} d \g^{j+1} = \int_{\p \a^{j}} \g^{j+1},$$ 
$$\int_{\a^{j}} d \star \g^{n-j-1} = \int_{\perp \p \a^{j}} \g^{n-j-1}$$ and
$$\int_{\a^{j}}   \star d  \g^{n-j+1} = \int_{\p \perp  \a^{j}} \g^{n-j+1},$$ respectively.

 \section{Discrete exterior calculus}\label{discretecalc}
    We next consider {\itshape \bfseries discrete} chains and forms supported in finitely many points. 
  All of our operators and products will have discrete versions in this category.    Convergence to the smooth continuum will be satisfied as long as the  wedge product is not involved.  If wedge product is used, we obtain continuity on pairs in certain subspaces of chainlets.

\subsection*{Discrete fields of elements and coelements}  The algebras $\cal{A}$ and $\cal{C}$ are based at  the origin, say.  Their members are chainlets and may therefore be translated to any $p \in \R^{n}$  via translation.   We denote these translated algebras by $$\cal{A}(p) := T_{p*}\cal{A}$$ and $$\cal{C}(p) := T_{p}^*\cal{C}.$$  That is,  $\a \in \cal{A}$ if and only if  $\a(p) \in \cal{A}(p)$ where  $\a(p) := T_{p*} \a$ and $\g  \in \cal{C}$ if and only if  $\g(p) \in \cal{C}(p)$ where  $\g(p) := T_{p}^*\g$.

  In this section, we let $K$ denote a finite set of points in $\R^n$.   Define $$\cal{A}_k^j(K) := \left\{ \sum a_i \a^j(p_i): \a^j(p_i) \in \cal{A}_k^j(p_i), p_i \in K\right\},$$ $$\cal{A}_k(K) = \oplus_j\cal{A}_k^j(K), \mbox{ and } \cal{A}(K) = \oplus_k\cal{A}_k(K).$$

Define the {\itshape \bfseries $j$-difference norm} of a discrete $j$-difference $k$-chain $A^j  \in \cal{A}_k^j(K)$  as $$\|A^j\|_j :=  \inf\left\{ \sum_{i=1}^m |a_i|\|(\Delta_{U^j}\a)_i\|_j: A^j =  \sum_{i=1}^m a_i  (\Delta_{U^j}\a)_i,(\Delta_{U^j}\a)_i \in \cal{A}_k^j(K) \right\}.$$

 \subsection*{Differential forms}\label{sectionforms}   A {\itshape \bfseries $k$-form } $\o$ defined on a subset $U \subset \R^n$   is a  $\cal{C}$-valued function\symbolfootnote[1]{Forms may also be defined by taking sections of appropriate bundles.}  on $U$.  Note that $U$ is not required to be open. 
   
The form is  {\itshape \bfseries discrete} if $U =K,$ a finite set of points.    It is a {\itshape \bfseries differential} form if $U$ is open in $\R^n$.
We sometimes write $\o(\a) =  \o(p)(\a)$ where $\a \in \cal{A}(p).$
The space of discrete $k$-forms $\o$ of order $j$, defined over $K$, is denoted $\cal{C}_k^j(K).$  Define $\cal{C}_k(K) := \oplus_j \cal{C}_k^j(K)$ and $\cal{C}(K):= \oplus_k \cal{C}_k(K).$

  \subsection*{Jet of a form} Suppose $\o \in \cal{C}_k(p).$     For $0 \le j \le r$, define $\g^j \in \cal{C}_k^j(p)$ by $\g^j(\a^j) := \o(\a^j)$ for $\a^j \in \cal{A}_k^j(p), p \in K$.  Then $\o = \sum_{j = 0}^r \g^j$.  We call $\sum _{j = 0}^r \g^j$  the {\itshape \bfseries $r$-jet} of $\o$ at $p$.   Recall that if orders of a coelement $\g^j$ and an element $\a^i$ do not match, then $\g^j(\a^i) = 0.$   Thus, $\o(\a^j) = \g^j(\a^j).$ However, the dimensions do not need to match to obtain nontrivial pairing, as seen in the interior product.   (See also \S \ref{preint}.)

The infinitesimal calculus at a single point of the previous section immediately extends to discrete forms $\o$ defined over discrete chains in $\cal{A}_k(K)$ giving us a simple version of {\itshape \bfseries discrete exterior calculus}.  This might be the end of the story, were we not also looking to  use differential forms defined on open sets $U$ to generate our coelements.  Eventually, we wish  to  
 achieve convergence to the smooth continuum.

We mention the main ingredients of this simple discrete theory.  
 All of our products and operators in the infinitesimal theory extend to the discrete theory. 

\subsection*{Integration over discrete chains} The integral of a discrete form $\o \in \cal{C}_k(K)$ over a discrete chain $A = \sum a_i \a_i \in \cal{A}_k(K)$ is defined by   
$$ \int_{A} \o   := \sum a_{i} \o(\a_{i}) = \sum_i \sum_j a_i \g^j(\a_i^j).$$

\subsection*{Boundary operator}   The infinitesimal boundary operator is defined for elements $\a$ supported at any $p \in K$.   By linearity, we may therefore define $\p A$ for any discrete chain $A \in \cal{A}(K).$ Then $\p: \cal{A}_k^j(K) \to \cal{A}_{k-1}^{j+1}(K)$ is a naturally defined linear mapping with $$\p\left(\sum a_{i}\a_{i}\right) := \sum a_{i}\p \a_{i}.$$

\subsection*{Exterior derivative of forms}  The infinitesimal exterior derivative extends to elements supported at a point $p$:  $$d \g^j(\a^{j-1}):=  \g^j(\p \a^{j-1}).$$    Define $d: \cal{C}_k^j(K) \to \cal{C}_{k+1}^{j-1}(K)$ as follows:  If $\o = \sum \g^j$ is the jet of $\o$, define $$d\o := \sum d\g^j.$$       It follows that $d \o(\a) = \o(\p \a), \a \in \cal{A}_{k+1}.$
Observe that  $\p \circ \p = 0$ implies $d \circ d = 0.$

\subsection*{Discrete Hodge star}  For $\a \in \cal{A}(p)$, define $$\perp \a(p) = \perp T_{p*}\a := T_{p*} \perp \a.$$   Extend by linearity to  obtain an operator $\perp:   \cal{A}_k^j(K)\to  \cal{A}_{n-k}^j(K)$ with 
$$\perp\left(\sum a_{i}\a_{i}\right) := \sum a_i \perp  \a_{i}.$$

  Define $\star: \cal{C}_k^j(p) \to \cal{C}_{n-k}^j(p)$ by $$\star \g(\a^j) := \g(\perp \a^j).$$
   Let $\o = \sum \g^j$ be the jet of $\o  \in \cal{C}_k(K)$.  Define the Hodge star operator  $$\star: \cal{C}_k(K) \to \cal{C}_{n-k}(K)$$  by 
 $$\star\o(\a):= \sum  \g^j(\perp\a)$$  for all $\a \in \cal{A}(K).$        
 
 \subsection*{Remark}  Assume $\o = \o^1 \wedge \cdots \wedge \o^k$ is a simple $k$-form.  We leave it as an exercise to show  
$(\star\o)(v^1 \wedge \cdots \wedge v^{n-k})$ is the volume of the $n$-parallelepiped spanned by  $(v^1, \cdots, v^{n-k}, w^1, \cdots w^k).$

\subsection*{Discrete pushforward}

Let $f:U \subset \R^{n} \to V \subset \R^{n}$.  For $u \in \R^{n}$ and $p \in U$, define  the {\itshape \bfseries pushforward} 
$$f_{p*}u_{}:= \lim_{h \to 0^+}\frac{f(p+hu)-f(p)}{h}.$$

 We assume that this limit exists and call such $f$ {\itshape \bfseries differentiable}.   In this section, we assume all mappings are differentiable. 

 For a list $U^k = (u_{1}, \dots, u_{k})$, define $$f_{*}U^k := (f_{*}u_{1},\dots, f_{*}u_{k}).$$ 
 Define $$f_{*} T_{p}\s(V^{k}) := T_{f(p)}\s(f_{*}V^{k})$$ and
$$f_{*} T_{p}\a(V^{k}) := T_{f(p)}\a(f_{*}V^{k}).$$ 

Define {\itshape \bfseries pushforward} $$f_*: \cal{A}(K) \to \cal{A}(f(K))$$     by  $$f_*(\nabla_{U^j}\a(V^k)(p)) :=  \nabla{f_*(U^j)} \a(f_*(V^k))(f(p))$$   where  $\nabla_{U^j}\a(V^k)(p) \in \cal{A}(p)$, and extending linearly. 
Define {\itshape \bfseries pullback} $$f^*:\cal{C}(f(K)) \to \cal{C}(K)$$ by $$f^* \g(\a) := \g(f_* \a), \a \in \cal{A}(K).$$

In a similar fashion,  the operators $\nabla_u$, $\e_{\b}$, $\cal{L}_v$  defined at a single point, extend to discrete chains and differential forms by linearity.  The products defined at a single point extend, as well, as long as the discrete chains and forms are supported in the same finite set of points.  
\subsection*{Wedge product of element chains}  Elements $\a_1, \a_2 \in \cal{A}$ have a wedge product defined: $\a_1 \wedge \a_2.$  This extends to pairs of element chains $A = \sum a_i \a(p_i), A' = \sum b_i \a'(p_i)$ based at the same finite set of points $P = \{p_i\}$ by linearity. 
$$A  \wedge A' := \sum a_ib_i \a  \wedge \a'.$$

\subsection*{Cross product} We may extend the cross product of elements to cross product of discrete chains in $\cal{A}(K)$.

 Define {\itshape \bfseries cross product}  $$A \times B := \perp(A \wedge B).$$
 
This may be extended to  discrete chains supported in the same finite set of points $K$.  $$\times: \cal{A}(K) \times \cal{A}(K) \to \cal{A}(K)$$ by $$\sum a_i \a_i(p_i) \times \sum b_i \b_i(p_i) := \sum a_i b_i \perp(\a_i \wedge \b_i).$$   Observe that the two chains have independent dimension and order. 

An important example arises when we develop EM theory as the {\itshape \bfseries Poynting vector} $$E \times H = E \times \frac{1}{\mu}B.$$

In a similar way we may extend  wedge product, interior product, exterior product, cross product, intersection product,  and projection to pairs of discrete chains supported in the same set of points.
  Inner product is sufficiently important to warrant its own section and we postpone it temporarily.

\subsection*{Summary of discrete calculus}  We now have the ingredients for calculus at finitely many points $K$\symbolfootnote[2]{This list is not complete, for the sake of brevity, and does not include all products and operators discussed in this paper.  For example,  Lie derivative is defined later, but also has meaning at finitely many points.  All operators combining these basic operators, such as Laplace and Dirac operators, have immediate discrete versions.}    Let $A = \sum a_i \a_i   \in \cal{A}_k(K)$  and $\o \in \cal{C}_k(K)$ 
 
Define  $$\int_{A} \o := \sum_ja_i \int_{\a_i} \o.$$

Discrete versions of Stokes', divergence, curl, change of variables theorems are given by $$\int_{A} d \o = \int_{\p A} \o,$$ 
$$\int_{A} d \star \o = \int_{\perp \p A} \o,$$ and
$$\int_{A}   \star d \o = \int_{\p \perp A} \o,$$ and
$$\int_{A} f^*\o = \int_{f_*A} \o,$$ respectively.
We also have
$$\int_{A} \nabla_v \o = \int_{\nabla_v A} \o,$$
$$\int_{A} i_{\b} \o = \int_{e_{\b}A }  \o,$$

 \newpage
  \section{Chainlet exterior calculus}\label{integralsofchainlets}
   \subsection*{Norms on forms}      We give a coordinate free definition of the degree of smoothness of a form $\o$ defined on an open subset $U$ of  $\R^n$.    Define $$\|\o\|_0 := \sup \left\{\frac{\o( \a)}{\|\a\|_0}:   \a \subset \cal{A}_k^0(K), K \subset U
 \right\}.$$
  
 For all $j > 0$, define  the family of seminorms
$$\|\o\|_j := \sup \left\{\frac{\o(\Delta_{U^j}\a)}{\|\Delta_{U^j}\a\|_j}:  \Delta_{U^j}\a \subset \cal{A}_k^j(K), K \subset U
 \right\}.$$

Define $$(\D_{U^j} \o) (\a) := \o (\D_{U^j} \a).$$
 
\begin{lemma}\label{36}  Let $A^j  \in \cal{A}_k^j(K)$ and $\|\o\|_j < \i$,  where $K \subset U$, and $\o$ is defined on $U$. Then  $$\left| \int_{A^j} \o\right|  \le \|A^j\|_j\|\o\|_j.$$ 
\end{lemma}

\begin{proof}   
 Let  $\e > 0.$  There exists $A^j =    \sum_{i=1}^m a_i  (\D_{U_i^j} \a)_i$  such that $$\|A^j\|_j >  \sum_{i=1}^m |a_i| \|(\D_{U_i^j} \a)_i\|_j - \e.$$
By the definition of $\|\o\|_j$,  
$$\begin{aligned} \left| \int_{\Sigma a_i(\D_{U_i^j} \a)_i}\o\right|& \le
\sum |a_i| \left| \o((\D_{U_i^j} \a)_i)\right| \\&\le \|\o\|^j \sum |a_i| \|(\D_{U_i^j} \a)_i \|_j \\&< \|\o\|^j(\|A^j\|_j + \e)
\end{aligned}$$

   The result follows since this inequality holds for all $\e > 0.$
\end{proof}

\begin{lemma}\label{boundaries}  If $\|\o\|_j < \i$ and  $\a^j \in \cal{A}_k^j(p)$, with $\a^j = \lim_{h \to 0}\frac{ \Delta_{hu_j} \a^{j-1}}{h}$, then $$  \o(\a^{j}) = \lim_{h \to 0} \o\left(\frac{\D_{h u_j} \a^{j-1}}{h}\right).$$

\end{lemma}

\begin{proof}   The sequence $\left(\frac{\D_{2^{-i}u_j} \a^{j-1}}{2^{-i}}\right)$ is Cauchy in $\| \cdot \|_j.$   Lemma \ref{36} implies the result.
 \end{proof}
 

This implies that if $\o$ is defined over an open subset $U$  with $\|\o\|_s < \i$,   $0 \le s \le r$,  then the $r$-jet of $\o = \sum \g^j$ is well defined and continuous at each $p \in U$.

 \subsection*{Forms of class $B^r$}

Define $$|\o|^{\natural_{0}} := \|\o\|_0$$ and  for $r \ge 1$, 
$$|\o|^{\natural_{r}} = \max\{\|\o\|_o, \cdots, \|\o\|_r, \|d\o\|_0 , \cdots, \|d\o\|_{r-1} \}.$$
  
   We say that $\o$ is of class $B^r$ if $|\o|^{\natural_{r}} < \i.$  Let $ \cal{B}_k^r$ denote the space of differential $k$-forms of class $B^r.$
Say $\o$ is of class $B^{\i}$ if there exists $C > 0$ such that $|\o|^{\natural_{r}} < C$ for all $r \ge 0.$  Let $ \cal{B}_k^r$ denote the space of differential $k$-forms of class $B^r, 0 \le r \le \i.$  If $\o$ is of class $B^r$, the jet of $\o$ is well defined:  $\o =\sum_{j=0}^r \g^j$ and each $\g^j$ may be computed recursively by taking limits.  Forms of class $B^{\i}$ are equivalent to $C^{\i}$ forms with a global bound on all derivatives, depending on the given form.  Forms of class $B^{\i}$ are not scale invariant.  For example, $sin xdx$ is a form of class $B^{\i}$, but $sin (2x) dx$ is not.

\begin{lemma}\label{smoother}  If $\o$ is a differential $k$-form of class $B^r$, then $d\o$ is a differential $(k+1)$-form of class $B^{r-1}$ and
 
 $$ |d\o|^{\natural_{r-1}} \le |\o|^{\natural_{r}}.$$
 
\end{lemma}
\begin{proof}
This   follows by the definition of the norm, and since $d \circ d = 0.$

\end{proof}

\begin{lemma}
If $\o$ is of class $B^r$, then $\star \o$ is of class $B^r$ and $|\o|^{\natural_{r}} = |\star \o|^{\natural_{r}}.$
\end{lemma}

\begin{proof}
This follows since $\|\perp \a^j\|_j = \|\a^j\|_j$ and $\o(\perp \a^j) = \star \o(\a^j).$
\end{proof}
 
\subsection*{The integral over polyhedral chains}
\begin{lemma} Suppose $Q$ is a $k$-parallelepiped in $U \subset \R^n$ and $A_{i} \in \cal{A}_k(K)$ with $A_i \to Q$ in the norm $\|\cdot \|_{1}$.  If $\o \in B_k^1(U)$,   then  
  $\int_{A_{i}}\o$ forms a Cauchy sequence.
  \end{lemma}

\begin{proof}   Choose $A_i \to \s$ in the norm $\|\cdot\|_1.   By Lemma \ref{36} 
  $$\left|\int_{A_{i} -{A_{\ell}}}\o\right |  \le \|A_{i}-A_{\ell}\|_{1}\|\o\|_{1}.S$  The result follows.  
 
\end{proof}
   
 Define $$\int_{Q}\o := \lim \int_{A_{i}}\o$$ 
 and $$\int_{ \sum a_{i}Q_{i}} \o := \sum a_{i}\int_{Q_{i}}\o.$$
 
 If $\s$ is a $k$-cell, write $\s = \sum_{i = 1}^{\i} Q_i$ as its Whitney decomposition into parallelepipeds.   This series converges in the mass norm.  It follow that
 $$\int_{\s} \o : = \sum \int_{Q_i} \o$$    is well defined if $\o$ is of class $B^0.$

\begin{lemma}
Suppose $S \sim S'$ are algebraic $k$-chains supported in $U$ and $\o$ is a differential form defined  over $U$.  Then 
$$\int_S \o = \int_{S'} \o.$$
\end{lemma}

\begin{proof}
It suffices to assume that $S$ is a $k$-cell  $\s$ and $S'$ is a subdivision of $\s.$   The result follows since $k$-element chains approximating $S$ also approximate $\s$ and vice versa. 
\end{proof}

The integral $$\int_P \o := \int_S \o,$$ where $P = [S]$ is a polyhedral $k$-chain, is therefore well defined for all $\o \in B_k^1(U)$, assuming $supp(P) \subset U$.    This is equivalent to the classical integral of differential forms over polyhedral chains.  This integral stands halfway between the infinitesimal theory and the full theory on chainlets, neither of which was known before.

\begin{proposition}\label{7.9}
 $$\left|\int_P \o \right| \le |P|^{\natural_{0}}|\o|^{\natural_{0}}.$$
\end{proposition}
  
\begin{proof}
It suffices to prove this for cells $\s$.  Let $A_i \to \s$ be a sequence of $k$-element chains converging to $\s$ in the $1$-natural norm with $|A_i|^{\natural_{0}} \to |\s|^{\natural_{0}}.$  Then 
$$\left|\int_{\s} \o\right| \le \liminf\left\{\left|\int_{A_i} \o\right|\right\} \le \liminf\left\{|A_i|^{\natural_{0}}  |\o|^{\natural_{0}}\right\}.$$ The result follows since $|A_i|^{\natural_{0}} \to |\s|^{\natural_{0}}.$ 
\end{proof}
  \subsection*{Equivalent norms for forms}  We next show that $|\o|^{\natural_{r}}$ may be defined using $k$-cells $\s$  in place of  $k$-elements $\a.$
  
  For all $j \ge 0$, define 
$$\|\o\|_{[j]} := \sup \left\{\frac{\o(\Delta_{U^j}\s)}{\|\Delta_{U^j}\s\|_j}:  \Delta_{U^j}\s \subset \cal{P}_k^j(U)
 \right\}.$$   
  
Define $$|\o|^{[{0}]} := \|\o\|_{[0]}$$ and  for $r \ge 1$, 
$$|\o|^{[r]} = \max\{\|\o\|_{[o]}, \cdots, \|\o\|_{[r]}, \|d\o\|_{[0]} , \cdots, \|d\o\|_{[r-1]} \}.$$

\begin{theorem}\label{newnorm}
$$|\o|^{\natural_{r}} = |\o|^{[r]}.$$
\end{theorem}

\begin{proof} 
  This follows from continuity arguments:
By definition, $\a = \lim \s_i$ in the $1$-natural norm with $M(\s_i) \to M(\a).$  Thus 
$$\frac{\int_{\a}\o}{M(\a)} = \lim_{i \to \i}\frac{\int_{\s_i}\o}{M(\s_i)}.$$   It follows that
$$\sup \frac{\left|\int_{\a}\o \right |}{M(\a)} \le\sup \frac{\left|\int_{\s}\o \right |}{M(\s)} .$$
 Conversely, since chains of $k$-elements are dense in the $1$-natural topology
$$\s =  \lim_{\ell \to \i} \sum_i \a_{\ell,i}$$ with $$M(\s)  = \lim_{\ell \to \i} \sum_i M(\a_{\ell,i} ).$$  Hence 
$$\frac{\left|\int_{\s}\o\right|}{M(\s)} \le \sup \frac{\left|\int_{\a_{\ell,i}}\o \right |}{M(\a_{\ell,i})} \le 
\sup \frac{\left|\int_{\a}\o \right |}{M(\a)}.$$ 
 
We next show $\|d\o\|_i$ can be equivalently defined using $k$-elements $\a$  in place of  $k$-cells $\s .$
If $\a = \lim \s_i$ in the $1$-natural norm with $M(\s_i) \to M(\a)$, then $\p \a = \lim \p \s_i$ in the $2$-natural norm.  Thus $$\frac{\int_{\p \a}\o}{M(\a)} = \lim_{i \to \i}\frac{\int_{\p \s_i}\o}{M(\s_i)}.$$  It follows that
$$\frac{\left|\int_{\p \a}\o\right|}{M(\a)} \le \sup \frac{\left|\int_{\p \s}\o\right|}{M(\s)}.$$ Conversely, suppose
$\s =  \lim_{\ell \to \i} \sum_i \a_{\ell,i}$ with $M(\s)  = \lim_{\ell \to \i} \sum_i M(\a_{\ell,i} ).$  Then 
$$\p\s =  \lim_{\ell \to \i} \sum_i \p \a_{\ell,i}.$$  Thus
$$\frac{\left|\int_{\p \s}\o\right|}{M(\s)} \le \sup \frac{\left|\int_{\p \a_{\ell,i}}\o \right |}{M(\a_{\ell,i})} \le 
\sup \frac{\left|\int_{\p \a}\o \right |}{M(\a)}.$$ 

Higher order equivalences are similar and we omit them.  
\end{proof}
  \begin{lemma}\label{omega2} If $\o \in \cal{B}_k^r$, $r \ge 1$, then $$  \max\{\|\o\|_o, \cdots, \|\o\|_r\} \le  |\o|^{\natural_{r}}   \le k\max\{\|\o\|_o, \cdots, \|\o\|_r\}.$$  
\end{lemma}
\begin{proof} We rely on  Theorem \ref{newnorm}.  Observe that $\p \a^i$ is a chain of $(i+1)$-elements with $k$ terms.  Thus   $\|\p \a^i\|_{i+1} \le k\|\a^i\|_i.$  Hence
$$\|d\o\|_i = \sup \frac{\left|\int_{\a^i} d\o\right|}{\|\a^i\|_i} =
 \sup \frac{\left|\int_{\p \a^i} \o\right|}{\|\a^i\|_i} \le k \|\o\|_i.$$ 
\end{proof}

\begin{theorem}
$\left|\int_{\p P}\o \right| \le k|P|^{\natural_{0}}\|\o\|_1$
\end{theorem}

\begin{proof}
Write $P = \sum a_i \s_i$ as a non-overlapping sum.  Then 
$$\left|\int_{\p P}\o \right|  \le \sum |a_i| \left|\int_{\p \s_i}\o \right|  \le \sum k|a_i|M(\s_i) \|\o\|_1$$
\end{proof}

\begin{theorem}[Stokes' theorem for polyhedral chains]\label{polystokes}  If $P$ is a polyhedral $k$-chain and $\o \in \cal{B}^1_{k-1}(U)$ is a differential $k$-form,  then
$$\int_{P} d\o = \int_{\p P}\o.$$ 
\end{theorem}   

\begin{proof}  It suffices to prove $$\int_{\s} d\o = \int_{\p \s}\o.$$ 

Write $\s = \sum Q_{i}$, the Whitney decomposition of $\s$ using homothetic replicas of a parallelepiped with the same direction as $\s$.  Since convergence is in the mass norm    we can use Proposition \ref{7.9} to deduce $$\int_{\s} d\o = \lim_{m \to \i} \int_{\Sigma_{i = 1}^m Q_i} d\o = \lim_{m \to \i} \int_{\Sigma_{i = 1}^m  \p Q} \o.$$   We wish to show this is the same as $\int_{\p \s} \o.$   

Claim:
$$\left| \int_{\p C}\o\right | \le kM(C) \|\o\|_{1}.$$  This reduces to proving this for a cell $\s$ which in turn reduces to proving it for a parallelepiped $Q$.  But the boundary of $Q$ can be approximated by $(k-1)$-elements which can be split into matching parallel elements with opposite orientations.  
This establishes the claim.

  It follows that 
$$
\begin{aligned}
\left|\int_{\p (\sum Q_{i}) - \p \s}\o\right | = 
\left|\int_{\p (\sum Q_{i}  -  \s)}\o\right | \le kM \left(\sum Q_{i}  -  \s\right)\|\o\|_{1} \to 0 \mbox{ as } m \to \i.
\end{aligned}$$  Thus $$\int_{\s} d\o = \int_{\p \s} \o.$$

\end{proof}

 \subsection*{Integration over chainlet domains}\label{chainletintegral}
 
 The next result  generalizes the standard integral inequality of calculus:
$$\left|\int_P \o\right| \le M(P)|\o|^{\natural_{0}} = |P|^{\natural_{0}}|\o|^{\natural_{0}}$$ where $P$ is polyhedral and $\o$ is a bounded, measurable form.

\begin{theorem}[Fundamental integral inequality of chainlet geometry]\label{oldintegral} Let $P \in \cal{P}_k$, $r \in \Z^+,$ and $\o \in \cal{B}_k^r$ be defined in a neighborhood of $supp(P).$ Then
$$\left|\int_P \o\right| \le |P|^{\natural_r}|\o|^{\natural_{r}}.$$
\end{theorem}
 
\begin{proof}    Recall $ \left|\int_{\s^{j}} \o \right| \le \|\s^j\|_{j}\|\o\|_{j}.$

  By linearity $$\left|\int_{D^j} \o\right| \le \|D^j\|_j \|\o\|_j$$
for all   $D^j \in \cal{D}_k^j$.

  We again use induction to prove $\left|\int_P \o\right| \le  |P|^{\natural_r}|\o|^{\natural_{r}}.$ 
We know   $\left|\int_P \o\right| \le |P|^{\natural_0}|\o|^{\natural_{0}}.$  
Assume the estimate holds for $r-1.$ 

  Let $\e > 0$. There exists $P = \sum_{j=0}^r  D^j  + \p C$ such that $|P|^{\natural_r} >
\sum_{j=0}^r \|D^j\|_j + |C|^{\natural_{r-1}} - \e$. By Stokes' theorem for polyhedral chains Theorem \ref{polystokes}, Lemma \ref{smoother}, and
induction
 $$\begin{array}{rll}  \left|\int_P \o\right| &\le \sum_{j=0}^r \left|\int_{D^j} \o \right| + | \int_C d\o| \\& \le \sum_{j=0}^r \|D^j\|_j\|\o\|_j + |C|^{\natural_{r-1}} |d\o|^{\natural_{r-1}}\\& \le (\sum_{j=0}^r
\|D^j\|_j  + |C|^{\natural_{r-1}}) |\o|^{\natural_{r}}\\&\le  (|P|^{\natural_r} +
\e)|\o|^{\natural_{r}}.

\end{array} $$
Since the inequality holds for all $\e > 0$ the result follows.
\end{proof} 

For forms $\o$ of class $B^{\i}$ the inequality becomes
$$\left| \int_P \o \right| \le |P|^{\natural_{\i}}|\o|^{\natural_{\i}}$$ since $|\o|^{\natural_{\i}} = \lim_{r \to \i} |\o|^{\natural_{r}}$ and $|P|^{\natural_{\i}} = \lim_{r \to \i} |P|^{\natural_{r}}.$
 
\begin{corollary}[Continuity of the integral]
  Given $\o_i$ and $P_i, i = 1, 2$,  then $$\left|\int_{P_1} \o_1 - \int_{P_2} \o_2\right| \le |P_2 - P_1|^{\natural_{r}}\|\o_1\|_r + |P_2|^{\natural_{r}}\|\o_1 -\o_2\|_r.$$ 
\end{corollary}

\begin{corollary} $ |P|^{\natural_r}$ is a norm on the space of polyhedral chains $\cal{P}_k$. 
\end{corollary}
\begin{proof} Suppose $P \ne 0$ is a polyhedral $k$-chain. .    There exists a non-overlapping representation $P = \sum a_i \s_i$.  Fix $p $ in the interior of $\s_1$, say.  Since the cells are non-overlapping, there exists an $n$-ball $B(p)$ that misses all other $\s_i, i > 1.$ Choose a smooth bump function $\phi$ that is supported in $B(p)$.   Assume wlog that the $k$-direction of $\s_i$ is parallel to the coordinate $k$-plane with basis $e_1, \dots, e_k.$  Let $\o_0 = \phi dx_1\cdots dx_k.$ Then $\o_0$ is of class $B^r$  with $\int_P  \o_0 \ne 0$. Then
$0 <  \left|\int_P \o_0 \right| \le |P|^{\natural_r}|\o_0|^{\natural_{r}}$ implies $|P|^{\natural_r} > 0.$   
\end{proof}

\begin{corollary}\label{normproof} $ |P|^{\natural_{\i}}$ is a norm on the space of polyhedral chains $\cal{P}_k$. 
\end{corollary}

\begin{proof}
Consider the same form $\o_0$ as in the previous proof.  There exist $\o_i \to \o_0$, each $\o_i$ with uniformly bounded derivatives, the bound increasing with $i$.  By Theorem \ref{oldintegral} $\int_P \o_i \to \int_P \o_0.$  Since $\int_P \o_0 \ne 0$, then $\int_P \o_i \ne 0$ for sufficiently large $i$.  
\end{proof}

 \begin{corollary}\label{normj}
$\|D^j\|_j$ is a norm on  $\cal{D}_k^j.$
\end{corollary}

\begin{proof}  If $D^j \ne 0$, then $|D^j|^{\natural_{}} \ne 0.$  The result follows since $0 < |D^j|^{\natural_{j }} \le \|D^j\|_j.$  
 
 \end{proof}
 
\subsection*{Integration over chainlets}
It follows from Lemma \ref{basicinequalities} that the boundary $\p J$ of a k-chainlet $J$ of class $N^r$ is well defined as a $(k - 1)$-chainlet of class $N^{r+1}$.   If $P_i \to J$  in the $r$-natural norm define $$\p J:= \lim_{i \to \i} \p P_i.$$   By Theorem \ref{oldintegral} the integral $\int_J \o$ is well defined for k-chainlets $J$ of class $N^r$ and differential $k$-forms of class $B^r$.  If $P_i \to J$  in the $r$-natural norm define $$\int_J \o:= \lim_{i \to \i} \int_{P_i} \o.$$ 

The following extension to Stokes' theorem is valid for all chainlet domains $J$ and sufficiently smooth forms $\o$.  We remind the reader that chainlets include
smooth submanifolds, rough domains, discrete domains, soap films, fractals, cell complexes, algebraic varieties, and foliations.  They may have locally infinite mass and no tangent spaces defined anywhere.   

\begin{theorem}\label{stokes}[Generalized Stokes' theorem] If $J$ is a $k$-chainlet of class $N^r$ supported in an open set $U$ and $\o$ is a $(k-1)$-form of class $B^{r+1}$ defined on $U$,   then 
$$\int_{\p J} \o = \int_J d\o.$$
\end{theorem}

\begin{proof}  Choose polyhedral $P_i \to J$ in the $r$-natural norm.  By Stokes' theorem for polyhedral chains
$$\int_J d\o = \lim \int_{P_i} d\o = \lim \int_{\p P_i} \o = \lim \int_{\p J} \o.$$
\end{proof}

Before we continue extending basic operators to chainlets, we use the exterior derivative and boundary operator to prove a de Rham isomorphism theorem on chainlets and cochainlets that will make some of the proofs simpler.
 \section{Cochainlets and the Isomorphism Theorem}\label{isosection} A {\itshape \bfseries $k$-cochainlet} $X$ of class $N^r$ is a member of the dual space
  $(\cal{N}_k^{r})'.$   
  The $r$-natural norm of   $X \in ({\cal N}_k^{r})^{\prime}$ is defined by
$$|X|^{\natural_r} := \sup_{P \in {\cal P}_k}\frac{|X \cdot P|}{|P|^{\natural_r}}.$$
The differential operator $d$  on cochainlets is defined as the dual to the
boundary operator $dX( J) := X ( \p J).$

We study how cochainlets relate to integration of differential forms and how the operator $d$   relates to the standard exterior derivative of differential forms.   
  If
$X \in ({\cal N}_k^{r})^{\prime}$, then $dX \in ({\cal N}_{k+1}^{r-1})^{\prime}$
by Lemma \ref{basicinequalities}.

\begin{lemma}
$$|dX|^{\natural_{r-1}} \le |X|^{\natural_{r}}.$$
\end{lemma}

\begin{proof}
$$|dX|^{\natural_{r-1}}  = \sup \frac{dX(P)}{|P|^{\natural_{r-1}}} \le \sup \frac{X( \p P)}{|\p P|^{\natural_{r}}}  \le |X|^{\natural_{r}}.$$
\end{proof}

\begin{lemma}
$$|\a^{i}|^{\natural_{i}} \le \|\a^{i}\|_{i } = M(\a).$$
\end{lemma}

\begin{proof}
We have seen that $\a^{i} = \lim \s^{i}_{\ell}$ in the $(i+1)$-natural norm with $M(\s_{\ell}) \to M(\a).$  Hence
$$|\s^{i}_{\ell}|^{\natural_{i}} \le M(\s_{\ell}) \to M(\a) .$$      
\end{proof}
\begin{theorem}\label{theorem.iso} For $r \ge 1$, 
$ (\cal{N}_k^r)'\buildrel {\phi} \over \cong \cal{B}_{k}^{r} $  and the isomorphism is norm preserving.    
\end{theorem} 

\begin{proof} 
Suppose $X \in (\cal{N}_k^r)'.$    If $r \ge 1$, define $\g(\a):= X(\a).$  This is well defined coelement since $\a$ is of class $N^{1}.$   Define the differential form $\o$ by requiring that $\o_{p} = \g$ for all $p \in \R^{n}.$    Then 
$$\begin{aligned}
\left|\frac{\int_{\a^{i}} \o }{\|\a^i\|_{i}}\right | 
\le \left|\frac{X(\a^{i})}{|\a^i|^{\natural_{i}}}\right |
  \le |X|^{\natural_{i}}  \le |X|^{\natural_{r}} 
  \end{aligned}$$
  
  Similarly, $$\left|\frac{\int_{\p \a^{i}} \o }{\|\a^i\|_{i}}\right | 
\le \left|\frac{dX( \a^{i})}{|\a^i|^{\natural_{i}}}\right |
  \le |dX|^{\natural_{r-1}}  \le |X|^{\natural_{r}}.$$

Hence $ |\o|^{\natural_{r}}\le   |X|^{\natural_{r}}.$   Define $\phi(X) := \o.$

Conversely, let $\o \in \cal{B}_{k}^{r}.$  Define $X(J) := \int_{J}\o.$  By Theorem \ref{oldintegral}   
$$\begin{aligned}
|X|^{\natural_{r}} := \sup \frac{X(J)}{|J|^{\natural_{r}}} = 
\sup \frac{\int_{J}\o}{| J|^{\natural_{r}}} \le  |\o|^{\natural_{r}}. 
\end{aligned}$$  In particular $X \in  (\cal{N}_k^r)'.$  

 It follows that $\phi$ is   a norm preserving isomorphism. 
\end{proof}

Therefore, we conclude that all three definitions of the natural norm are the same.

\begin{corollary} \label{theorem.char} If $J_1, J_2 \in {\cal N}_k^{r}$
satisfy $$\int_{J_1} \o =
\int_{J_2}
\o$$ for all $\o \in {\cal B}_k^{r}$, then $J_1 = J_2$.
\end{corollary}

\begin{proof}  Let $X = \phi^{-1}(\o)$ be the cochainlet $X(J) = \int_J \o$.  By Theorem \ref{theorem.iso} $X \in  ({\cal N}_k^{r})^{\prime}. $   Hence
 $$X \cdot (J_1-J_2) = \int_{J_1-J_2}\phi(X) = 0.$$ It
follows that $J_1 =J_2.$
\end{proof}

\begin{corollary} \label{theorem.norm} If $J \in   \cal{N}_k^r$, then

$$|J|^{\natural_r} = \sup\left\{\int_J \o :   \o \in B^r_k, |\o|^{\natural_{r}} \le 1\right\}.$$

\end{corollary}

\begin{proof}  By Theorem \ref{theorem.iso}

$$
\begin{array}{rll}
|J|^{\natural_r} &= \sup\left\{\frac{|X \cdot J|}{|X|^{\natural_r}}: X \in  (\cal{N}_k^r)^{\prime}\right\} \\&= 
\sup\left\{\frac{|\int_J \phi(X)|}{|\phi(X)|_{r}}: \phi(X)
\in   {\cal B}_k^{r}\right\} \\&=   \sup\left\{\frac{|\int_J
\o|}{|\o|^{\natural_{r}}}: \o\in   {\cal
B}_k^{r}\right\}.
\end{array}
$$
\end{proof}

This result does not imply that chainlets form a reflexive space.  
In \cite{currents} it is shown that chainlets are isomorphic to {\itshape \bfseries summable currents}, a class of currents first identified at the level of distributions by Schwarz \cite{schwarz}.  For experts, these are  currents with distributional coefficients that are the $r^{th}$ distributional derivative of $L^1$ functions.   

\subsection*{Support of a form}     The {\itshape \bfseries support} of $\o$ is the closed set $supp(\o)$ such that $p \in \R^n \backslash supp(\o)$ if and only if there exists a neighborhood $B$ of $p$ in  $\R^n \backslash supp(\o)$ such that $\o(q)$ is the zero coelement for all $q \in B.$

From Theorem \ref{theorem.iso} we know 
$\cal{N}_k^r \subset (\cal{N}_k^r)^{**} = (\cal{B}_k^r)^*.$ The RHS  is a proper subspace of $(\cal{B}_{k}^r)_0^*$ which is the subspace of forms of class $B^r$ with compact support.  We conclude that chainlets of class $N^r$ form a proper subspace of currents defined as dual to smooth forms with compact support.    Thus  chainlets identify a proper subspace of   currents for which topological methods are naturally applicable.      In the space $\cal{N}_k^{\i}$ the forms do not have compact support.

\subsection*{Cup product} Given a $k$-cochainlet $X$ and a $j$-cochainlet $Y$, we
define their {\itshape \bfseries cup product} as the $(j+k)$-cochainlet  
$$X\cup Y := \phi^{-1}(\phi(X) \wedge \phi(Y)).$$
The next result follows directly from Theorem
\ref{theorem.iso}.

\begin{lemma}  Given  $X \in  (\cal{N}_k^r)^{\prime}$ and $Y \in  (\cal{N}_j^r)^{\prime}$  the cochainlet $X\cup Y \in  ({\cal
N}_{k+j}^{r}(R^n))^{\prime}$  with
                            $$|X\cup Y|^{\natural_r} = |\phi(X) \wedge \phi(Y)|_{r}.$$
Furthermore
                                $$\phi(X \cup Y) = \phi(X) \wedge \phi(Y).$$

\end{lemma}

\begin{theorem} If
  $X \in
({\cal N}_k^{r})^{\prime}, Y \in
({\cal N}_j^{r})^{\prime}, Z \in
({\cal N}_{\ell}^{r})^{\prime},$  and  $f
\in \cal{B}_0^{r+1}$, then
\begin{itemize}
\item[(i)] $|X \cup Y|^{\natural_r}
\le |X|^{\natural_r}|Y|^{\natural_r}$;
 \item[(ii)]    $d(X \cup Y) = dX \cup Y + (-1)^{j+k }  X \cup
dY;$
\item[(iii)]  $(X \cup Y) +
(Z \cup  Y) = (X+Z)\cup Y;$ and
\item[(iv)] $a(X \cup Y) = (aX \cup Y) = (X \cup aY), a \in \mathbb{F}.$

\end{itemize}
\end{theorem}

\begin{proof} These follow by using the isomorphism of
differential forms and cochainlets Theorem \ref{theorem.iso} and then applying corresponding results for
differential forms and their wedge products.

\end{proof}

Therefore the isomorphism $\phi$ of Theorem \ref{theorem.iso} is one on graded algebras.  This theorem have proved to be useful for finding lower bounds for the natural norm.  Since upper bounds are readily found from the definition of the norm, we may now calculate the $r$-natural norm precisely for simple examples such as a simplex $\s$. 
\begin{lemma}
$|\s|^{\natural_{r}} = M(\s).$ 
\end{lemma} 

\begin{proof}
This can be seen by letting $\o$  be the volume form associated to the $k$-direction of $\s$.  Then $\int_{\s} \o = M(\s)$  and $|\o|^{\natural_{r}} = 1.$ 
\end{proof}

\subsection*{Scale invariance}    The chainlet Banach spaces $\cal{N}_k^r$  are unchanged under change of scale.   The norms are comparable, leading to the same topologies.  

Let $\phi_{\l*}$ denote linear contraction (or expansion) by a constant $\l > 0.$  

\begin{theorem}\label{scale} Suppose $P$ is a polyhedral $k$-chain, $ r \ge 0,$ and $\l > 0$.  Then
$$|\phi_{\l*} P|^{\natural_{r}} \le \max\{1, \l^{k+r} \}|P|^{\natural_{r}}.$$
\end{theorem}

\begin{proof}  The proof is by induction on $r$.   
The inequality follows for $r = 0$ since  $|\phi_{\l*} P|^{\natural_{0}} = \l^k|P|^{\natural_{0}}.$   

Suppose the inequality holds for $r-1.$  Observe that 
$\|\phi_{\l*}\s^i\|_i \le \l^{k+j}\|\s^i\|_i$ for each $i$-difference $k$-cell $\s^i$.  Hence $\|\phi_{\l*}D^i\|_i \le \l^{k+j}\|D^i\|_i.$
Let $\e > 0.$  There exists $P = \sum_{i = 0}^r D^i + \p B$ such that 
$$|P|^{\natural_{r}} > \sum_{i = 0}^r \|D^i\|_i + |B|^{\natural_{r-1}} - \e.$$  Then  
$$ \begin{aligned} |\phi_{\l*}P|^{\natural_{r}}  &\le \sum_{i = 0}^r |\phi_{\l*}D^i|^{\natural_{r}} + |\p \phi_{\l*} B|^{\natural_{r}}  \\&\le  \sum_{i = 0}^r \|\phi_{\l*}D^i\|_i +  
 |\phi_{\l*} B|^{\natural_{r-1}}  \\&\le \sum_{i = 0}^r  \l^k \|D^i\|_i   + \max\{1, \l^{k+r}\} |B|^{\natural_{r-1}}
 \end{aligned} 
$$  If $\l \ge 1$, then the last term is bounded by $\l^{k+r}( \sum_{i = 0}^r  \|D^i\|_i   +  |B|^{\natural_{r-1}}).$  If $\l < 1$, the last term is bounded by  $ \sum_{i = 0}^r  \|D^i\|_i   +  |B|^{\natural_{r-1}}.$   Hence $$ |\phi_{\l*}P|^{\natural_{r}} \le   \max\{1, \l^{k+r} \} (\sum_{i = 0}^r  \|D^i\|_i   +  |B|^{\natural_{r-1}}) < \max\{1, \l^{k+r} \}(|P|^{\natural_{r}} + \e).$$  Since this inequality holds for all $\e >0$, the result follows.
\end{proof}

   The reader may also be concerned that we appear to be adding quantities of different units, e.g.,  inches and square inches.    Our units are suppressed, as is the tradition with these norms, but in our direct definition, we are dividing the last term by the basic unit, thereby solving the immediate problem.  In the natural norm, one may think of $|P|^{\natural_{}}$ and $\|D^j\|_j$ as having distance to the $k^{th}$ power and $\rho$ as having dimension of distance.     The next norm shows what happens under a change of scale for the higher dimensional term.  

Define $|P|^{\natural_{0}}_{\rho}  = M(P).$ Let $\rho >0.$  Define
$$|P|^{\natural_{r}}_{\rho} := \inf_r\left\{\liminf \sum_{j=0}^r \|D^j\|_j + \frac{|B|^{\natural_{r-1}}}{\rho}: P = \sum_{j=0}^r D^j + \p B\right\}.$$

It follows immediately from the definition that if $\rho_2 \le \rho_1$, then
$$|P|^{\natural_{r}}_{\rho_1} \le |P|^{\natural_{r}}_{\rho_2} \le \frac{\rho_1}{\rho_2}|P|^{\natural_{r}}_{\rho_1}.$$

For each $\rho > 0$ the space of polyhedral $k$-chains in the natural $\rho$-norm has a completion $\cal{N}_{\rho,k}.$  The inequalities lead at once to the following 

\begin{theorem}\label{rho}
The elements of $\cal{N}_{\rho,k}$ are independent of $\rho$, only the norms differing.  For each $\rho$, the function $\phi_{\rho}(J) = |J|^{\natural_{r}}_{\rho}$ of natural $k$-chains $J$ is continuous in   $\cal{N}_{\rho',k}$ for each $\rho'.$  
\end{theorem}

\begin{proposition}\label{epsilonversion} The seminorm $|\cdot|^{\natural_{r}}_{\rho}$ is the largest   seminorm in the space of translation invariant seminorms $|\quad|_s$ satisfying
\begin{enumerate}
\item $ |P|_s \le M(P);$
\item $|\p P|_s \le |P|^{\natural_{r-1}}/\rho;$
\item $|T_vP - P|_s \le |v| |P|^{\natural_{r}}.$
\end{enumerate}
\end{proposition}

The proof follows along the lines of the proof for $\rho = 1$ and we omit it.

\begin{theorem}
$$ |X|^{\natural_{r}}_{\rho} = \sup_{0 \le j\le r}\{\rho |dX|^{\natural_{r-1}}, \|X\|_j\}.$$
\end{theorem}

\begin{proof}  $(\ge):$ From Proposition \ref{epsilonversion}(2)
$$\rho\frac{X \cdot \p C}{|C|^{\natural_{r-1}}} \le \rho \frac{|X|^{\natural_{r}}_{\rho}|\p C|^{\natural_{r}}_{\rho}}{|C|^{\natural_{r-1}}} \le |X|^{\natural_{r}}_{\rho}.$$  Since
$$|dX|^{\natural_{r-1}} = \sup_C\frac{dX \cdot C}{|C|^{\natural_{r-1}}} =\sup_C\frac{X \cdot \p C}{|C|^{\natural_{r-1}}} $$ we conclude
$\rho |dX|^{\natural_{r-1}} \le  |X|^{\natural_{r}}_{\rho}.$

Similarly, by Proposition \ref{epsilonversion}(1) and (3),
 $$\frac{X \cdot D^j}{\|D^j\|_j} \le \frac{|X|^{\natural_{r}}_{\rho}|D^j|^{\natural_{r}}_{\rho}}{\|D^j\|_j} \le |X|^{\natural_{r}}_{\rho}.$$
 
 It follows that $$\|X\|_j = \sup \frac{X \cdot D^j}{\|D^j\|_j} \le |X|^{\natural_{r}}_{\rho}.$$
 
 $(\le):$   
 $$ \frac{\sum X \cdot D^j + dX \cdot B}{\sum \|D^j\|_j + |B|^{\natural_{r-1}}/\rho } \le \sup\{\|X^j\|_j, \rho|dX|^{\natural_{r-1}}\}.$$  The inequality follows since 
$$|X|^{\natural_{r}}_{\rho} = \sup \frac{X \cdot P}{|P|^{\natural_{r}}}$$ and $|P|^{\natural_{r}}$ can be approximated by  sums of the form 
$\sum \|D^j\|_j + |B|^{\natural_{r-1}}$ where $P = \sum D^j + \p B.$

\end{proof}
\begin{corollary}\label{rhoX}
For each $r$-natural cochainlet $X$ there is $\rho_0 > 0$ such that 
$$ |X|^{\natural_{r}}_{\rho} =  \sup_{0 \le j \le r}\{\|X\|_j\}$$ if $0 < \rho \le \rho_0.$
\end{corollary}

\begin{theorem}
For any polyhedral chain $P$
$$\lim_{\rho \to 0} |P|^{\natural_{r}}_{\rho} = \inf\left\{\sum_{j=0}^r \|D^j\|_j : P = \sum D^j\right\}.$$
\end{theorem}

  \subsection*{The supports of a  cochainlet and of a chainlet}   The
support $supp( X)$ of a cochainlet $X$ is the set of points $p$ such that for each $\e > 0$ there is a cell $\s  \subset
U_{\e}(p)$ such that $X\cdot \s \ne 0.$

 The support $supp(J)$ of a chainlet $J$ of class $N^r$  is the set of points $p$ such that for each $\e > 0$ there is a cochainlet $X$ of class $N^r$ such that $X \cdot J \ne 0$ and $X \cdot \sigma = 0 $ for each $\sigma$  supported outside $U_{\e}(p).$   We prove that this coincides with the definition of  the support  of $P$ if $P$ is a polyhedral chain. Assume $P = \sum_{i=1}^m a_i \sigma_i$ is non-overlapping and the $a_i$ are nonzero.  We must show that $supp(P)$ is the union $F$ of the $supp(\sigma_i)$ using this new definition.    Since $X \cdot P = \int_{P}  \phi(X)$  it follows that $supp(P) \subset F.$   Now suppose $x \in F$; say $x \in \sigma_i.$  Let $\e > 0.$  We find easily a smooth differential form $\omega$ supported in $U_{\e}(p)$, $\int_{\sigma_i} \omega \ne 0$, $\int_{\sigma_j} \omega = 0, j \ne i$.  Let $X$ be the cochainlet determined by $\omega$ via integration.   Then $X \cdot P \ne 0$     and $X \cdot \sigma = 0 $ for each $\sigma$  supported outside $U_{\e}(p).$  
  
\begin{lemma}
$$supp(\p J) \subset supp(J),$$ 
\end{lemma}

\begin{proof}
Suppose $\p \in supp(\p J).$  Let $\e > 0.$ There is a cochainlet $X$ of class $N^r$ such that $X \cdot \p J \ne 0$ and $X \cdot \sigma = 0 $ for each $\sigma$  supported outside $U_{\e}(p).$   Then $dX \cdot J \ne 0$ and $dX \cdot \sigma = X \cdot \p \sigma = 0$ for each cell $\sigma$  supported outside $U_{\e}(p).$   It follows that $p \in supp(J)$. \end{proof}
\begin{proposition}\label{supp}  If $J$ is a chainlet  of class $N^r$ with  $supp(J)= \emptyset$, then $J = 0.$  If $X$ is a cochainlet of class $N^r$ with   $supp(X) = \emptyset $, then $ X = 0.$
\end{proposition}

\begin{proof}  By Corollary \ref{theorem.norm} suffices to show $X \cdot J = 0$ for any cochainlet $X$ of class $N^r$.  Each $p \in supp(X)$ is in some neighborhood $U(p)$ such that $Y \cdot J = 0$ for any $Y$ of class $N^r$ with $\phi(Y) = 0$ outside $U(p)$.    Choose a locally finite covering  $\{U_i, i \ge 1\}$  of $supp(X)$.  Using a partition of unity $\{\eta_i\} $  subordinate to this covering we have $$X = \sum \eta_i X$$ and $\phi(\eta_i X) = \eta_i \phi(X) = 0$ outside $U_i$.  Hence $$X \cdot J = \sum (\eta_iX \cdot J) = 0.$$

For the second part it suffices to show that $X \cdot \s = 0$ for all simplexes $\s$.  Each $p \in \s$ is in some neighborhood $U(p)$ such that $X \cdot \t = 0$ for all $\t \subset U(p).$  We may find a subdivision $\sum \s_i$ of $\s$ such that each $\s_i$ is in some $U(p)$.  Therefore $X \cdot \s = \sum X \cdot \s_i = 0.$
\end{proof}  

\begin{lemma}
$$supp(J+K) \subset supp(J) \cup supp(K).$$
$$supp(\l J) \subset supp(J).$$
\end{lemma}

\subsection*{Change of variables}
Let $f:U \subset \R^{n} \to V \subset \R^{n}$ be differentiable.    
 
Recall $$f_{*} T_{p}\s(V^{k}) := T_{f(p)}\s(f_{*}V^{k}).$$  
 
Define $$|f|^{\natural_{r}} := \sup\frac{|f^* \o|^{\natural_{r}} }{|\o|^{\natural_{r}}}.$$  We say $f$ is of class $B^r$ if $|f|^{\natural_{r}} < \i.$

Pushforward of a $k$-cell is defined by writing it as a sum of parallelepipeds and summing their pushforwards.  This is well defined, if $f$ is of class $B^1$.  
By linearity, this leads to the definition of the pushforward of an algebraic $k$-chain $S$.

\begin{lemma}  If $f$ is of class $B^1$ and 
 $S \sim S'$, then $f_*S \sim f_*S'.$
\end{lemma}
\begin{proof}
This can be seen by taking a common cellular subdivision. 
\end{proof}

This leads to a definition of pushforward of a polyhedral chain $f_*P$.
$$f_*P := f_*S$$ where $P = [S].$
Define $$f^*X(P):= X(f_*P).$$

Since we can define the integral of $f^* \o$ over   chains of $k$-elements, we may define its norm $|f^* \o|^{\natural_{r}}.$  (It might be infinite.)  

\begin{theorem}
$$|f^{*}\o|^{\natural_{r}} \le |f|^{\natural_{r}} |\o|^{\natural_{r}}$$
\end{theorem}
\begin{corollary}
$$|f^{*}X|^{\natural_{r}} \le |f|^{\natural_{r}} |X|^{\natural_{r}}.$$
\end{corollary}
\begin{corollary}
$|f_*P|^{\natural_{r}} \le |f|^{\natural_{r}}|P|^{\natural_{r}}.$ 
\end{corollary}

\begin{proof}
$$|f_*P|^{\natural_{r}}  = \sup \frac{X(f_*P)}{|X|^{\natural_{r}}} = \sup \frac{f^*X(P)}{|X|^{\natural_{r}}}   \le 
|f|^{\natural_{r}}\sup \frac{f^*X(P)}{|f^*X|^{\natural_{r}}}   \le |f|^{\natural_{r}}|P|^{\natural_{r}}.$$
\end{proof}
We may now define $f_*J$ for a chainlet $J$ by taking limits in the natural norms.  

\begin{lemma}
 $$supp(f_*J) \subset f(supp J).$$
\end{lemma}

The proof is straightforward and we omit it.

\begin{theorem}[Change of variables]\label{variables}  Suppose $J$ is a $k$-chainlet, $f$ is a mapping of a neighborhood of $supp(J)$   and $\o$  is a $k$-form defined in a neighborhood of $supp(f_*J)$, all of class $N^r$.  Then 
$$\int_{J} f^* \o = \int_{f_*J} \o.$$
\end{theorem}

For example, let $f(x) = sin(x)$ and $J = [0, 2\pi].$   Then $f_*J = 0.$  
\begin{theorem}\label{boundthm}  If $f$ is differentiable, then
 $f_* \p   = \p f_*  .$
\end{theorem}
\begin{proof}
It follows from the definitions that $f_* \p \s = \p f_* \s$.  Therefore, $f_* \p A = \p f_* A$ for element chains $A \in \cal{A}_k^0.$  By continuity, $$f_* \p J  = f_* \lim \p A_i = \lim f_* \p A_i = \lim \p f_* A_i = \p f_* \lim A_i = \p f_* J.$$
\end{proof}
\begin{theorem}\label{commute}
$df^* = f^* d.$
\end{theorem}

 This extends the well known classical result.  
 
 \begin{theorem}[Classical change of variables]\label{classicchange} If $P$ is a polyhedral  $k$-chain, $\o$ is a smooth $k$-form and $f$ is an orientation preserving diffeomorphism  defined in a neighborhood of $P$, then
$$\int_{f P} \o = \int_P f^* \o.$$

\end{theorem}

  Using these results, we may now define chainlets on abstract manifolds.

  \section{Star theorem}    In this section, we define an operator $\perp$ on chainlets (relative to an inner product on vectors) that appears central to mathematics.   It is dual to the Hodge star operator $\star$ on forms. It is well known that $d$ and $\star$ may be combined to form other useful operators such as $\d = \star d \star$, $\D = d \d + \d d$, $d \star$ = divergence, $\star d$ = curl.     We combine $\perp$ with boundary $\p$ to form geometric duals to these operators.  But $\perp$ is even more foundational.  First of all, it restricts to multiplication by $i$. One may combine it with wedge product to extend inner product, intersection product, and cross product to element chains.

\subsection*{Star theorem}
  
\begin{proposition}
$$|\perp A|^{\natural_{r}} =  |A|^{\natural_{r}}$$ for all $A \in \cal{A}_k(K).$
\end{proposition}

\begin{proof}
We use the isomorphism theorem to deduce
$$|\perp A|^{\natural_{r}} = \sup\frac{\int_{\perp A} \o}{|\o|^{\natural_{r}}} = \sup\frac{\int_{  A} \star \o}{| \star \o|^{\natural_{r}}}  \le  |A|^{\natural_{r}}.$$
\end{proof}

We may therefore define $\perp J$ for any chainlet $J$ by taking limits:
Suppose $J = \lim A_i$ in the $r$-natural norm.  Define
$$\perp J := \lim \perp A_i.$$  Since $\{A_i\}$ forms a Cauchy sequence, so  does $\{\perp A_i\}$ and thus the limit converges.

  \begin{theorem}[Star theorem]\label{thm.star}
$\perp  : {\cal N}_k^{r} \to{\cal N}_{n-k}^{r} $ is a
   linear  
operator adjoint to the Hodge star operator on forms.  It  satisfies $\perp  \perp  =  (-1)^{k(n-k)} I$ and 
  $$\int_{\perp  J} \o =   \int_J \star \o$$   for all $J \in \cal{N}_k^r$ and
 all $(n-k)$-forms $\o$ of class $B^r, r \ge 1,$ defined in a neighborhood of $supp(J).$
\end{theorem}
This result was first announced in \cite{madeira} and will appear in \cite{hodge}

\begin{proof} Use Theorem \ref{approx} to find $J = \lim A_i$ in the $r$-natural norm.  Then
$$\int_{\star  J} \o =  \lim_{i \to \i} \int_{\star  A_i} \o =
 \lim_{i \to \i} \int_{ A_i} \perp \o =  \lim_{i \to \i} \int_{J} \perp \o.$$
\end{proof}
 \begin{lemma}     $$supp(\perp J) = supp(J).$$  
 \end{lemma}

\subsection*{Extensions of   theorems of Green and Gauss} It is not necessary for  tangent spaces  to exist for $J$ or $\p J$ for our optimal extensions of the classical theorems of Gauss and Green.     In \S \ref{classical} we show how these general theorems lead to the classical versions.

\begin{theorem}[Chainlet curl theorem]  Let $J$ be a  $k$-chainlet of class $N^r$ and $\omega$ a differential $(k-1)$-form of class $B^r$ defined in a neighborhood of $supp(J).$  Then $$\int_{\perp J} \star  d \omega =
\int_{\p J}
\o.$$
\end{theorem}  
 
\begin{proof}
This is a direct consequence of  Theorems \ref{stokes} and \ref{thm.star}.
\end{proof}

\begin{theorem}[Chainlet divergence theorem]  
Let $J$ be a  $k$-chainlet of class $N^r$   and $\omega$ a differential $(n-k+1)$-form of class $B^{r+1}$    defined in a neighborhood of $supp(J)$
then
$$\int_{\perp  \partial J} \omega = \int_{J} d\star \o.$$
\end{theorem}

\begin{proof}
This is a direct consequence of  Theorems \ref{stokes} and \ref{thm.star}.
\end{proof}

As before, tangent vectors need not be defined for the theorem to be valid and it holds in all dimensions and codimensions.

 \subsection*{Geometric representation of
differentiation of distributions}
 
  An {\itshape \bfseries $r$-distribution} on $\R^1$ is a bounded
linear functional on functions $f \in {\cal B}_0^{r}(\R^1)$  with compact support. 
Given a one-dimensional  chainlet $J$ of class $N^r$ 
define the  $r$-distribution
$\th_J$  by  $\th_J(f) := \int_J f (x)dx$,   for
$f \in {\cal B}_0^{r}(\R^1).$

\begin{theorem}  $\theta_J$ is linear and injective.  Differentiation in the sense of distributions
corresponds geometrically to the operator $ \star \p $.   That is,  $$\th_{\perp \p J} = (\th_J)^{\prime}.$$\symbolfootnote[2]{Since this paper was first submitted in 1999, the author has extended this result to currents. \cite{currents}} 
\end{theorem}

\begin{proof}
Suppose $\th_J = \th_{J'}$.  Then $ \int_J
f(x)dx = \int_{J'} f(x)dx$ for all functions $f \in
{\cal B}_0^{r}.$  But all $1$-forms $\omega \in {\cal
B}_1^{r}$ can be written $\omega = f dx.$  By Corollary \ref{theorem.char}
chainlets are determined by their integrals and    thus  
   $J = J'$.
 
  We next show that $\th_{\perp \p J} =
(\th_J)^{\prime}.$  Note that $\star (f(x) dx) = f(x).$  Thus
$$
\begin{array}{rll}
\th_{\perp \p J}(f) &= \int_{\perp \p J} f(x)dx  = \int_{\p J}
f = \int_J df \\&= \int_J f^{\prime}(x)dx =
\th_J(f^{\prime}) = (\th_J)^{\prime}(f).
\end{array}
$$
\end{proof}
\subsection*{Extension of products on pairs of elements to certain pairs of chainlets}  (Under development) Let $J$ and $J'$ be chainlets   Since wedge product of elements in $\cal{A}$ is well defined, we may extend the products defined for $\cal{A}$ such as inner product, cross product, and intersection product, to   $Vec(J)$  and $Vec(J').$   
For example, chainlets $J$ of class $N^r$ and  chainlets of the form $Ch(\o)$ where $\o$ is a differential form of class $B^r$.  (See \S\ref{xxx}.)   Since $\perp$ is defined for all chainlets, the various products on elements can be extended.  Later, we will see how to define the part of a chainlet $J$ in the cell $\s_i$ of a mesh. Then   $\sum Vec(J\lfloor \s_i)$ limits to $J$ in the natural norms as the mesh size tends to zero.  
\subsection*{Geometric coboundary of a chainlet}
Define the {\itshape \bfseries geometric  coboundary } operator $$\diamondsuit: \cal{N}_k^r\to \cal{N}_{k+1}^{r+1}$$  by $$\diamondsuit  : =
\star  \p
\star.$$ Since $\p^2 = 0$  and $\star \star = \pm I$ it follows that $\diamondsuit^2 = 0.$

 \begin{figure}[b]
\begin{center}
\resizebox{3.5in}{!}{\includegraphics*{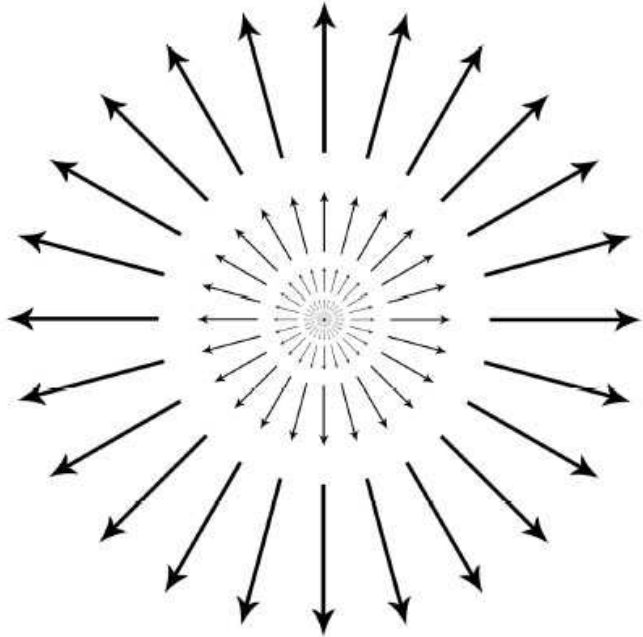}}\label{figure4}
\caption{Geometric coboundary of a  point  as a limit of polyhedral chains}
\end{center}
\end{figure}

The following theorem follows immediately from properties of boundary $\p$ and star $\star $.  Let $\d:=  (-1)^{nk +n +1}\star d \star$ denote the coboundary  operator on differential forms.  

 \begin{theorem}[Coboundary theorem] $\diamondsuit : \cal{N}_k^r \to {\cal N}_{k+1}^{r+1}$ is a nilpotent linear operator
satisfying
 \begin{itemize}
\item[(i)] $\int_{\diamondsuit J} \omega = (-1)^{n+1} \int_J  \delta \omega$  for all $\omega$ defined in a neighborhood of $supp(J)$;
\item[(ii)] $\star  \p  = (-1)^{n+k^2 +1}\diamondsuit\star;$ and
\item[(iii)] $|\diamondsuit J|^{\natural_r} \le | J|^{\natural_{r-1}}$ for all chainlets $J$. 
\end{itemize} 
 \end{theorem}
 
\subsubsection*{Geometric interpretation of the coboundary of a chainlet} This has a geometric
interpretation seen by taking  approximations by polyhedral chains.   

\subsection*{Examples}
1.  The
coboundary
of a $0$-chain $Q_0$ in $\R^2$ with  unit $0$-mass and supported in
a single point $\{p\}$  is the ``starlike'' limit of $1$-chains $P_k$   as depicted in Figure 7.

2. The coboundary of a $1$-dimensional unit cell $Q_1$   in $\R^3$  is approximated by a ``paddle wheel'',
 supported in a neighborhood of  $\s$.

3.  If  $Q_2$ is a unit $2$-dimensional square in
 $\R^3$, then its coboundary $\diamondsuit Q_2$ is approximated by the sum of
 two weighted sums of oppositely oriented pairs of small $3$-dimensional
 balls, 
 one collection slightly above
 $Q_2$, like a mist, the other collection slightly below $Q_2.$     A snake approaching the
 boundary of a lake knows when it has arrived.  A bird
 approaching the coboundary of a lake knows when it has
 arrived.
 
\subsection*{Geometric Laplace operator}

The geometric Laplace operator $$\Delta: {\cal N}_k^{r}
\to {\cal N}_k^{r+2}$$ is defined on chainlets
by $$\square  := (\p + \diamondsuit)^2  = (\p \diamondsuit + \diamondsuit \p).$$

    Let $\D$ denote the Laplace operator on  differential forms. 

\begin{theorem}[Laplace operator theorem] Suppose $J \in {\cal N}_k^{r}$ and  $\o \in {\cal
B}_k^{r+2} $ is defined in a neighborhood of $supp(J).$ Then $\square J \in  {\cal N}_k^{r+2}$,  $$|\square J|^{\natural{r+2}} \le |J|^{\natural_r},$$ and $$\int_{\square J} \o = (-1)^{n-1}\int_{J} \Delta \o.$$ 
\end{theorem}

The geometric Laplace operator on chainlets requires at least the
$2$-natural norm.   Multiple iterations of
$\square$ require   the $r$-natural norm for larger and larger $r$.
For spectral analysis and applications to dynamical systems the normed linear space
${\cal N}_{k}^{{\i}}$  with the operator  $$\square: {\cal N}_{k}^{{\i}} \to {\cal
N}_{k}^{{\i}}$$ should prove useful.  

A chainlet is   {\itshape \bfseries harmonic} if   $$\square J = 0.$$  It should be of considerable interest to 
study the spectrum of the   geometric Laplace operator $\square$ on chainlets.\symbolfootnote[1]{The
geometric Laplace operator was originally defined by the author with the objective of developing a geometric Hodge theory.}  
 
\subsection*{Examples} 1.  If we apply the geometric Laplace operator to our three coboundary examples, we get the zero chainlet.  

2.   If $J$ is a cup, then $\square J$ produces a chainlet which is a kind of smeared out monopole, matching the curvature.   If $J$ is a cap we get a similar monopole, but with opposite orientation.  
4.  If $S$ is a soap film, including branches, then $\square S = 0.$  This measures mean curvature.    

\subsection*{Geometric Dirac operator} We find a geometric dual to the Dirac operator   $G = -i(\p + \diamond).$  Then  $$D^2 = \square.$$  Furthermore, $$\int_{GJ} \o = \int_{-i(\p + \diamond)} \o = \int_J -i(d + \d) \o =   \int_J D \o.$$

\section{Chainlets in Manifolds} (Sketch, to be expanded) Suppose $M^n$ is a smooth manifold of class $C^{\i}$.  
We create a Banach space of chainlets in $M$ by forming sums of local chainlets in charts.  A Borel measure on $M$ may be used  to define the chainlet norms on $k$-elements and thus on forms in each chart, leading to a definition of $|\o|^{\natural_{r}}$ for each $r$, as well as $|\o|^{\natural_{}}$ in each chart $U$.  Since the Borel measure maintains uniformity, and by our change of variables theorem, the norms are well defined over $M$.      
$$|\o|^{\natural_{r}} : = \sup_{U}\{|\o|_{U}|^{\natural_{r}}\}$$     If $J_i$ is a local chainlet in $U_i$, then $J= \sum a_i J_i$ is a chainlet in $M$       where $$|J|^{\natural_{r}} := \sup \frac{\int_J \o}{|\o|^{\natural_{r}}}.$$

The boundary of a chainlet in $M$ is well defined as $\p J = \sum a_i \p J_i.$  Since $\p$ commutes with pushforward, then boundary of a local chainlet is well defined.     Stokes' theorem for chainlets in manifolds is valid as well as our change of variables result.

If $v$ is a vector field and $ \a$ is a field of $k$-vectors    we obtain a Lie derivative of $\a$ in the direction $v$.  Let $f_t$ be the flow of $v$.  Fix $x_0$ and define $$\cal{L}_v (\a)_{x_0} := lim_{t \to 0} \frac{f_{-t*} \a_{x_t} - \a_{x_0}}{t}.$$     If $v$ is smooth, then the limit exists.   

We similarly define $\cal{L}_v$ on fields of differential $k$-elements or order $s$.   The limit exists in the $s$-natural norm as a field of differential $k$-elements or order $s$.    

\begin{lemma}
$$\cal{L}_v \p  = \p \cal{L}_v.$$
\end{lemma}
\begin{proof}
$$\cal{L}_v \p \a = \lim \frac{f_{-t} \p \a_{x_t} - \p \a_{x_0}}{t} = \p \lim\frac{f_{-t}  \a_{x_t} -  \a_{x_0}}{t} = \p \cal{L}_v \a.$$
\end{proof}

Define $$\cal{L}_v \o(p;\a) := \o(p; \cal{L}_v \a).$$  

\begin{proposition}
The above definition coincides with the standard definition of Lie derivative of a differential form.  Furthermore, $$d \cal{L}_v = \cal{L}_v d.$$
\end{proposition}

\begin{proof}
$$d \cal{L}_v \o(p; \a) = \cal{L}_v \o(p; \p \a) = \o(p; \cal{L}_v \p \a) = \o(p; \p \cal{L}_v \a) = \cal{L}_v d\o(p; \a).$$ 
\end{proof}

Lie derivative of a chainlet with respect to a vector field $v$ on $M$ is naturally defined using our pushforward operator.   We  define $T_v J$ first by defining $T_v \a$ for a $k$-element $\a$, then define $T_v A$ for a $k$-element chain $A$, and finally we take limits to define $T_v J$ for a $k$-chainlet $J$.  Define
$$\cal{L}_v J:= \lim_{t \to 0+} \frac{T_{tv} J - J}{h}.$$   
 If the chainlet itself corresponds to a vector field, this leads to the   Lie bracket.  Wedge product of forms is well defined on a smooth manifold, and so, therefore, are exterior and interior products of certain pairs of chainlets.

\begin{theorem}
$$\int_{\cal{L}_v J} \o = \int_J \cal{L}_v \o$$ for smooth forms $\o.$
$$\cal{L}_{f_* v} f_* \a = f_* \cal{L}_v \a$$ if $\a$ is a $k$-element and $f$ is a smooth function defined in a neighborhood of $\a$.
$$\p \cal{L}_v = \cal{L}_v \p.$$
\end{theorem}

 If our manifold has a Riemannian structure, we may define the operator $\perp$ and prove the Star Theorem \ref{thm.star}.  We may then derive the divergence and curl theorems and define all the operators and products defined in Euclidean space, using $\p, \wedge$ and $\perp.$   The covariant derivative of a $k$-element chain w.r.t a vector field $u$ on $M$ is defined using a connection.   We believe this extends to a continuous operator on chainlets.  In that case, we may extend the Ricci curvature transformation to chainlet domains in a Lie algebra $$R(u,v)(J) := \nabla_u \nabla_v J - \nabla_v \nabla_u J - \nabla_{[u,v]} J.$$

 Clifford allows different definitions of product.  Instead of $\a \wedge \a = 0$, he permits $\a \wedge \a = -Q(\a)$ where $Q$ is  a quadratic form.   This idea may be applied to  the algebra $\cal{A}$ and its derived algebras.  This is expected to lead to a rich extension of the theory of chainlets, such as extensions fo semi-Riemannian manifolds.

\subsection*{Abstract chainlets} (Under development)
Recall that  a {\itshape \bfseries topological n-manifold with boundary} is a Hausdorff space in which every point has an open neighborhood homeomorphic to either an open subset of $\R^n$   or an open subset of the closed half of $\R^n$.

An {\itshape \bfseries abstract $k$-chainlet} of class $N^r$  is a Hausdorff space in which every point has an open neighborhood homeomorphic to a $k$-chainlet in $\R^n$.   We require that the transition maps be restrictions of diffeomorphisms of $\R^n$ of class $B^r.$  One may also work through a  geometric Mayer-Vietoris argument.

 Let $X$ be a closed subset of $\R^n$.  Let $\cal{N}_k^r(X)$ denote the set of all chainlets supported in $X$.    The next lemma follows since $supp(J+K) \subset supp(J) \cup supp(K).$
 
\begin{lemma}
$\cal{N}_k^r(X)$ is a subspace of $\cal{N}_k^r$.
\end{lemma}

Since the boundary operator satisfies $supp(\p J) \subset supp(J)$ we conclude that  
$$\p : \cal{N}_k^r(X) \to \cal{N}_{k-1}^{r+1}(X).$$
$$\perp:\cal{N}_k^r(X) \to \cal{N}_{n-k}^r(X).$$
$$f_*: \cal{N}_k^r(X) \to \cal{N}_k^r(\overline{fX}).$$
Therefore, each closed set $X$ has a restricted calculus, if we restrict ourselves to functions mapping $X$ into itself for our change of variables results.  

An example of interest is the topologist's sine circle $X$.   We may compute homology and cohomology groups of $X$ since we have a boundary operator defined.   

{\itshape \bfseries Betti numbers} can be calculated and Euler characteristic defined for chainlets.  
$$\chi(J) = b_0 - b_1 + b_2 - \cdots + b_n.$$    

{\itshape \bfseries Norms on homology groups} can be defined by infinimizing the natural norms of representative cycles.    If $h$ is a homology class define
$$|h|^{\natural_{r}} := \inf\{|J|^{\natural_{r}}: [J] = h\}.$$  With this, we may study continuity properties of homology classes, and thus certain topological invariants.   Dual norms exist for cohomology classes.

   \subsection*{$s$-weighted norms}  (draft)     
  There are a variety of ways of modifying the $r$-natural norms to find alternate norms.   For example, we may replace $|B|^{\natural_{0}}$ with $|B|^{\natural_{s}}, s > 0$ in the definition of the $r$-natural norm.  That is, set $|B|^{\natural_{s,0}} := |B|^{\natural_{s}}$  and recursively define
  $$|P|^{\natural_{s,r}} = \inf\{ \sum_{i=0}^r \|D^i\|_i + |B|^{\natural_{s, r-1}: P = \sum D^i + \p B}.$$
   This leads to smaller norms, and thus more chainlets.

 This class of chainlets should be useful in stochastic analysis where rough paths are so important.  Our work is valid, of course, for domains of all dimension and codimension.

\subsection*{H\"older conditions}
In the definitions of the natural norms, we may replace the norm of a translation vector $|v|$ with $|v|^{\a}$ where $0 < \a \le 1.$  This leads to dual spaces of forms satisfying H\"older conditions.
  \section{Inner product of discrete chains}   There are several inner products that arise in chainlet geometry.  The most basic are those for the algebras $\cal{A}_k$ and $\cal{C}_k.$      It is not possible to find an inner product for all pairs of chainlets.  It is well known that the space of $L^1$ functions is not a Hilbert space and these functions are naturally embedded in the chainlet spaces.  

 In \cite{ravello} there is defined an inner product over pairs of chainlets $J$ and chainlets of the form $Ch(\o)$ where $\o$ is a differential form.  $$<J, Ch(\o)>:= \int_{J \cup \perp Ch(\o)} dV = \int_J \o.$$   Chainlets of this form are dense in the space of chainlets and this is probably the best that can be done.   It is shown that this inner product restricts to the standard one given form differential forms as follows:
 $$<Ch(\a), Ch(\b)> = \int_{Ch(\a) \cup \perp Ch(\b)} dV = \int \a \wedge \star \b.$$

 This inner product restricts to a simpler expression for   chains of elements as follows.

 Let $\cal{A}_k^0(P)$ denote sections of the bundle of elements over the base set $P$.  
We form chains of elements $A = \sum a_i \a_i(p_i)$.  If $A$ and $B = \sum b_i \a_i'(p_i)$ are supported in the same set of points $P = \{p_i\}$ we may form the wedge product $$A \wedge B :=  \sum a_i b_i \a_i\wedge\a_i'(p_i).$$  Define the inner product (w.r.t. $P$) by 
$$<A,B>:= \int_{ A \wedge \perp B} dV.$$     
\begin{proposition}
$$\sqrt{<A,A>} = M(A).$$
\end{proposition}

This makes each $\cal{A}_k^0(K)$ into a Hilbert space.   An orthonormal basis in the algebra $\cal{A}_k^0$ leads to a basis of  $\cal{A}_k^0(K)$.    This inner product leads to simple matrix methods in the discrete theory for it makes $\cal{A}_k^0$ into a Hilbert space.    Since this is dense in the space of chainlets, we may take limits and apply operators and products, each time projecting into a discrete subspace.   By making use of the inner product on $\cal{A}_k = \oplus \cal{A}_k^j$ we can define an inner product,  by linearity, on $\cal{A}_k(K)$.  

As $K$ becomes more and more dense, that is, the distance between points of $K$ tends to zero,  finite matrix methods converge to the smooth continuum.    For example, the boundary operator is represented by a matrix (w.r.t a basis) since it maps $\cal{A}_k^j \to \cal{A}_{k-1}^{j+1}$ which are finite dimensional Hilbert spaces.  The matrix dimension is not large, there will just be very many  matrices, to solve simultaneously.   By continuity of the boundary, $\p A_i \to \p J$  if $A_i \to J$.   An orthonormal basis may be chosen, by way of the Gram-Schmidt process, for each $\cal{A}_k^j(p)$, for all $i, j \ge 0$ and $p \in K$.  With an orthonormal basis, coefficients for the basis are found by way of the inner product.  (This section will be expanded, but is completely standard and well known.)

 \subsection*{Inner product on discrete chains in $\cal{A}_k^j(K)$}  Of special importance is the inner product for discrete chains $A$ and $A'$, each supported in the same finite set of points $K$.  If $A = \sum a_i \a_i$ and $A' = \sum a_m' \a_m'$ define
 $$<A, A'> := \sum_i \sum_m a_i \a_m'<\a_i, \a_m'>.$$  Even though the space of chainlets is not a Hilbert space, the subpaces of discrete chainlets supported in a fixed finite set of points $K$ is a Hilbert space.   This leads to standard applications of inner product spaces including
 \begin{itemize}
 \item Gram-Schmidt methods for finding orthonormal bases;
 \item Coefficients can be calculated for discrete chains, and estimated for arbitrary chainlets.  
 \item Operators are represented as matrices.  
 
 \end{itemize}
 
 \subsection*{Numerical solutions of differential equations} (Draft) We first give a geometric interpretation of exterior derivative of a coelement: Each simple differential coelement $\g^i$ corresponds to a simple differential element $\sharp \g_k^j = \nabla_{U^j} \a(V^k).$  Its exterior derivative $d \g_k^j$ is a $(k+1)$-coelement of order $j-1$ that is a ``filled in'' version of $\g_k^j$.  That is, $\sharp d \g_k^j =  \frac{1}{k}\sum_{i=1}^j \nabla_{(U^{j-1})_i} \a(V^k) \wedge \a(u_i).$  (Here, we let   $(U^{j-1})_i$ refer to the list $U^j$ without its $i^{th}$ term $u_i$.)  

This process may be reversed, up to a constant, by way of the boundary operator.  If $\g_{k+1}^{j-1}$ is a  simple $(k+1)$-coelement of order $j-1$, then $\sharp(\g_{k+1}^{j-1})$ is a simple $(k+1)$-element of order $j-1$.  Write this as $\nabla_{U^{j-1}} \a(V^{k+1}).$  Its boundary    $\p\nabla_{U^{j-1}} \a(V^{k+1})$ has a simple expression using sums of $\nabla_{U^{j-1}} \p \a(V^{k+1}).$  Apply the flat operator to obtain a $k$-coelement  of order $j$. Now $d$ of any simple coelement $\g^0$ of order zero is zero. This lets us solve  differential equations $d \g = \eta$, roughly speaking, as $$\g = \p \eta + C.$$  More precisely, $$d\flat(\p \sharp \g + \sharp \g^0) = \g$$  and $$\p \sharp (d \g) =\sharp \g + C.$$  The star operator is invertible.  Thus, we may solve equations involving any combinations of $\star$ and $d$, at a single point, up to a choice of constants.   For example,
(omitting the $\sharp$ and $\flat$ symbols) $$d \star d \g = \eta \implies \g = \p (\star(\p \eta +C_1) +C_2.$$    Since $\D = \star d \star d + d \star d \star$, the same technique solves   the second order equation $$\D \g = \eta,$$ at a point, up to a choice of constants.  This methods extends, by linearity, to discrete chains supported in a finite set $K$.  The important question of solving such equations  for arbitrary chainlets, with   initial conditions specified, is under study.    The idea is simple:  We approximate $J$ with discrete chains, making sure that all points with conditions specified are included in the set $K$.      We may solve the equation at each $p \in K$, choosing constants   close to the constraints.  Errors should be bounded by the mesh size.     This process will converge in dimension one.  For example, this method shows  $f'(x) = f(x) $ implies $f(x) = e^x.$ In higher dimensions, there is a serious question of holonomy obstructions, and it might not be possible to find a solution.   This topic is under further study.

\subsection*{Numerical methods}(Draft)  Discretization of a manifold has many important applications -- computer-aided
design, starting points for NewtonÕs method, statistical
sampling,   finite element tessellationsÑto name but a few.   We are naturally guided by the application we have in mind.  A beautiful discretization of a manifold is given by the minimum energy points of Hardin and Saff \cite{hardinsaff}.  When combined with the discrete theory of chainlets, we have at our disposal geometric operators and the full theory of calculus.

  Suppose $M$ is a smooth, oriented $k$-submanifold with boundary.  Discretize $M$ at finitely many points $p_i,$  Form a Delauney triangulation $\t$  approximating $M$ with vertices $p_i$.  Each oriented simplex  $\s_j$ determines a $k$-direction and $k$-mass.  Let $q_j \in \s_j$ and $\a(q_j)$ be the $k$-element with the same $k$-direction and $k$-mass as $\s_j$ and supported in $q_j.$  Then the $k$-element chain $A_{\t} = \sum \a_i(q_i)$ approximates $M$.  That is, as the distance between the sample points tends to zero, the resulting element chains limit to $M$ in the chainlet topology.   Observe that we no longer use the original discretization or Delauney triangulation, but instead work with the $k$-element chain $A_{\t}.$

On the surface, this may look quite familiar, but it introduces new possibilities not seen before.  For example, we may apply the boundary operator to the element chain $A_{\t}$ and approximate the boundary of $M$.  We may integrate a form $\o$ over $A_{\t}$ and approximate the integral of $\o$ over $M$.  The flux of a vector field can be estimated by looking only at the elements of the chain $A_{\t}.$
Pushforward operator leads to dynamics of the discretization.  If $f_t$ is the flow of a vector field, then $f_{t*}J$ is well defined for all chainlets $J$.   This leads to geometrical methods for computer graphics, for example.

The element chains are finite dimensional.  When we apply the boundary operator, we obtain higher order element chains, each of which is in a finite dimensional space.  That is, the algebra $\cal{A}_k^j$ is finite dimensional for each $j$.   We have defined an inner product on these spaces.  This immediately leads to matrix methods on our element chains of any order since we have a Hilbert space.   This leads directly to numerical methods involving millions of small parallel matrices, rather than one large, sparse matrix.  Convergence is guaranteed by the chainlet continuity theorems, as long as the distance between the sample points tends to zero.  

Standard ways of speeding up convergence may be applied.  Sinai-Ruelle-Bowen measures can be used to implement chainlet geometry.  The orbit of a point under a flow with an SRB invariant measure has the convenient property that the asymptotic number of iterates in an open set is proportional to the volume of the set.  This would be optimal if the domain and form are both homogeneous.  However if either offers complexity in one region over another, it might pay to sample more points in the first region.  (This is a project under development. )

\subsection*{Meshes and chainlets}  In this section, we work with an orthonormal basis of $\R^n$ which generates an orthonormal basis $(\a_j^j)$ of each $\cal{A}_k^j$.   Suppose $\R^n = \sum Q_i$ is a mesh.  If $J$ is a $k$-chainlet with finite mass, then $$J = \sum_i  J\lfloor_{Q_i}.$$  Then $A = \sum_i Vec( J\lfloor_{Q_i})$ is a  discrete approximation to $J.$  
   Numerical methods to estimate the terms of this sum are being developed for a later version of these notes.

\subsection*{Virtual models}  We may use discrete chainlets to create virtual models for experimentation, computer graphics, movies, etc.  There are numerous ways so simulate an object.  For example, one can scan a patient's heart  with magnetic resonance imaging, and use this to create a discrete chainlet model $J_0.$   Dynamics of the pumping heart may be simulated from the pumping heart, using  the pushforward operator,   and we find a one-parameter family models in time $J_t$.   One may calculate pressure, net flux of blood flow and do virtual experiments with surgical modifications.  (See the section on solving chainlet differential equations below.)  

Fluid flows may also be modeled using the pushforward operator.  $$J_t := f_{t_*} J_0.$$    


\subsection*{ Remark}  Vector fields are modeled   on an $n$-element chain  $\sum a_i \a_i$ as $1$-elements elements supported within the $n$-elements $\a_i$.  This in marked contrast to the difficulty of modeling vector fields using discrete models based upon meshes.  $k$-vector fields are modeled similarly.  (See, for example,  \cite{marsden}.)

\section{Multiplication of a chainlet by a function}\label{multiply}
 
 Let $\a_p $ be a $k$-element and $f$ a smooth function defined in a neighborhood of $p$.   Define $$f \a_p :=  f(p) \a_p.$$  Extend to $k$-element chains $A = \sum a_i \a_i$ by $$f A := \sum a_i f \a_i.$$   Define $$f \o(\a):= \o(f \a).$$  A discrete theorem follows immediately.  
\begin{theorem}
$\int_{fA} \o = \int_{A}f \o.$
\end{theorem}
     
\begin{theorem}  Let $A$ be a $k$-element chain and $f$ a smooth function defined in a neighborhood of $supp(A)$.  JJJ  Then 
$$|f A|^{\natural_{r}} \le 2^r|f|_r | A|^{\natural_{r}}.$$
\end{theorem}

\begin{proof}  By the chain rule and the Fundamental Integral Inequality of chainlet geometry, 
 $$ \left|\int_{f A} \o\right|   = \left|\int_{A}f \o\right| \le |A|^{\natural_{r}}|f \o|^{\natural_{r}} \le   |A|^{\natural_{r}}2^r |f|_r|\o|^{\natural_{r}}.$$     By \ref{theorem.norm} 
 $$|f A|^{\natural_{r}} = \sup\frac{|\int_{f A} \o|}{|\o|^{\natural_{r}}} \le 2^r |f|_r |A|^{\natural_{r}}.$$
 \end{proof}

If  $J = \lim A_i$ is a chainlet and $f$ a smooth function  define $$fJ:= \lim f A_i.$$

\begin{theorem}
$$\int_J f \o = \int_{fJ} \o.$$
\end{theorem}

The proof follows by taking limits.
\section{Vec(J)}  

\begin{lemma}\label{Riemann}
Suppose $P$ is a polyhedral chain   and $\|\o\|_0 < \i.$   If $\o(p) = \o_0$ for a fixed coelement $\o_0$ and for all $p,$ then $$\int_P \o = \o_0 (Vec(P)).$$
\end{lemma} 

\begin{proof}
This follows for element chains $A \in \cal{A}_k^0$ since $\o_0(\a) = \o_0(Vec(\a)).$   Let $\s$ be a $k$-cell.  Choose $A_i \to \s$ such that $Vec(A_i) =  Vec(\s).$  Then $\o_0(Vec(\s) = \o_0(Vec(A_i)) = \int_{A_i} \o \to \int_\s \o.$   The result follows for polyhedral chains by linearity.    
\end{proof}

 \begin{theorem}\label{massr}
If $P$ is a polyhedral $k$-chain and  $r \ge 1$, then  $$M(Vec(P)) \le |P|^{\nat_r}. $$
If $supp(P) \subset B_{\e}(p)$  for some $p \in \R^n$ and  $ \e > 0$, then 
$$|P|^{\nat_1} \le M(Vec(P)) + \e M(P) .$$
\end{theorem}

\begin{proof} Let $\eta_0$ be a coelement such that $|\eta_0|_0 = 1$, and $\eta_0 (Vec(P)) = M(Vec(P))$. Define the $k$-form $\eta$ by $\eta(\alpha_p)  := \eta_0(\alpha) .$  Since $\eta$ is constant it follows that $\|\eta\|_r = 0$ for all $r > 0$  and $\|d\eta\|_r = 0$ for all $r \ge 0.$  Hence $|\eta|_r = |\eta|_0  = |\eta_0|_0= 1.$
  By Lemma \ref{Riemann} and Theorem \ref{oldintegral} it follows that 
$$M(Vec(P)) = \eta_0 (Vec(P) ) = \int_P \eta \le    |\eta|_r |P|^{\natural_{r}} = |P|^{\natural_{r}}.$$

For the second inequality we use Corollary \ref{theorem.norm}.  It suffices to show that $\frac{|\int_P \o|}{|\o|^{\natural_{1}}}$    is less than or equal the right hand side for any $1$-form $\omega$ of class $B^1$.   Given such $\omega$ define the $k$-form        $\omega_0(\alpha_q)  := \o(\alpha_p) $ for all $q$. By Lemma \ref{Riemann} 
$$
\begin{aligned}
 \left|\int_P \o\right| &\le \left|\int_P \omega_0\right| + \left|\int_P \omega -\omega_0\right |  \\&\le
 |\o(p)(Vec(P))| + \sup_{q\in supp(P)}|\o(p) -\o(q)| M(P) \\& \le
   \|\o\|_0M( Vec(P))  +  \e \|\o\|_1 M(P) \\& \le
   |\o|^{\natural_{1}}( M(Vec(P)) + \e M(P))    \end{aligned}
$$
\end{proof}

If $J = \lim_{i \to \infty} P_i$ in the $r$ natural norm, then $\{P_i\}$ forms a Cauchy sequence in the $r$-natural norm.   By Theorem \ref{massr} $\{Vec(P_i)\}$ forms a Cauchy sequence in the mass norm on $\cal{A}_k.$   Define $$Vec(J) := lim Vec(P_i).$$ This is independent of the choice of approximating $P_i$, again by Theorem \ref{massr}.
\begin{corollary} \label{massrcor}  $$Vec: \cal{N}_k^r \to \cal{A}_k$$ is linear and continuous. 
\end{corollary}
   
\begin{corollary} \label{massrcor2}
Suppose $J$ is a chainlet of class $N^r$ and $\omega$ is a differential form of class $B^r.$ If $\o(p) = \omega_0$ for a fixed coelement $\omega_0$ and for all $p$,  then $$\int_J \omega = \omega_0 ( Vec(J)).$$
\end{corollary}

\begin{proof}  This is merely Lemma \ref{Riemann} if $J$ is a polyhedral chain.
   Theorem \ref{massr} lets us take limits in the $r$-natural norm.  If $P_i \to J$ in $\cal{N}_k^r$, then by Corollary \ref{massrcor} $Vec(P_i) \to Vec(J).$  Therefore
   $$ \int_J \o = \lim_{i \to \infty} \int_{P_i} \o = \lim_{i \to \infty} \o_0 ( Vec(P_i)) = \o_0 ( Vec(J)).$$
\end{proof}

 \begin{proposition}\label{vecprop}  For each  nonzero simple  $k$-element $\alpha$ and $p \in \R^n$  there exists a unique chainlet $\alpha_p \in \cal{N}_k^1$ such that   $Vec(\alpha_p) = \alpha,$ $supp(\alpha_p) = \{p\}$ and $\int_{\alpha_p} \o = \o(p; \alpha)$ for all forms $\o$ of class $\cal{B}_k^1.$
\end{proposition}

\begin{proof} Let $\a_p = \lim Q_{\ell}$ be as in the construction of $\a.$.   It is unique by  Corollary \ref{theorem.char} since 
 $\int_{\alpha_p} \o = \o(p; \alpha)$ for all forms $\o$ of class $\cal{B}_k^1.$   Since  $Vec(\alpha_p) = \alpha$ we know $\alpha_p \ne 0.$  Since $supp(Q_{\ell}) \subset B_p(2^{-\ell})$, then 
  $supp(\alpha_p)$ is either the empty set or the set $\{p\}.$  By Proposition \ref{supp} $supp(J) = \emptyset \implies J = 0.$    Hence $supp(J) = \{p\}.$   
\end{proof}

\begin{theorem} Fix $p\in \R^n.$ The operator $$Vec: \cal{N}_k^r \to \cal{A}_k$$ is one-one on chainlets supported in   $p$.
 \end{theorem}

\begin{proof}    By Proposition \ref{vecprop} and 
Theorem \ref{vec}   we only need to show that if $J \in \cal{N}_k^r$ which is supported in $p$ and satisfies $Vec(J) = 0$, then $J = 0.$  Let $X$ be an $r$-natural cochainlet.  Define $X_0$ by $$\phi(X_0)(q) := \phi(X)(p)  \mbox{ for all } q.$$     By Corollary \ref{massrcor2} 
$$X \cdot J = X_0 \cdot J = \phi(X)(p) \cdot Vec(J) = 0$$ implying $J = 0.$ 
\end{proof}

\newpage
\section{Lower semicontinuity of the $s$-natural norm in the $r$-topoplogy}    
 The celebrated example of Schwarz is that of a sequence of polyhedral chains $P_i$ converging in the $1$-natural norm     to the cylinder $C$.      However the area of $P_i$ tends to infinity.   Thus we cannot hope that the mass of a chainlet $J$ can be defined as the limit of any $M(P_j)$ such that $P_j \to J.$

    Suppose $J \in \cal{N}_k^r$ and $s \ge 0 $.  Define 
$$|J|_{s,r} := \inf\{\liminf |P_i|^{\natural_{s}}: P_i \buildrel ^{\natural_{r}} \over \to J\}.$$   Then $|J|_{s,r}$ is monotone decreasing as $s, r \to \i.$    If $s \ge r$, then $J$ is naturally included in $\cal{N}_k^s$ with $|J|^{\natural_{s}}  \le |J|_{s,r}.$

A real valued function $f:\cal{N}_k^r \to \R$ is said to be {\itshape \bfseries lower semicontinuous at $x_0$} if $$\liminf_{x \buildrel {\natural_{r}} \over \to x_0} f(x) \ge f(x_0).$$ 

\begin{theorem}\label{lowersemi}  The function  $|\cdot|_{s,r}: \cal{N}_k^r \to \R$ is lower semicontinuous. 
\end{theorem}

In particular, if $s = 0$  the quantity $|J|_{0,r}$  denotes the {\itshape \bfseries mass} of a chainlet $J$ of class $N^r$.

The proof will follow from a sequence of Lemmas.

By Corollary \ref{theorem.norm}        if $P$ is a polyhedral chain $P$, $0 \le s \le r$ and $\e > 0$ 
 there exists a form $\o_1$ of class $B^{s}$ such that $$|P|^{\natural_{s}} \le \int_P \o_1 +\e $$ and $|\o_1|^{\natural_{s}} = 1.$

By the Weierstrass approximation theorem,  for each $ \e > 0$ there exists a form $\o_0$ of class $C^{\i}$ such that $|\o_0 -\o_1|^{\natural_{s}} < \e.$ 

\begin{lemma}\label{infinity}
 If $P$ is a polyhedral chain, $0 \le s \le r$, and $\e > 0,$ 
 there exists a form $\o_0$ of class $C^{\i}$ such that $$|P|^{\natural_{s}} \le \int_P \o_0 +\e $$ and $|\o_0|^{\natural_{s}} = 1.$
\end{lemma}

\begin{lemma}\label{massofP}  If $P$ is a polyhedral chain, then
$$|P|_{s,r} = |P|^{\natural_{s}}$$ for all $0 \le s \le r.$
\end{lemma} 

\begin{proof}
We know $|P|_{s,r} \le |P|^{\natural_{s}}$ by choosing $P_i = P.$

Conversely,   suppose there exists $P_i \buildrel {\natural_{r}} \over \to P$ such that $\liminf |P_i|^{\natural_{s}} < |P|^{\natural_{s}} - C$ for some $C > 0.$   
 By Lemma \ref{infinity}   there exists a form $\o_0$ of class $C^{\i}$ such that $$|P|^{\natural_{s}} \le \int_P \o_0 +C/2  $$ and $|\o|^{\natural_{s}} = 1.$ 
By continuity of the integral \ref{oldintegral} 
 $\int_{P_i} \o_0 \to \int_P \o_0.$  
 Therefore, for sufficiently large $i$, $$|P_i|^{\natural_{s}} < \int_{P_i} \o_0 -3C/4 \le \sup \frac{\int_{P_i} \o}{|\o|^{\natural_{s}}} -C/4 = |P_i|^{\natural_{s}} -C/4.$$    This contradicts our assumption.

\end{proof}

\begin{proof}[Proof of \ref{lowersemi}]  Let $P_j \buildrel {\natural_{r}} \over \to J.$  Then  $$\liminf |P_j|_{s,r}=  \liminf |P_j|^{\natural_{s}} \ge |J|_{s,r}.$$
\end{proof}

\begin{corollary}\label{convergemass}  Suppose $J$ is a chainlet of class $N^r$ with $|J|_{s,r} < \i.$  
There exists $P_i \buildrel {\natural_{r}} \over \to J$ such that $|P_i|^{\natural_{s}} = |P_i|_{s,r} \to |J|_{s,r}.$
\end{corollary}

Later we will show how to choose $P_i$ supported in a neighborhood of $supp(J)$ but at this stage, such a result is not obvious.   
  \begin{theorem}  If $J$ is a chainlet with finite mass, then
$J \wedge \perp J$ is well defined with   $$M(J)^2 =  \int_{J \wedge \perp J} dV.$$
\end{theorem}    

\begin{proof}
Approximate $J$ with element chains $A_i \to J$ and $M(A_i) \to M(J)$.  The result follows since $M(\a)^2 =  \int_{\a \wedge \perp \a} dV.$ and therefore $M(A)^2 =  \int_{A \wedge \perp A} dV.$
\end{proof}

\subsection*{Dimension}   The {\itshape \bfseries topological dimension} of a chainlet $J \in \cal{N}_k^r$ is defined as 
$$dim_T(J) := k.$$ This is an intrinsic dimension and not especially related to properties of $supp(J)$.  For example, we have seen that a $k$-element $\a$ is supported in a single point, but has topological dimension $k$. 

 The function $\phi_J(s) = |J|_{s,r}: [0,\i) \to [0,\i]$ with values in the extended reals for a fixed $J \in \cal{N}_k^r$   provides tools for classifying chainlets.   We have seen that this is monotone decreasing.  It might start with $\phi_J(0) = \i$.  We know that $\phi_J(r) < \i.$  Roughly speaking, at the point $d$ where it first becomes finite, we define $dim_E(J) : = dim_T(J) +d.$

 More precisely,  the {\itshape \bfseries extrinsic dimension} $$dim_E(J) := k+ \inf\{s: \phi_J(s) < \i\}.$$ This quantity turns out to have important implications for calculus.   Since it is defined using norms rather than measures,
extrinsic dimension is sensitive to both algebraic and geometric properties of the domain
and takes into account orientation and multiplicity of approximating polyhedral chains.      It is a dimension tied to algebraic properties of a domain, not just set theoretic properties, as is the case with other well known dimensions such as Hausdorff and box dimensions. 

 Extrinsic dimension acts as a real-valued function on the space of chainlets. Unlike the $|\,|_{s,r}$ norm, it is not semi-continuous in the
$r$-natural topology.  Consider two examples.  The first is a sequence of fractal curves,
with constant dimension $1 +
\a$,
  converging in the
$1$-norm to a line segment.  All one needs to find such an example is to construct
``snowflake-type'' arcs that have the same endpoints as the line segment such that the
area between them tends to $0$.  The extrinsic dimension drops suddenly, at the limit,
showing that $dim_E$ is not upper semi-continuous.  On the other hand, consider any
sequence of polyhedral  chains converging to the snowflake in the $1$-norm.  The
extrinsic dimension increases suddenly, at the limit, showing $dim_E$ is not lower
semi-continuous.

\begin{proposition} $$dim_T(\p J) = dim_T(J) -1.$$
$$dim_E(\p J) \le dim_E(J).$$
\end{proposition}

\begin{proof}
This follows since $|\p P|^{\natural_{r}} \le |P|^{\natural_{r-1}}.$
\end{proof}
 One could imagine an embedded disk in
$\R^3$ with smooth boundary and fractal interior so that  $dim_E(\p A) = 1$ and
$dim_E(A)  > 2.$ On the other hand, one could have a fractal boundary and smooth
interior so that $dim_E(\p A) = dim_E(A) = 2.$

\medskip \noindent  {\bf Examples} 

\noindent \medskip (i)  $k$-elements $\a$.     $$dim_E(\a) = dim_T(\a) = dim_S(\a) = k.$$  

(ii)  $k$-elements of order $j$  $$dim_E(\a^j) = dim_S(\a^j) = k+j.$$

 After this, various possibilities arise for the qualitative behavior of the graph of $\phi_J(s)$ beyond monotonicity.  We will see examples such as the Sierpinski Gasket for which $\phi_J(s) = 0$ for all $s > dim_E(J) - dim_T(J)$  and $\phi_J(s) = \i$ for all $s < dim_E(J) - dim_T(J).$\symbolfootnote[1]{This looks very much like the graph of Hausdorff $s$-measure.}   As a chainlet, the Sierpinski Gasket barely exists.  If we integrate sufficiently smooth forms around the limits of tiny cycles, we get zero.  Rougher forms can give nonzero answers, even infinity.\symbolfootnote[2]{It turns out that  the critical degree of smoothness is the   Hausdorff dimension of the Sierpinski Gasket.}   
 
 Now consider a chainlet of the opposite extreme, a $k$-cell $\s$.  We have seen that $|\s|^{\natural_{r}} = M(\s)$ for all $r \ge 0.$   Thus $\phi_{\s} (s) = M(\s)$ for all $s \ge 0.$   The same feature holds for $\phi_{\a} = M(\a)$ where $\a$ is a $k$-element.    However, if $\a^1$ is a $k$-element of order $1$, then 
 $\phi_{\a^1}(s) = \|\a^1\|_1$ for all $s \ge 1$ and $\i$ for all $s < 1.$  
 
Exercise:  Prove that if $\a^j$ is a $k$-element of order $j$, then $\phi_{\a^j}(s) = \|\a^j\|_j$ for all $s \ge j$ and $\i$ for all $s < j.$  (Hint:  Use the isomorphism theorem.)

It follows from the exercise that extrinsic dimension of a chainlet may exceed the topological dimension of its ambient space $\R^n$. 

This brings us to a third natural dimension for chainlets, the {\itshape \bfseries stabilizing dimension} 
 $$dim_S (J) :=  dim_T(J) + \inf\{ s: \phi_J(t) = \phi_J(s),  \forall t \ge s\}.$$

We have $dim_T(J) \le dim_E(J) \le dim_S(J).$

\newpage
\section{Integration of measurable forms over nonsmooth domains} The real importance of extrinsic dimension is that it tells us how smooth a form has to be for integration to be possible.   Much of mathematical analysis deals with forms that are not smooth but defined over smooth domains.   Our emphasis has been on the study of smooth forms defined over nonsmooth domains.    Now we consider nonsmooth forms over nonsmooth domains.

\begin{proposition}\label{dog} If
 $\o \in \cal{B}^s$, $0 \le s \le r$, and $P$ is a  polyhedral chain,  then $\left| \int_P \o\right| \le |P |_{s,r} |\o|^{\natural_{s}}.$  
\end{proposition}

\begin{proof}
Apply Theorem \ref{oldintegral} and Lemma \ref{massofP}.   
\end{proof}

If $J \in \cal{N}_k^r$ with $|J|_{s,r} < \i$ and $\o \in \cal{B}^s$  define  
$$\left| \int_J \o \right| :=  \inf\left\{ \liminf \left|\int_{P_j} \o \right| : P_j \to J \right\} $$

We next remove the absolute value signs.   Suppose $P_j \buildrel {\natural_{r}} \over \to J$ and $P_j' \buildrel {\natural_{r}} \over \to J$ with  
$\int_{P_j} \o \to \left| \int_J \o \right|$ and $\int_{P_j'} \o \to -\left| \int_J \o \right|$.  Then $\int_{(P_j + P_j')/2} \o \to 0.$  Since  $\frac{P_j + P_j'}{2} \buildrel {\natural_{r}} \over \to J$
it follows that  $\left| \int_J \o \right| = 0.$  We may therefore define
$\int_J \o  = \pm \left| \int_J \o \right|$, accordingly.

\begin{theorem} If $J \in \cal{N}_k^r$ with $|J|_{s,r} < \i$ and $\o \in \cal{B}^s$, then $$\left| \int_J \o\right| \le |J|_{s,r} |\o|^{\natural_s}.$$
\end{theorem}

\begin{proof}   Suppose $P_j \buildrel {\natural_{r}} \over \to J$.  By Proposition \ref{dog} $$\left|\int_J \o\right| \le \liminf \left| \int_{P_j} \o \right| \le \liminf |P_j|_{s,r}|\o|^{\natural_s}.$$  Therefore,
$$\left|\int_J \o\right|    \le \inf\{\liminf |P_j|_{s,r}|\o|^{\natural_s}: P_j \buildrel {\natural_{r}} \over \to J\}|\o|^{\natural_s} =  |J|_{s,r} |\o|^{\natural_s}.$$
\end{proof}

 \begin{corollary} If $\o \in \cal{B}^s$ where $s \ge \max\{0, dim_E(J)\}$, then $$\left|\int_J \o\right| \le |J|_{s,r} |\o|^{\natural_{s}}.$$
\end{corollary}

Whitney's celebrated paper, ''A function nonconstant on a connected set of critical points has been modernized by \cite{norton}.   We take a subarc $\g$ of the Van Koch curve with endpoints $p$ and $q$ as a specific example.  It has Hausdorff dimension $\l = ln4/ln/3.$  The Hausdorff measure of $\g$ is finite within its dimension.  For $x \in \g$, define $f(x)$ to be the Hausdorff  $\l$-measure of the arc from $p$ to $x$.  This extends to a function of class $C^{\l}$ defined in a neighborhood of $\g$.   Furthermore, both $\p f/\p x$ $\p f/\p y$ are zero for each point in $\g$.   (See \cite{norton}.) 
The oriented curve $\g$ is represented canonically by a chainlet $G$ with finite $r$-natural norm for $r > \l-1$.   It is not, therefore, a domain of integration in the full theory of calculus for the 1-form $df$ which is of class $C^{\l -1}.$    However, the  $\l-1$-norm is finite.  Thus the extended integral is defined and half of Stokes' theorem is valid.  It remains to verify that $\int_{\g}df = 0.$    Thus if the differentiability class of a form drops below $dim_E(J) - dim_T(J)$, then calculus fails.   Either the integral is not defined or there is no Stokes' relation of any kind.  
   
\begin{theorem}[Stokes' inequality] Let $J \in \cal{N}_k^r.$  If $\o \in \cal{B}^s$, $s \ge 1$, and $dim_E(J)  - dim_T(J) \le  s$, then $$\left|\int_{\p J} \o\right| \le \left|\int_{ J} d \o \right|.$$
\end{theorem}

\begin{proof}  By Stokes' theorem 
$$
\begin{aligned} \left| \int_{J} d\o \right| &= \inf \left\{ \liminf \left|\int_{P_j} d\o\right|: P_j \buildrel \nat_r \over \to J\right\} \\&= \inf \left\{ \liminf \left|\int_{\p P_j} \o\right|: P_j \buildrel \nat_r \over \to J\right\} \\&\ge
\inf \left\{ \liminf \left|\int_{Q_j} \o\right|: Q_j \buildrel \nat_{r-1} \over \to \p J\right\}
\\&= \left|\int_{\p J} \o \right|.
\end{aligned}
$$
\end{proof}

Problem.  Find an example where there is a strict inequality.  That is, 
$\left|\int_{\p J} \o\right| < \left|\int_{ J} d \o \right|.$

A similar result holds for the change of variables result.

\begin{theorem} Let $J \in \cal{N}_k^r.$  Suppose $f$ is a mapping of class $B^s$ defined in a neighborhood of $supp(J)$.  If $\o \in \cal{B}^s$, $s \ge 1$, and $dim_E(J)  - dim_T(J) \le  s$, then $$\left|\int_{f_* J} \o\right| \le \left|\int_{ J} f^* \o \right|.$$
\end{theorem}
  
 The proof is much like that of the preceding result and we omit it.  If $f$ is invertible, then we obtain an equality.

 \begin{theorem} Let $J \in \cal{N}_k^r.$  Suppose $f$ is a diffeomorphism of class $B^s$ defined in a neighborhood of $supp(J)$.  If $\o \in \cal{B}^s$, $s \ge 1$, and $dim_E(J)  - dim_T(J) \le  s$, then $$\left|\int_{f_* J} \o\right| = \left|\int_{ J} f^* \o \right|.$$
\end{theorem}

The star theorem carries over to this category as an equality.
\begin{theorem} Let $J \in \cal{N}_k^r.$    If $\o \in \cal{B}^s$, $s \ge 1$, and $dim_E(J)  - dim_T(J) \le  s$, then $$\left|\int_{\perp J} \o\right| = \left|\int_{ J} \star \o \right|.$$
\end{theorem}

As corollaries we obtain versions of the divergence and curl theorems.  

\begin{corollary}[Divergence inequality] Let $J \in \cal{N}_k^r.$   If $\o \in \cal{B}^s$, $s \ge 1$, and $dim_E(J)  - dim_T(J) \le  s$, then $$\left|\int_{\perp \p J} \o\right| \le \left|\int_{ J} d \star \o \right|.$$
\end{corollary}

\begin{corollary}[Curl inequality] Let $J \in \cal{N}_k^r.$   If $\o \in \cal{B}^s$, $s \ge 1$, and $dim_E(J)  - dim_T(J) \le  s$, then $$\left|\int_{ \p  \perp J} \o\right| \le \left|\int_{ J}   \star d \o \right|.$$
\end{corollary}
\newpage

Our generalized deRham's theorem  \ref{theorem.iso} has direct application to specific examples for it leads
to a relatively simply method for computing the fractal dimension of chains.

\noindent \medskip (ii) {\itshape \bfseries polyhedral  chains} It follows from Lemma \ref{xxx}
that
$dim_S(\s) = dim_T(\s) = k$ for all $k$-cells $\s$.  However, if
$P$ is a polyhedral  chain, it might satisfy $|P|^{\natural_{1}} < M(P).$  An example would be a
polyhedral  chain $P$ approximating the Van Koch snowflake $S$.  Its $1$-norm is
bounded by the   area inside $S$, the
$0$-norm is the length of $P$.   If $P$ is a polyhedral  chain  $dim_E( P) = dim_T(P)$ 
since its mass is finite.

\medskip \noindent  {\bf Problem}  Show the stabilizing dimension of a polyhedral 
chain is finite.

This leaves open the possibility that the extrinsic dimension of the snowflake itself
strictly exceeds $1$.   (See below.) 

\medskip \noindent  {\bf Upper bounds for dimension}  

We show how the isomorphism theorem for forms and cochainlets (Corollary
\ref{theorem.iso}) gives a practical method for finding lower bounds.  Upper bounds,
however, can be found by considering well-chosen decompositions for $P =\sum D^i + \p B.$
\medskip \noindent  {\bf Examples} 

(iv) {\itshape \bfseries Middle third Cantor set $\G$}  Let $I_k$ denote the sum of the $2^k$
oriented middle third intervals that have been ``removed'' to form $\G$ at the
$k^{th}$ stage and define the chain
$C_k = I - 
\sum I_k.$  Let
$\d >
\frac{ln2}{ln3}.$  The Cantor set $\G$ naturally supports a   $1$-chainlet of class $\cal{N}^{1+d}$
$C$ that is the limit  of the $C_k.$    The boundary $\p C$ is
therefore a  $0$-chainlet of class $\cal{N}^{d}$, also with support $\G$.  This gives additional algebraic and
geometric structure to the Cantor set.  It can be viewed either as a natural
$0-$chain or
$1-$chain.  In both cases, the extrinsic dimension is bounded above by
$\frac{ln2}{ln3}$.

(v) {\itshape \bfseries Van Koch snowflake $S$}  Let $P_k$ be the oriented $1$-dimensional polyhedral 
approximators to
$S$ that satisfy $P_k - P_{k+1} = \p(T_k)$ where $T_k$ is a sum of 
$4^k$ oriented triangles of diameter $3^{-k}$.  By considering the weighted area of
$T_k$ it is easy to verify that $S$ supports a   $1$-chainlet $P$ with $dim_E(P)
\le
\frac{ln4}{ln3}.$  As with the Cantor set, one could view
$S$ as a
$2$-chainlet with the same upper bound for its extrinsic dimension.
(vii) {\itshape \bfseries Graphs of H\"older continuous functions}   The graph of any measurable
function is in ${\cal N}_1^1$ since the Lebesgue area of its subgraph is finite.   

We show the graph of an $\a$-H\"older function is represented in ${\cal N}_1^{\b}$
where
$\b > 1-\a$.   Thus $dim_E(\G) \le 2 - \a.$  The graphs of approximating step
functions form a Cauchy sequence in the norm.  This follows directly from the
definition of the
$\b$-norm.  The small rectangles between the approximating curves have a base $B$ and
height $H$.
$H$ is on the order of $B^{\a}$. Consider the $\b$-weighted mass by raising $B$ to the
power
$\b$. If
$\b > 1 -
\a$, then  the series is small, tending to 0 as the partition size tends to 0.  This
makes sense because if $\a = 1$, then $\b = 0$, corresponding to the Lip case where
the graph is rectifiable and so is represented in $A_{1,0}$.  If $\a = 0$, then $\b =
1$,  corresponding to the continuous case that is part of the measurable case above.  

(viii) {\itshape \bfseries Graphs of measurable functions}

(ix)  {\itshape \bfseries Curves in $\R^3$ with extrinsic dimension > 2}

\medskip \noindent {\bf Lower bounds for extrinsic dimension}\label{sec.bound}

 dimension since box dimension is bounded by the topological dimension of the ambient
space.   

\begin{theorem} If $\g$ is an oriented Jordan curve in the plane with measure $0$, then 
$$dim_E(\g) = dim_B(\g) \ge dim_H(supp(\g)).$$
\end{theorem}
Extrinsic dimension can be roughly thought of as box dimension with
multiplicity. 
\begin{proof} Suppose  $dim_B(\g) = d$.  We show $d$ is an upper bound for
$dim_E(\g)$ by finding $\a = d-1$-weighted spanning chains  for $\g$.  Consider the
Whitney decomposition of the domain $D$  interior to $\g$.  Since the box dimension of
$\g$ is $d$, it follows that there are $\le 2^{kd}$ Whitney squares $\s_i$ of side
length
$2^{-k}.$ Hence for $\b > \a$, the weighted mass $|D|_{\b-1} < \i.$  Hence 
$dim_E(\g) \le
\a.$  

In order to show $d$ is a lower bound, we find a differential $1$-form
$\o$ of class $C^{\b}$ such that $\int_{\g} \o = \i.$  Let $A_k$ denote the boundary
of the Whitney cubes of size $\ge 2^{-k}$.  Then $A_k \to \g$ in the $\b$-norm.  (The
$\b$-weighted area between $\g$ and $A_k$ tends to $0$.   Hence    $\int_{\g} \o
=\lim^{\b}
\int_{A_k} \o.$   Let $\o$ be the $1$-form corresponding (via the Euclidean inner
product structure) to the vector field tangent to the $\p A_k$ and with length
$2^{-k\b}.$  It follows from the Whitney extension theorem that $\o$ extends to a
neighborhood of $D$ and is of class $C^{\b}.$  Furthermore, $ lim \int_{A_k} \o =
\i.$  It follows that 
$dim_E(\g)
\ge \a.$
\end{proof}

   DICTIONARY 

Hausdorff dimension  $\leftrightarrow$ extrinsic dimension

$s$-measure $\leftrightarrow$ $s$-norm

measurable sets $\leftrightarrow$ normable chainlets
    
Conjecture:  $\cal{H}_s(supp(J)) \le |J|_{s,r}.$  Or, $|J|_{s,r} < \i$ implies $\cal{H}_s(supp(J)) < \i.$
See Fleming for something related.  

\subsection*{Expansion and contraction}   Recall the exterior and interior products: $e_{\b}$ and $i_{\b}.$  These are sometimes called expansion and contraction operators.     
\begin{proposition}  If $A \in \cal{A}$ then
$$|e_{\b} A|^{\natural_{r}} \le |A|^{\natural_{r}}.$$
\end{proposition}
 

 Therefore,  the expansion operator extends naturally to chainlets $J$.  
 
\begin{theorem}
 $$\int_J i_{\b} \o = \int_{e_{\b} J} \o.$$
\end{theorem}

 \section{Borel measures associated to a chainlet}     
 
 Suppose $J $ is a $k$-chainlet with finite mass and $X \subset
\R^n$ a Borel set.     We show how to define a chainlet $J\lfloor_X$ and a Borel measure $\mu_J$  such that  $\mu_J(X) = M(J\lfloor_X).$ 
We   call $J\lfloor_X$ the {\itshape \bfseries part of} $J$ in $X$.   $Vec(J\lfloor_X)$ becomes a $k$-element valued finitely additive measure $\mu_J$.

We say a sequence   $P_j \buildrel
\natural_r\over\longrightarrow J$    is a {\itshape \bfseries mass sequence} if   $M(P_j) \to M(J).$    By lower semicontinuity of mass, such a sequence always exists, even if $M(J) = \i.$ 

  A polyhedral chain $P$ generates a measure $\mu$ satisfying  $\mu(Q) = M(P \lfloor_Q)$  for every  $n$-dimensional cell $Q$.  These are measures since for
pairwise disjoint
$\{Q_i\}$  we have   $M(P \lfloor_{\cup Q_i}) = M(\cup (P \lfloor_{Q_i}) = 
\sum_i M(P
\lfloor_{Q_i}).$    
 Hence $\mu$ generates a Borel measure.   
     
Let $P_j \buildrel \natural_r\over\longrightarrow  J$ be a mass sequence.  For $j = 0,1, \dots,$  let   $\mu_j$
  be the measure determined by    $\mu_j(Q) =
M(P_j \lfloor_Q)$.     
 Then      $\mu_j(\R^n) =
M(P_j)$ is  bounded.    Therefore the
collection  $\{M(P_j\lfloor_Q)\}$ is   uniformly bounded over all $Q$.  By taking   subsequences  we may assume
that $M(P_j
\lfloor_Q)$ converges to a
limit denoted $\mu_J(Q). $   Thus $\mu_j$ tends weakly to a limit  measure   $\mu_J.$ 
Later we show that   $\mu_J$
is independent of any mass sequence $P_j$.  We show   $\mu_J$ 
  defines  a compactly-supported  Borel
measure of finite total variation.   

By taking a further subsequence, we may assume $$\sum |P_{j+1}
- P_j|^{\natural_{r}} < \i.$$
    We say that  $Q$
is  {\itshape \bfseries exceptional} w.r.t $\{P_j\}$  if either
\begin{enumerate}
\item
  $\sum  |(P_{j+1} - P_j)\lfloor_Q|^{\natural_{r}} = \i$ or 
  \item $\mu_J(\p Q) > 0.$
  \end{enumerate}
Suppose $Q$ is nonexceptional w.r.t. $\{P_j\}$.   Then
the sequence $P_j \lfloor_Q$ tends to a limit denoted $J
\lfloor_Q $ in the $r$-natural norm.   
 Since $\mu_J(\p Q) = 0$ it follows that   $ \mu_j(Q) \to \mu_J(Q),$ and
$\mu_j(Q^c) \to \mu_J(Q^c)$.

We say that $Q$ is {\itshape \bfseries nonexceptional}  w.r.t. $J$ if there exists a mass sequence $P_j \to J$ and $Q$ is nonexceptional w.r.t.
$\{P_j\}.$
 
\begin{lemma} \label{lemexcep} If $Q$ is nonexceptional w.r.t. $J$,, then
$$M(J \lfloor_Q) =
\mu_J(Q), M(J - J
\lfloor_Q) = \mu_J(Q^c).  $$  
\end{lemma}  

\begin{proof}  Since $P_j \lfloor_Q \buildrel \natural_r\over\longrightarrow  J \lfloor_Q$ and $P_j -(P_j \lfloor_Q)
\buildrel \natural_r\over\longrightarrow J - (J \lfloor_Q)$, we use lower semicontinuity of $M$ in the $r$-natural norm to deduce $$M(J
\lfloor_Q) \le \liminf M(P_j \lfloor_Q) = \liminf \mu_j(Q) = \mu_J(Q).$$ Similarly
$M(J - (J\lfloor_Q)) \le \mu_J(Q^c).$  But
$$\mu_J(Q) + \mu_J(Q^c) = \mu_J(\R^n) = M(J) \le M(J \lfloor_Q|) + M(J - (J \lfloor_Q)|).$$  It follows that $\mu_J(Q) = M(J \lfloor_Q|)$ and $\mu_J(Q^c) = M(J-(J \lfloor_Q)|).$    
\end{proof}

\subsection*{The part of a chainlet in a Borel set}   
 Let $J $ be a chainlet with finite mass.
  If $X = Q_1 \cup Q_2 \cup \cdots Q_p$, where the $n$-cells $Q_i$ are non-overlapping and nonexceptional w.r.t.
$J$, then we set $J\lfloor_X = \sum_{i=1}^p J \lfloor_{Q_i}.$    Since each $ J \lfloor_{Q_i}$  is a chainlet, so is     $ J \lfloor_X.$
  Applying Lemma
\ref{lemexcep} for cells repeatedly, 
$$M(J\lfloor_X) = \sum \mu_J(Q_i) = \mu_J(X)$$ and  
$$M(J- (J\lfloor_X)) = \mu_J(X^c).$$  

Any $\mu_J$-measurable set $X$ is the limit in $\mu_J$-measure of a
sequence $X_j$ of
the above type.   Then $J\lfloor_X$ is defined as the limit in
$M$ of the sequence $J\lfloor_{X_j}$   Thus $J\lfloor_X$  is a chainlet. The
$\cal{N}_k^r $-valued set function
$J \lfloor_{\cdot}$   so defined is countably additive with $M(J\lfloor_X) = \mu_J(X)$   for every Borel set $X$.   
 
\begin{corollary}\label{exceptional}  If $J$  is a chainlet with   finite mass and $X$ is a $\mu_J$-measurable set  there exists
a unique  chainlet $J\lfloor_X$ such that 
   $M(J\lfloor_X) =
\mu_J(X)$ and  
$ M(J - J
\lfloor_X) = \mu_J(X^c).$  
\end{corollary}
     \subsection*{Fubini inequalitites} 
Fix $s \in \R$ and let $H_s$ be a half-space of $\R^n$, $H_s := \{(x_1, \dots, s, \dots, x_n) \in \R^n\}. $ 
For a $k$-cell $\sigma$, define the {\itshape \bfseries slice}
$\sigma_s = \partial (\sigma \cap H_s) - (\partial \sigma) \cap
H_s.$

Exercise.  Show that $\sigma_s$ is a $(k-1)$-cell.   Thus if $S = \sum a_i \s_i$ is an algebraic chain, then $S_s = \sum a_i \s_{is}$ is an algebraic chain, the slice of $S$.  

\begin{lemma}  If $S \sim S'$ are algebraic $k$-chains, then $S_s \sim S'_s$ are algebraic $(k-1)$-chains.

\end{lemma}

\begin{proof} The proof reduces to showing the following: If $S'$ is a non-overlapping subdivision of a simplex $\s$, then each slice $S'_s$ is a non-overlapping subdivision of the slice $\s_s.$  This follows from the definitions.    
\end{proof}

Define  the {\itshape \bfseries slice} of a polyhedral chain $P$ by
$$P_s = \partial (P \cap H_s) - (\partial P) \cap H_s.$$   It follows from the previous lemma that $P_s$ is well defined as a  polyhedral chain since $P$ is a polyhedral chain. Furthermore,   $(\p P)_s = \partial ( \p P \cap H_s) = \p(P_s).$
 
 \begin{lemma}[Fubini inequalities for   polyhedral chains] \label{lintegral}Let  $P$ be a polyhedral chain.  Then  
 $$\int_{-\infty}^{\infty} |P_s|^{\natural_{r}} ds \le |P|^{\natural_{r}}.$$    
\end{lemma}
 
\begin{proof}    Let $r = 0.$
First note the   $k$-chain estimate  $$\int_{-\infty}^{\infty} M(\sigma_s)ds \le M(\sigma)$$

 Suppose $S = \sum a_i \sigma_i$ is a non-overlapping algebraic chain.   Then
$S_s = \sum a_i  \sigma_{is} $  is a non-overlapping algebraic chain a.e. $s$ and 
$$
\begin{array}{rll}
\int_{-\infty}^{\infty}M(S_s)ds   =  \sum  |a_i| \int_{-\infty}^{\infty}M(\sigma_{is}) 
 \le \sum |a_i|M(\sigma_i)     =  M(S)  .
\end{array}
$$
  If $S \sim S'$ is a polyhedral chain, then $S_s \sim S'_s$ a.e. $s$.  
    Therefore, $$  \int_{-\infty}^{\infty} M(P_s) ds  
\le M(P).$$  
The proof proceeds by induction.  
Let $\e > 0.$  There exists $P = \sum_{i=0}^r D^i + \p B$ such that 
$$|P|^{\natural_{r}} > \sum_{i=0}^r \|D^i\|_i + |B|^{\natural_{r-1}} - \e.$$  Then $P_s = \sum_{i=0}^r D_s^i + \p B_s$  
Hence $$\int |P_s|^{\natural_{r}}ds \le \sum_{i=0}^r \int |D_s^i|^{\natural_{r}}ds +  \int |B_s|^{\natural_{r-1}}ds \le K(\sum_{i=0}^r \|D^i\|_i + |B|^{\natural_{r-1}}).$$
  
\end{proof}

\begin{lemma}\label{slices}
$ \int_{-\i}^{\i} |D^i \lfloor_{H_s}|^{\natural_{i}}ds \le K\|D^i\|_i.$

\end{lemma}
\begin{lemma}  Assume $P\in \cal{P}_k$ has compact support with diameter $K$.  Then  
$$\int_{-\i}^{\i} |P\lfloor_{H_s}|^{\natural_{r}}ds \le K|P|^{\natural_{r}}.$$
\end{lemma}

\begin{proof}
Let $\e> 0.$ There exists $P = \sum_{i=0}^{r} D^i + \p B$ such that 
$|P|^{\natural_{r}} >  \sum_{i=0}^{r} \|D^i\|_i  + |B|^{\natural_{r-1}} - \e.$  
Then $$\begin{aligned} \int_{-\i}^{\i} |P\lfloor_{H_s}|^{\natural_{r}}ds &\le    \sum_{i=0}^{r} \int_{-\i}^{\i} |D^i\lfloor_{H_s}|^{\natural_{i}}ds + \int_{-\i}^{\i} |\p B\lfloor_{H_s}|^{\natural_{r}}ds.
\end{aligned}$$ 

Let $K$ be the diameter of $supp(P).$  
By the Fubini inequality \ref{lintegral} and Lemma \ref{slices},  
$$\begin{aligned} 
 \int_{-\i}^{\i} |D^i \lfloor_{H_s}|^{\natural_{i}}ds &\le K\|D^i\|_i.
\end{aligned}$$
 $$\begin{aligned} 
\int_{-\i}^{\i} |(\p B)\lfloor_{H_s}|^{\natural_{r}}ds &\le  \int_{-\i}^{\i}|\p (B\lfloor_{H_s})|^{\natural_{r}}ds  + 
\int_{-\i}^{\i}|B_s|^{\natural_{r-1}}ds \\&\le  \int_{-\i}^{\i}|B\lfloor_{H_s}|^{\natural_{r-1}}ds + |B|^{\natural_{r-1}}\\&\le K|B|^{\natural_{r-1}}
\end{aligned}$$ The result follows.
\end{proof}

\begin{proposition}\label{gmt} If  
$P_j \in \cal{P}_k  $ with $\sum |P_j|^{\natural_{r}} < \infty$, then  a.e.
$s$ $$\sum |P_j \lfloor H_s|^{\natural_{r}} <
\infty.$$
 \end{proposition}
 \begin{proof}
 $$\int_{\i}^{\i}\sum |P_j \lfloor H_s|^{\natural_{r}}ds = \sum \int_{\i}^{\i} |P_j \lfloor H_s|^{\natural_{r}} ds \le K \sum |P_j|^{\natural_{r}}  < \i.$$
 \end{proof}
 
\begin{proof} (Old proof)  For each $j$, there exists a decomposition $P_j = \sum_{i =0}^{r_j} D^i_j + \partial B_j$ with $$\sum_j \sum_{i =0}^{r_j}  \|D^i_j\|_i + |B_j|^{\natural_{r-1}} <
\infty.$$  Then
$$P_j \lfloor H_s =  \sum_{i =0}^{r_j}  (D^i_j \lfloor H_s) - B_{js} +  \partial(B_j \lfloor H_s)$$ where
 $B_{js}$ is the slice   of $B_j$ as defined above.  Since $M(P\lfloor_{H_s}) \le M(P)$ we have
$$
\begin{array}{rll}
|P_j \lfloor H_s|^{\natural_{r}} &\le  
  |B_j \lfloor H_s|^{\natural_{r-1}} + |B_{js}|^{\natural_{r-1}} + \sum \|D^i_j \lfloor H_s\|_i \\&
\le
|B_j|^{\natural_{r-s}} + |B_{js}|^{\natural_{r-1}}+ \sum \|D^i_{js}\|_i.
\end{array}
$$
  By Lemma \ref{lintegral} $\sum\int_{-\infty}^{\infty} |B_{js}|^{\natural_{r-1}}ds \le \sum
|B_j|^{\natural_{r-1}} < \infty$  and  
$$\int |D^i\lfloor_{H_s}|^{\natural_{i}}ds \le \|D^i\|_i.$$
We conclude $\sum  |B_{js}|^{\natural_{r-1}} <
\infty$ and $\sum \|D_j^1\lfloor_{H_s}\|_1 < \i$  a.e. $s$.
  Since $\sum  \|D^i_j\|_i + |B_j|^{\natural_{r-1}}  < \infty$ it follows that for a.e. $s$, $\sum
|A_j \lfloor H_s|^{\natural_{r}} < \infty.$

\end{proof}

\begin{proposition} $\mu_J$ and $J \lfloor $  do not depend
on the particular
sequence $P_j$ used in their definitions.
\end{proposition}

\begin{proof}
If ${P_j}'$ is another sequence of mass polyhedra tending to $J$, then by taking subsequences we may assume that $\sum
 |P_j -
{P_j}'|^{\natural_{r}} < \i.$   
We apply \ref{gmt} and conclude that except for $n$-cells $Q$ whose faces lie on
hyperplanes of a
certain null set,  $$\sum |(P_j - {P_j}') \lfloor_Q|^{\natural_{r}}  = \sum |(P_j \lfloor_Q)  -  ({P_j}'  \lfloor_Q)|^{\natural_{r}} < \i.$$  
It follows that
$P_j \lfloor_Q$ and ${P_j}' \lfloor_Q$ tend to the
same limit.   Therefore $\mu_J$ and   $J \lfloor $  do not depend
on the particular
sequence $P_j$ used in their definitions.

\end{proof}

  \begin{lemma} \label{measure} Let $J_j \buildrel \natural_r\over\longrightarrow J$ with
  $M(J_j) \to M(J) .$  Then
$J_j \lfloor_X
\buildrel \natural_r\over\longrightarrow J
\lfloor_X$ and $M(J_j \lfloor_X) \to M(J\lfloor_X)$ 
for every measurable $X$ such that
$\mu_J( fr X)  = 0$.  
\end{lemma}

\begin{proof}   We first prove this for polyhedra $P_j \buildrel \natural_r\over\longrightarrow
J$ with  $M(P_j) \to M(J).$ 
By taking subsequences, it suffices to prove this for sequences such that
$\sum |P_{j+1} - P_j|^{\natural_{r}} < \i.$  Given $\e > 0$ there is $Y \subset X$
which is the
finite union of nonexceptional cubes  and $\mu_J(X-Y) < \e/3$ and $\mu_J(fr(X-Y)) = 0 $.  
    Then
 $$
 \begin{array}{rll}
 |(J\lfloor_X) - (P_j \lfloor_X)|^{\natural_{r}} \le &|(J \lfloor_Y) - (P_j \lfloor_Y)|^{\natural_{r}} \\&+ M(J \lfloor_{ (X-Y)}) + M(P_j \lfloor_{(X-Y)})   .
\end{array}
$$

   By \ref{gmt} $P_j \lfloor_Y \buildrel \natural_r\over\longrightarrow J \lfloor_Y$.  Thus  
$|(J \lfloor_Y) - (P_j \lfloor_Y)|^{\natural_{r}} \le \e/3 $ for sufficiently large $j$. 
We know  
  $\mu_j(X-Y) \to
\mu_J(X-Y)$ since $\mu_J(fr(X-Y)) = 0 $.    
Hence $|(J\lfloor_X) - (P_j \lfloor_X)|^{\natural_{r}} < \e$ for sufficiently large $j$.  The Lemma follows for polyhedra $P_j \to J$.  

This readily extends to any   chainlets $J_j \to J$ such
that $M(J_j) \to M(J).$
Approximate $J_j$ with polyhedra $P_{kj}$ essentially
missing the edges of $n$-cells defining
$X$.  Proposition \ref{gmt}  also applies to this case and shows that  $P_{kj}
\lfloor_X \buildrel
\natural_r\over\longrightarrow J \lfloor_{X}.$   Thus $J_j \lfloor_X \buildrel
\natural_r\over\longrightarrow J\lfloor_X.$  Since $\mu_j \to \mu_J$  
weakly and $X$ is measurable we conclude  $M(J_j \lfloor_X) \to M(J\lfloor_X).$  

\end{proof}

	   The {\itshape \bfseries support} of a measure $\mu$ is the smallest closed set $supp(\mu)$ whose complement is $\mu$-null.    Say that $\mu$ is a {\itshape \bfseries measure on $Y$} if $Y$ contains  $supp(\mu)$.

An open set $U$ is {\itshape \bfseries nonexceptional} w.r.t. $J$ if $U$ is the union of nonexceptional cubes $Q$.
\begin{theorem} \label{newsupport} Given   $J \in \cal{N}_k^r$  with $M(J) < \i$ and a nonexceptional neighborhood 
$U$ of $supp(\mu_J)$
there exist
$P_j
\buildrel \natural_r\over\longrightarrow J$  with  $supp(P_j) \subset U$ and $M(P_j) \to M(J).$ 
   
\end{theorem}  
 
\begin{proof} There exists $P_j'  \buildrel \natural_r\over\longrightarrow J  $ with   $M(P_j') \to M(J)$.    Let
$U$ be any nonexceptional open set containing $supp(\mu_J)$.   
   Then $M(P_j' \lfloor_{U^c}) = \mu_j(U^c) \to
\mu_J(U^c) =  0$,  and  $P_j' \lfloor_U \buildrel \natural_r\over\longrightarrow J \lfloor_U = J.$ Hence $M(P_j' \lfloor_U) \to M(J).$  For
each   cell  $\s$ of 
$P_j'$, replace
$\s \lfloor_U$  a chain approximation  with support contained in $ supp(\t \lfloor_U)$.   This gives a sequence
$P_j$  supported in $U$.  It tends to $J $ in the $r$-natural norm since $$M(P_j - P_j' \lfloor_U) \le M(P_j - P_j')   + \e$$
and $$M(P_j' - P_j' \lfloor_U) \le M(P_j' \lfloor_{U^c}) \to 0.$$   Therefore we can choose the
$P_j$ so that
$M(P_j - P_j'\lfloor_U)
\to 0$ as
$j
\to
\i$.    By lower semicontinuity of mass, 
$M(J) \le \liminf_{j \to \i} M(P_j) =\liminf_{j \to \i} M(P_j' \lfloor_U) = M(J). $ Thus $M(J)  = \lim_{j \to \i} M(P_j).$
 \end{proof}
 
 Note:  This may be extended to show that if $J$ is a cycle, then the $P_j$ may be chosen to be cycles.
 
 We say a closed set
$F$ {\itshape \bfseries supports} a chainlet  
$J$ if for every open set
$U$ containing $F$ there is a
sequence $\{P_i\}$ of  polyhedra tending to $J$ in the $r$-natural norm such
that   $| P_i|\subset U$ for each $j$.  If there is a smallest set $F$
which supports $J$, then $F$ is called the {\itshape \bfseries support} of $J$ and
denoted $supp(J)$.  The next theorem shows that every chainlet with finite energy has a well defined support.

\begin{theorem}\label{support}  If $J \in \cal{N}_k^r$ with $M(J) < \i$, then $supp(J) =
supp(\mu_J).$
\end{theorem}
\begin{proof}    It follows from \ref{newsupport} that $supp(J)
\subset supp(\mu_J).$

Now suppose $supp(J)\subset F.$   
If $x \in supp(\mu_J) - F$, then there exists
a sequence $P_j$ tending to $J$ and a nonexceptional cube $Q$ containing
$x$ such that
$P_j \lfloor_Q = 0.$  Since  $P_j
\lfloor_Q \buildrel \natural_r\over\longrightarrow J \lfloor_Q $ we deduce $J \lfloor_Q = 0.$  Then $\mu_J(Q) = 0$, contrary to the
assumption that $x \in
supp(\mu_J).$  Hence $supp(\mu_J) \subset F.$  It follows that $supp(\mu_J)  \subset supp(J).$

\end{proof}

\begin{corollary} \label{intersect} The set $supp(J)$ is $\mu_J$-measurable and 
$\mu_J(supp(J\lfloor_Q)) = M(J \lfloor_Q).$
\end{corollary}

\begin{corollary} \label{newsupportcorollary} Given $J \in \cal{N}_k^r$ with $M(J) < \i$, $supp(J)$ compact and a nonexceptional
neighborhood 
$U$ of $supp(J)$
there exist
$P_j \buildrel \natural_r \over
\to J$   with  $supp(P_j)  \subset U$ and
$M(P_j) \to M(J). $  
\end{corollary}  

\begin{proof}  This combines \ref{newsupport} and \ref{support}.
\end{proof}
 
\begin{theorem}  Let $J$ be a chainlet and $\R^n = \sum Q_i$ non-overlapping cells that are nonexceptional wrt $J$.  Then if $J$ is a chainlet $$J = \sum J\lfloor_{Q_i}.$$
\end{theorem}

\begin{proof}
It suffices to verify the integrals of forms are the same.
But $\int_{J \lfloor_{Q }} \o = \int_J \o|_Q.$  Hence
$$\int_J \o = \sum \int_J \o|_{Q_i} = \sum \int_{J \lfloor_{Q_i }} \o.$$  The result follows
\end{proof}

\subsection*{Chainlets supported in a point}  If $J$ is a chainlet of class $N^r$ denote $M(J) := |J|_{0,r}$.
 \begin{corollary}\label{massvec}
If $J$ is a   $k$-chainlet of class $N^r$,  $r \ge 1,$ then  $$M(Vec(J)) \le |J|^{\nat_r}. $$
\end{corollary}
\begin{proof}
By Theorem \ref{massr} the result holds for polyhedral chains.  Choose $P_i \to J$ with $M(P_i) \to M(J).$  Then $Vec(P_i) \to Vec(J)$ by continuity of $Vec$.  By lower semicontinuity of mass, $$M(Vec(J)) \le \liminf M(Vec(P_i)) \le  \liminf |P_i|^{\natural_{r}} = |J|^{\natural_{r}}.$$

\end{proof}
 
\begin{theorem}\label{point}
Let $J_i$ be a sequence of $k$-chainlets of class $N^r$ such that for some $C > 0$  and $k$-element $\a$   the following hold: $$M(J_i) < C, \, supp(J_i) \subset B_{\e_i}(p),  \, Vec(J_i) \to \a \mbox{ as }  \e_i \to 0.$$  Then $J = \lim J_i$ exists and $Vec(J) = \a.$
\end{theorem}

\begin{proof}  Suppose $\e_j \le \e_i$ for $j > i$.   Choose $P_s \to J_i -J_j$ with $M(P_s) \to M(J_i - J_j)$ and $supp(P_s) \subset B_{\e_j}(p)$.   By Theorem \ref{massr} 
 $$|P_s|^{\nat_1} \le M(Vec(P_s)) + \e M(P_s) .$$  Hence
$$|J_i - J_j|^{\natural_{r}} \le |J_i - J_j|^{\natural_{1}} \le 
 \liminf M(Vec(P_s)) + \e_i M(J_i -J_j).$$
 But $Vec(P_s) \to Vec(J_i - J_j) \to 0$ by continuity of $Vec$.   
Hence $J = \lim J_s$ exists and $Vec(J) = \a.$
 \end{proof}

 \section{$k$-vector-valued additive set functions and chainlets}  We generalize Whitney's theorem in XI of \cite{whitney} relating sharp chains to $k$-vector valued additive set functions.  We outline our result here and will provide full details in a separate paper. Roughly, the theorem states there is a one-to-one correspondence between $k$-vector valued additive set functions $\g$ and $k$-chainlets $J$ with finite mass.  It is a norm-preserving isomorphism.  The norm on chainlets is the $r$-natural norm.  The norm on $\g$ is defined by:
$$|\g|^{\natural_{r}} : = \sup_{\o} \frac{\int \o \cdot d\g}{|\o|^{\natural_{r}}}.$$
Define $M(\g)$   as the variation of $\g$ over $\R^n$ (See \cite{whitney}, XI \S 2.)
Let $M_k$ denote all $k$-vector valued additive set functions $\g$ on $\R^n$ with norm $|\g|^{\natural_{r}} < \i.$
Let $\cal{J}_k$ denote all chainlets in $\cal{N}_k^{\natural_{r}}$ with finite mass.  We say $J \in \cal{J}_k$ and $\g \in M_k$ correspond if,
$$\g(X) = Vec(J \lfloor X)$$ for all $J$-measurable sets $X$.  
It follows that if $J$ and $\g$ correspond, then 
 for every $k$-form $\o$ of class $N^{\i}$,   
$$\int_J \o = \int_{\R^n} \o \cdot d\g.$$  
\begin{theorem}
The above correspondence $J \mapsto \g_J$ is a one-one linear mapping of $\cal{J}_k$ onto $M_k$ such that
$$|J|^{\natural_{r}} = |\g_J|^{\natural_{r}}, M(J) = M(\g_J), Vec(J) = \g(\R^n).$$
\end{theorem}
 
 \begin{proof} Since $\g_J(X) = Vec(J \lfloor X)$   the correspondence is linear and one-one.   We next show it is onto.
 
  Given   $\g \in M_k$, we can subdivide $\R^n$ into cells $\s_i$of a grid $g$.  Consider $\g$ as a  $k$-element valued set function, rather than $k$-vector valued.   Choose a point $p_i$ in each cell $\s_i$ and let $A_g : = \sum \g(\s_i)_{p_i}.$  Since $\g$ is additive, it can be shown that the sequence $A_g$ converges in the natural norms as the mesh size of $g$ tends to zero.  
 \end{proof}

  \subsection*{Intermediate Value Theorem}    We first modify our change of variables result for exact forms preserving the boundary.  

   \begin{theorem}[Boundary Preserving Change of Variables]\label{change}
Let $A$ and  $B$ be $k$-chainlets of class $N^r$.   Assume $g$ is a mapping of class $N^r$ defined in a neighborhood of $supp(A)$  with $g_* (\p A) =  \p B.$    Let $\o$ be an exact $k$-form of class $N^r$ defined in a neighborhood of $supp(B)$.  Then
    $$\int_{B} \o = \int_A g^* \o.$$

\end{theorem}

\begin{proof} Suppose $d \eta = \o.$  By Stokes' theorem for chainlets \ref{stokes} and   change of variables for chainlets \ref{variables}
$$\int_{B} \o = \int_{B} d \eta = \int_{\p B} \eta = \int_{g_*(\p A)} \eta  = \int_A dg^* \eta = \int_A g^* \o.$$
\end{proof}

This implies $\int_{B} \o = \int_{g_*A} \o$ but we may not deduce that $ g_* A = B.$   For example, $g_*A$ and $B$  might be smooth arcs with the same endpoints.  

\subsection*{Degree of a mapping}
\begin{theorem} Suppose   $A, B \subset \R^n$ are $n$-chainlets determined by positively oriented open sets with compact support. 
Assume $f:supp(A) \to \R^n$ is differentiable and $f_*(\p A) = \p B.$  Then there exists an integer $deg f$ such that 
$$\int_A f^* \o = deg f \int_B \o$$ for all $n$-forms $\o$ supported in $supp(B).$  
\end{theorem}

This will follow from a sequence of lemmas.

\begin{lemma}\label{zero} Suppose  $C$ is an $(n+1)$-chainlet in $\R^{n+1}$, $f:supp(C) \to \R^n$ is differentiable.  Then $$\int_{\p C} f^* \o = 0$$ for every $n$-form $\o$ on $\R^n.$
\end{lemma}

\begin{proof}  Note that all $(n+1)$-forms vanish in $\R^n$ and apply Theorems \ref{stokes} and \ref{commute}:
$$ \int_{\p C} f^* \o = \int_C f^*d\o = 0.$$
\end{proof}

\subsection*{Linear homotopies} Let $f_0$ and $f_1$ be Lipschitz from an open set $U$ into a bounded
subset of $\R^n$, and let $h(t,x) = (1 - t) f_0(x) + tf_1(x), 0 \le t \le 1$. Let $I$ be the interval
$[0, 1 ]$, regarded as a polyhedral 1-chain. Then
$$f_{1*}J -f_{0*}J = \p h_*(I \times J) + h_*(I \times \p J).$$
since this is true when $J$ is a polyhedral chain. 

\begin{corollary}
If $f_0, f_1: A \to \R^n$ are homotopic maps where $A$ is an $n$-chainlet.  Then for every $n$-form $\o$ on $\R^n$ 
$$\int_A f_0^* \o = \int_A f_1^* \o.$$
\end{corollary}
 \begin{proof}
Let $F: I \times supp(A) \to \R^n$ be a differentiable homotopy such that $F|_{0 \times supp(A)} = f_0$ and $F|_{1 \times supp(A)} = f_1$.  Now $\p(I \times A ) = A_1 - A_0.$ By Lemma \ref{zero} 
$$0 = \int_{\p(I \times A)} F^* \o = \int_{A_1} F^* \o - \int_{A_0} F^* \o=  \int_{A} f_1^* \o - \int_{A} f_0^* \o.$$ 
\end{proof}

We next see that the natural norms vary continuously under a smooth homotopy.  The algebra $\cal{A}(\R^n)$ is naturally included in $\cal{A}(\R^{n+1})$ via the identity $\a(V^k) \mapsto \a(V^k, \frac{\p}{\p x_{n+1}}).$  If $\a \in \cal{A}(\R^n)$, we denote $\a$ for its image in $\R^{n+1}$, as well.   Suppose $\a$ is a simple $k$-element in $\R^n$.     If $\b$ is a simple $j$-element in $\R^n$, then $\a \times \b: = \a \wedge \b$ is a simple $(k+j)$-element in $\R^{n+j}.$

According to Proposition \ref{Cartesian},  if $J$ is a $k$-chainlet in $\R^n$, then $[0,t] \times J$ is a $(k+1)$-chainlet in $\R^{n+1}$ with 
$$|[0,t] \times J|^{\natural_{r}} \le  2t(k+1)|J|^{\natural_{r}} .$$    
 
\begin{theorem}
Suppose $F: I \times supp( J)  \to \R^n$  is a smooth homotopy where $J$ is a $k$-chainlet.   Then $$||F_{t*}J|^{\natural_{r}}-  |F_{0*}J|^{\natural_{r}}| \le 2t(k+1) |F|^{\natural_{r}}|J|^{\natural_{r}}.$$
\end{theorem}
 
\begin{proof}
Denote $J_t = \{t\} \times J.$
Since $\p([0,t] \times J ) = J_t - J_0$   we have
$$ \int_{[0,t] \times J} dF^* \o = \int_{\p([0,t] \times J)} F^* \o   = \int_{J_t} F^* \o - \int_{J_0} F^* \o =  \int_{F_{t*}J} \o - \int_{F_{0*}J} \o.$$ 
Thus $$|F_{t*}J|^{\natural_{r}} = \sup \frac{\left| \int_{F_{t*}J} \o \right|}{|\o|^{\natural_{r}}} \le \sup \frac{\left| \int_{F_{0*}J} \o \right| +  \left|\int_{[0,t] \times J} dF^* \o\right| }{|\o|^{\natural_{r}}}.$$ Recall that $|d \o|^{\natural_{r}} \le |\o|^{\natural_{r}}.$
It remains to estimate
$$\frac{\left|\int_{[0,t] \times J} dF^* \o\right| }{|\o|^{\natural_{r}}} \le  \frac{|[0,t] \times J|^{\natural_{r}}|F^*d\o|^{\natural_{r}}}{|\o|^{\natural_{r}}} \le 2t(k+1)|J|^{\natural_{r}} |F|^{\natural_{r}}.$$
\end{proof}

We next establish a local version of the degree formula around regular values.  
This is the standard proof, modified for chainlets.  
\begin{lemma}   Let $J$ be an $n$-chainlet in $\R^n$ determined by an open subset of $\R^n.$  Let $y$ be a regular value of the map $f:supp(J) \to \R^n.$    There exists a neighborhood $U$ of $y$ and open set $V \subset supp(J)$ such that $f(V) = U$ and the degree formula 
$$\int_{V} f^* \o = (deg f) \int_{\R^n} \o$$ is valid for every $n$-form $\o$ with support in $U$.
\end{lemma}

\begin{proof}
Since $f$ is a local diffeomorphism  at each point in the preimage $f^{-1}(y)$, $y$ has a neighborhood $U$ such that $f^{-1}(U)$ consists of disjoint open sets $V_1, \dots, V_N$, and $f:V_i \to U$ is a diffeomorphism for each $i = 1, \dots, N$.  
If $\o$ has support in $U$, then $f^{-1}\o$ has support in $f^{-1}(U).$  Hence
$$\int_J f^*\o = \sum_{i=1}^N \int_{V_i} f^* \o.$$  Since $f:V_i \to U$ is a diffeomorphism, we have
$$\int_{V_i} f^* \o = \s_i \int_U \o,$$ the sign $\s_i = \pm.$ depending on whether $f:V_i \to U$ preserves or reverses orientation.  By definition, $deg(f) = \sum \s_i.$  Let $V = \sum V_i.$

\end{proof}

Proof of the degree formula:

This uses  the isotopy lemma which relies on some linear algebra and the inverse function theorem. 
 
Lang in Calculus on Manifolds has a proof to this using the contraction mapping theorem.  
\begin{corollary}[Intermediate Value Theorem] Let $A, B \subset \R^n$ be $n$-chainlets determined by positively oriented open sets with compact support.   Assume $g: supp(A) \to \R^n$ is differentiable.  Then $$g_* (\p A) =  \p B \iff g_*A = B.$$ 
\end{corollary}

\begin{proof} $( \implies)$  
Assume by contradiction that either the set $U = supp{B} -  supp(g_*A)$ or $U = supp(g_*A) - supp(B)$ has nonempty interior.   Let $\o = \chi_UdV$ in the   Theorem \ref{change}.   In the first case we deduce
$$0 \ne \int_U \chi_U dV = \int_B \chi_U dV = \int_A g^*\chi_U dV = 0.$$

In the second case,  we use Theorems \ref{change} and \ref{variables} 
$$0 = \int_B \chi_U dV = \int_A g^* \chi_U dV = \int_{g_*A} \chi_U dV \ne 0.$$  Thus $supp(g_*A) = supp(B).$

$(\impliedby)$  Suppose $ B =  g_*A.$  Then $\p B = \p g_* A = g_* \p A.$
\end{proof}
This is a   extension of the classical result for $\R^n$.   
\begin{theorem}[Classical Intermediate Value Theorem] Let $g$ be continuous map of the unit ball in $\R^n$ into $\R^n$ that is the identity on the boundary.  The the image of $g$ covers every point in the unit ball.  
\end{theorem}

\begin{corollary} Let $J$ be an $n$-chainlet determined by a positively oriented open set with compact support.   Then there is no smooth retraction of $J$ onto $\p J.$
\end{corollary}

\begin{corollary}
If $g$ is a continuous map of the closed $n$-ball $B$ into itself, then it has a fixed point.
\end{corollary}

\begin{proof} Approximate $g$ with a smooth map.  
Suppose $g(x) \ne x$ for each $x \in B.$ Let $T(x)$ be the intersection of $\p B$ with the line passing through $x$ and $g(x).$   (The point $x$ should lie between $g(x)$ and $T(x)$.)  Then $T$ maps $B$ onto $\p B$ and is homotopic to the identity.  
\end{proof}

   \section{Applications to Classical Calculus}\label{classical}
 
 \subsection*{One variable calculus}

 Consider $k =1$ and $n =1.$  A $1$-form is of the form $f(x) dx$.   $\int_J f(x)dx$  is defined for $1$-chainlets $J$.  Examples of $J$ would include the Cantor set, but also something as simple as the interval $a \le x \le b.$  The definition of integral we give is equivalent to  the Riemann   integral.  We divide $[a,b]$ into subintervals.  $[a,b] = \sum [a_i, a_{i+1}]$.  Choose $p_i \in  [a_i, a_{i+1}]$ and let $\a_{p_i}$ denote the $1$-element supported in $p_i$ with mass $a_{i+1} -a_i.$  Then $dx(\a_{p_i}) = M(\a_{p_i}) = a_{i+1} -a_i.$  The sum 
$$\sum  fdx(\a_{p_i}) = \sum f(p_i) (a_{i+1} -a_i),$$ the usual Riemann sum.   It converges to $\int_a^b f(x)dx$ if $J = [a,b].$  Furthermore, $\int_a^b f'(x) dx = f(b) - f(a)$ follows from our general Stokes' theorem.  

   Define $f'(x) := Vec f_*(\a^1)_x$ where $\a^1$ denotes the $0$-element of order $1$, supported in $x$.   

\begin{proposition}
$f'(x) = \star df.$
\end{proposition}

\begin{proof}  Let $\b$ be a $1$-element.
$$f'(x) dx(\b) = Vec f_*(\a^1)_xdx(\b) = Vec f_*(\a^1)x(\p \b)$$  On the other hand, 
$$df(\b) = f(\p \b) = \int_{\p \b} f  $$
\end{proof}

Theorem \ref{change} implies the standard change of variables result in 1-variable since $f(x) dx = d(\int_a^x f(t) dt).$

$$\int_c^d f(x)dx = \int_a^b f(g(t))g'(t) dt.$$

If $f$ is a function of two variables we can define directional derivatives by apply the $0$-form $f$ to the $0$-element of order $1$ $$f\left(\frac{\p \a_p}{\p u}\right) = \frac{\p f}{\p u}(p).$$

The fundamental theorem of calculus for smooth curves follows directly from the general Stokes' theorem.  

Rolle's theorem is an immediate consequence of the Intermediate Value Theorem and the Fundamental Theorem of Calculus.  The Mean Value theorem follows readily. 

\subsection*{Integration by parts}  

$$\int_J d(\a \wedge \b) = \int_J d\a \wedge \b + \int_J \a \wedge d \b.$$ 
 \subsection*{Vector calculus}   Explicit formulas in the classical sense may be derived easily, using the operators $\star$ and $d$.   (This will shortly be expanded.)
  
  The usual way to calculate divergence of a vector field
$F$ across a boundary of a smooth surface
$D$ in $\R^2$  is to integrate the dot product of $F$ with
the unit normal vector field of
$\partial D$.   According to
Green's  Theorem, this quantity equals  the
integral of the divergence of $F$ over $D$.  That is, 
$$\int_{\partial D} F \cdot nd\sigma = \int_D divF dA.$$   Translating this into the language of differential forms and chainlets with an appropriate sign adjustment, we
  replace the unit normal vector field over $\p D$ with the chainlet $\perp  \p D$ and $div F$ with the differential form $d \star \o.$ 
  
  \subsection*{Curl of a vector field over a chainlet}

Let $S$ denote a smooth, oriented surface with boundary in
$\R^3$ and $F$ a smooth vector field defined in a neighborhood of $S$.  The usual way to integrate the curl of a vector
field $F$ over $S$ is to integrate the Euclidean dot product of
curl$F$ with the unit normal vector field of $S$ obtaining    $\int_S curlF \cdot n dA$.  By the curl theorem this integral equals
$\int_{\partial S} F \cdot  d\s.$ 

We translate this into the language of chainlets and differential forms.

Let $\omega$ be the unique differential $1$-form associated to $F$
by way of the Euclidean dot product.   The differential form version of $curl F $ is    $\star d\o.$  
The  unit normal vector field of $S$ 
 can be represented as the chainlet $\perp S$.  Thus the net curl of $F$ over
$S$ takes the form  $\int_{\perp S} \star  d\o.$ By the Star theorem \ref{thm.star} and Stokes' theorem for chainlets \ref{stokes}  this integral equals $\int_S d\omega = \int_{\p S}
\o.$ The vector version of the right hand integral is  $\int_{\partial S} F \cdot  ds. $ 

  \subsection*{Applications to Differential Topology}

\subsection*{Brouwer Fixed Point Theorem}  
The Brouwer Fixed Point Theorem is easily equivalent to noncontractibility of the n-sphere S, i.e.,  
the identity map f  of S is not homotopic to a constant map g.  Assume there 
is such a homotopy.  Approximate it by a smooth homotopy  
$H:S \times [0,1] \to S$ from f to g.   If w is a volume form on the sphere, then  
the integral of  $w= f*w$ is >0, but it equals  the integral  of $g*w =0$.  
Equality of the integrals follows from Stokes:  Integrate $H*w$  over 
$S \times  [0,1].$

\subsection*{Degree of a mapping of chainlets}

\begin{theorem}
Let $f: M \to N$ be local diffeomorphism of $n$-chainlets with compact support.    Then there exists an integer $deg f$ such that $$f_*M = degf N.$$  
\end{theorem}

This is a consequence of our change of variables formula and the degree formula of differential topology. We give a direct proof. 

Prove this for polyhedral chains first.   Consider $f_*\s$ where $\s$ is so small that the pushforward is a singular cell in $Y$.   Get sign of plus or minus one, according to orientation.  

\begin{theorem}Let $f: X \to Y$ be a smooth map of $n$-chainlets with compact support and finite mass.     Suppose    
 $X = \p W$ such that $W$ is compact and $f:X \to Y$ extends smoothly to all of $W.$ Then 
$\int_X f^*\o = 0$ for every $n$-form on $Y$.  
\end{theorem}

\begin{proof}
Let $F : W \to Y$ be an extension of $f$. Then
$$\int_X f^* \o = \int_{\p W} F^* \o = \int_W F^* d \o.$$   but $d\o = 0$ since it has dimension $k+1.$ 
\end{proof}

\begin{corollary}
 If $f_0, f_1: X \to Y$ are homotopic, then for every $n$-form $\o$ on $Y$
 $$\int_X f_0^* \o = \int_X f_1^* \o.$$ 
\end{corollary}

\begin{proof} Let $F: I \times X \to Y$ be a homotopy.  Since 
$$\p(I \times X) = X_1 - X_0$$ we have
$$0 = \int_{\p(I \times X)} (\p F)^* \o = \int_{X_1} (\p F)^* \o - \int_{X_0} (\p F)^* \o = 0$$   But $\p F$ becomes $f_0$ on $X_0$ and $f_1$ on $X_1$  

\end{proof}

Local degree.  Assume $f$ is a local diffeomorphism at each point.

Degree formula.  
\subsection*{de Rham theory}     Let $J$ be a $k$-chainlet.  Suppose $d E = 0$ for some $r$-form $E$ defined in a neighborhood of $supp(J)$.  We wish to find an $(r-1)$-form $\phi$, defined in a smaller neighborhood of $supp(J)$ such that $E = -d \phi$.  That is, we wish to show that each closed form is exact.      Assume that  $supp(J)$ is connected and $J$ is simply connected.  Then there exists a neighborhood $U$ that is connected and simply connected.   This means that any two arcs are homotopic by a homotopy in $U$.   Suppose $r =1$.  Fix a point $p \in supp(J).$  Let $q \in U.$   Choose an arc $\g_q$ in $U$ that connects $p$ and $q$.  Define $\phi(q) = -\int_{\g_q} E.$   If we choose another arc, we get the same answer by Stokes' theorem.     Then $d \phi = E$  Proof:  (standard)
This does not seem to use any deep methods of chainlet geometry since we assume the form is smooth in a neighborhood of $supp(J).$
 
 
 \subsection*{Relation to work of Grassmann} Grassmann's ideas are very much part of this work and inspired it.  His algebra corresponds to our algebra of $k$-elements of order $0$.     
 \subsection*{Relation to work of Hestenes}
 The products of \cite{hestenes} may be extended to our algebras $\cal{A}$ and $\cal{C}.$  He works with the Grassmann algebra, taking direct sums over the dimension of the $k$-vectors.      Hestenes' geometric product may be extended to 
  $$\a \b := \a/\b + \a \wedge \b$$  for all $\a, \b \in \cal{A}$, of all dimensions and orders,  if this should prove to be of value.
   In terms of the operators $\perp$ and $\wedge$ we have
 $$\a \b = \perp(\b \wedge \perp \a) + \a \wedge \b.$$  It seems problematic that his geometric product is not a derivation of the boundary operator and it is not clear how his theory behaves under change of metric.
  Quaternions are equivalent to rotors of \cite{hestenes} and these can be found through our operators $\perp$ and $\wedge,$ as above.

 \subsection*{Intersections of chainlet curves}    Let $P$ and $Q$ be staircase polyhedral chains in $\R^n$, of complementary dimension.  Consider a point $x$ of the intersection $supp(P) \cap supp(Q).$    Take a small cube $R$ centered at $x$, and consider the parts of $P$ and $Q$ in $R.$  These will be sums $\sum \s_i$, $\sum \t_j$, resp., depending on $R$.  Define 
 $$Vec(P \lfloor_Q, x) := \sum \frac{Vec(\s_i) \wedge Vec(\t_j)}{2^{-2k}} .$$
 Define $$Vec(P\lfloor_Q):= \sum_x Vec(P \lfloor_Q,x).$$
\begin{lemma}
$$M Vec(P\lfloor_Q \cap R) \le M(Vec(P \cap R)) M(Vec(Q \cap R))/ M(R).$$
\end{lemma} 
There exists $P_k \to J$ with $M(P_k) \to M_r(J)$ and $Q_k \to J'$ with $M(Q_k) \to M_r(J').$   Then $P_k \cap R \to J \cap R$ and $Q_k \cap R \to J' \cap R.$   We also want $M(P_k \cap R) \to M_r(J \cap R)$ and $M(Q_k \cap R) \to M_r(J' \cap R).$
We deduce that if $P_k \to J$ and $Q_k \to J'$, then  $Vec(P_k \lfloor_{Q_k})$ forms a Cauchy sequence in the mass norm.    Define the intersection $J \cap J'$ to be the limit of  $Vec(P_k \lfloor_{Q_k})$. 
Define the intersection number of $P$ and $Q$ to be $\pm MVec(P\lfloor_Q).$  The sign is chosen according to the orientation of the n-vector at each intersection point.

We conclude:  The sequence $Vec(P_k \lfloor_{Q_k})$ converges to a $2$-vector $\a_{J,J'}$ in the mass norm.

Conjecture:      The mass of $\a_{J,J'}$   is a homotopy invariant.

This would imply the mass $\a_{J,J'}$, that is, the intersection number, is an integer.  
 
 Suppose $J$ and $J'$ are two closed $1$-chainlets   in Euclidean space $\R^2$.  Approximate them both with staircase polyhedral chains $P_k \to J$, $Q_k \to J'$.  Now consider the intersection $P_k \cap Q_k.$   This will be supported in a finite set of points wherever the intersection is transverse.   In a small neighborhood $B_i$ of each of these points $p_i$ consider  the $2$-vector $Vec(P_k \cap B_i)  \wedge Vec(Q_k \cap B_i)$.      Let $$Vec(P_k \cap Q_k) = \sum_i Vec(P_k \cap B_i)  \wedge Vec(Q_k \cap B_i).$$ 
\begin{theorem}
   
\end{theorem}  
We call $\a_{J,J'}$ the {\itshape \bfseries intersection} of $J$ and $J'$.   Note that this does not coincide with the standard definition of intersection for transverse submanifolds, but it is continuous in the $1$-natural norm, whereas the standard definition lacks continuity when the intersection is not transverse.  
  The signed mass of $\a_{J,J'}$ is defined to be the {\itshape \bfseries intersection number} of $J$ and $J'$.    The intersection number coincides with intersection number of transverse submanifolds and thus to intersection number of homology classes.     This theorem and definition can be extended to closed chainlets of complementary dimension in $\R^n$.

\subsection*{Green's formula for chainlets}
 In the following theorem, the forms $\o$ and $\eta$ satisfy Sobolev conditions, $M$ is a smooth manifold with boundary.  $\bf{t}\o$ is the tangential component to $\p M;$  $\bf{n} \eta$ is the normal component.  
\begin{theorem}[Green's formula]
Let $\o \in W^{1,p} \Omega^{k-1}(M)$ and $\eta \in W^{1,q}\Omega^k(M)$ be differential forms on $M$ where $\frac{1}{p}+ \frac{1}{q} =1.$  Then
$$<d\o,\eta> = <\o, \d \eta> + \int_{\p M} \bf{t} \o \wedge \star \bf{n} \eta.$$
\end{theorem}
  In this section, we give a version of Green's formula that assumes smooth forms and chainlet domains.

\begin{theorem}[Chainlet Green's Formula]
Let $\o \in \cal{B}_{k-1}^{r+1}$ and $\eta \in \cal{B}_k^{r+1}.$ If $J \in \cal{N}^r(M)$, then
$$ \int_J d \o \wedge \star \eta - \int_J \o \wedge d \star \eta =  \int_{\p J} \o \wedge \star \eta.$$
\end{theorem}

\begin{proof}
This follows from Stokes' theorem for chainlets:
$$\int_J d(\o \wedge \star \eta) = \int_{\p J} \o \wedge \star \eta.$$  But 
$$\int_J d(\o \wedge \star \eta) = \int_J d \o \wedge \star \eta - \int_J \o \wedge d \star \eta.$$
\end{proof}

\begin{corollary}
Suppose $J$ is a chainlet without boundary and let $f$ be a smooth function on $supp(J)$ with compact support such that $d \star f \ge 0.$  Then $f$ is constant.  Thus every harmonic function with compact support is constant.
\end{corollary}

\begin{proof}
 By Green's formula  $\int_J d\star f = 0.$   Thus $d \star f = 0$   Now apply the formula to $f^2$ and get $\int_J d \star f^2 = 0$.   What is $d \star f^2$?    $$0 = \int_J 2f d\star f  - \int_J 2 (df)^2f.$$  Thus $df = 0.$  Hence $f$ is a constant.  s  
\end{proof}
Green's formula is often stated assuming Sobolev conditions on the form.  In the second part of these  lecture notes, we will give Sobolev $W^{1,p}$ versions of the natural norms by weighting the difference norms with the constant $p$.

\section{Poincar\'e duality}\label{duality}

In this chapter we establish Poincar\'e duality at the level of chainlets and cochainlets.   We begin by finding a chainlet representation $Ch(\o)$ of a differential form $\o$ which converts a contravariant form into a covariant chainlet.    This means that we will be able to do such things as pushforward differential forms under smooth mapping and define their boundaries.  We call these representatives of forms {\itshape \bfseries exterior chainlets}.  We have to exercise caution here.  Operators closed in the chainlet spaces may not preserve subspaces of exterior chainlets.  For example, the boundary of an exterior chainlet is not generally an exterior chainlet itself but is a chainlet of lower dimension.  The pushforward of an exterior chainlet is not generally an exterior chainlet unless the pushforward mapping is a diffeomorphism. (See Theorem \ref{variables}.)   

This leads us to a definition of cap product and a formulation of Poincar\'e duality at the level of chainlets and cochainlets  which passes to the classical duality for homology and cohomology classes.

  \subsection{Exterior chainlets}  
\subsection*{Inner products of $k$-elements}   Suppose $\a$ is a $k$-element   supported in a point $p.$    Since $\a$ is a  chainlet, the star operator applies.  Moreover, $\star \a$ is also a discrete  $(n-k)$-cell determined by the mass of $\a$ and star of the $k$-direction of $\a$. 
This leads to a natural definition of inner product of $k$-elements $\a$ and $\b$, supported in the same point.
$$<  \a,\b>vol:=   \a \wedge  \star \b.$$  

\begin{lemma}
$<\a,\b>$ is an inner product.
\end{lemma}

\begin{proof}  Bilinearity follows from linearity of star and bilinearity of wedge product.   Symmetry follows by symmetry of starred wedge product $\a \wedge \star \b = \b \wedge \star \a.$ By definition of star we know
$\a^\star\a = vol.$ Therefore for a unit $k$-vector $\a$, $<\a,\a>vol = vol$ implies $<\a,\a> = 1.$  If $\a$ = 0, then $<\a,\a>vol = 0$ implies $<\a.\a> = 0.$  $<\a,\b>vol = \a \wedge \star \b = \b \wedge \star \a = <\b,\a>vol.$   
  \end{proof}
 
A differential  $k$-form $\o$ of class $B^r$    determines a unique  $k$-chainlet  $Ch(\o)$ of class $N^r$, called an {\itshape \bfseries exterior chainlet}.  We show below that   $$\int_M <\eta,  \o> vol =  \int_{Ch(\o)} \eta$$ 
 for all $k$-forms $\eta$ of class $B^r$.   We construct  $Ch(\o)$ as a limit of polyhedral chains.   
  
 Remark on notation. Old notation for discrete chains was $\dot{P}.$  We now prefer to use $A$ and will eventually make the notes uniform. 
 \subsection*{Chainlet representations of differential forms}   Let $\o$ be a differential $k$-form of class $B^1$ defined in an open set $U \subset \R^n$.   Assume $U$ is bounded.  At each $p$ the form $\o$ determines a unique $k$-element $Vec(\o,p)$ via the inner product on $k$-element chains at a point.   That is, by the Riesz representation theorem there exists a unique $Vec(\o,p)$ such that
 $$\o(p,\a) = <Vec(\o(p)),\a>$$ for all $k$-vectors $\a.$      Take a cube $Q \subset U$ from the binary lattice and subdivide it into smaller binary cubes $Q_{k,i}$ with midpoint $p_{k,i}$  and edge $2^{-k}$.  Define $$\dot{P}_k = \sum_i M(Q_{k,i}) Vec( \o( p_{k,i})).$$  It is a straightforward exercise to show that  $\dot{P}_k$ forms a Cauchy sequence in the $1$-natural norm.   Denote the chainlet limit by $Ch(\o,Q)$.   Now use a Whitney decomposition to subdivide $U$ into cubes $U = \cup_{j=1}^{\i} Q^j$. Define $$Ch(\o) := \sum_{j=1}^{\i} Ch(\o, Q^j).$$  This converges in the $1$-natural norm since the total mass of the non-overlapping binary cubes is bounded. 
 
\subsection*{Remark}  It is worth noting that a parallel attempt to show the discrete form $\dot{\o}_k = \sum_i M(Q_{k,i})  \o( p_{k,i})$ converges in the $B^r$ norm will fail.   The sequence is not Cauchy as long as the supports are finite.  However, these important and useful forms arise naturally in chainlet geometry as the dual spaces to subspaces of element chains.   Here is a place where chainlets and cochainlets are quite different.

\begin{theorem} \label{Chtheorem} Suppose $\eta, \o$ are $k$-forms.  Then
$$\int_M <\eta, \o> dV = \int_{Ch(\o)} \eta.$$
\end{theorem}

\begin{proof}
It suffices to work with the discrete approximators to $Ch(\o)$ in a cube $Q$ for both integrals.
$$
\begin{aligned} \int_{M(Q^j_{k,i}) Vec(\o(p_{k,i}))} \eta &= \sum_i M(Q^j_{k,i}) \eta(p_{k,i}) \cdot Vec(\o(p_{k,i})) \\& =  \sum_i M(Q^j_{k,i})< Vec(\eta(p_{k,i})), Vect(\o(p_{k,i}))> 
\\&= \sum_i\int_{Vec(Q^j_{k,i})} <\eta, \o> dV. 
\end{aligned}
$$

Hence
$$
\begin{aligned}
\int_{Q^j} <\eta, \o> dV &= \lim_{k \to \i} \int_{\Sigma_i Vec(Q^j_{k,i})} <\eta, \o> dV \\&=   \int_{Ch(\o, Q^j)} \eta.
\end{aligned}
$$
Now take write $U$ as a sum of cubes in the Whitney decomposition to obtain $$\int_U <\eta, \o> dV = \int_{Ch(\o)} \eta.$$  Since the result holds in each coordinate chart, it is valid over a manifold $M$.

 \end{proof}

\begin{corollary} Suppose $M$ is compact and $r \ge 1$.  Then there exists a constant $C > 0$ such that
$$|Ch(\o)|^{\natural_{r}} \le C|\o|^{\natural_{r}}.$$ for all $k$-forms $\o$ of class $B^r.$  
\end{corollary}

\begin{proof}
This reduces to the following estimate:
$$\left| \int_M <\eta, \o> dV \right| \le |M|^{\natural_{r}} |<\eta, \o> dV |_r \le C |\eta|_r|\o|^{\natural_{r}}$$ where $C = vol(M).$
\end{proof}

\begin{theorem}\label{dense}
$Ch: \cal{B}_k^r(M) \to \cal{N}_k^r(M)$ is a one-one linear mapping of $k$-forms of class $B^r$ into $r$-natural $k$-chains with dense image.
\end{theorem}

\begin{proof}
The mapping $\o \mapsto Ch(\o)$ is clearly  one-one and linear.  To show the image is dense, it suffices to show that for any cell $\s$ and $\e > 0$ there is an $\o$ such that
$$\left|\int \phi(X) \cdot \o - X \cdot \s\right| \le |X|^{\natural_{r}} \e$$ for any $X$ of class $N^r$. This yields $$|Ch(\o) - \s|^{\natural_{r}} < \e.$$  
Choose an $(n-k)$-cell $Q'$ through $q_0 \in \s$ orthogonal to $\s$.  The simplexes $\s(q) = T_{q-q_0} \s$ with $q \in Q'$ form a cell $Q$.  Choose $Q'$ so small that $|\s(q) - \s|^{\natural_{r}} < \e/2, q \in Q'.$  Define $\b = Vec(\s)/M(Q)$in $Q$ and $\b = 0$ in $\R^n -Q$ it follows that
$$\left|\int \phi(X) \cdot \b - X \cdot \s\right| = \left|\int_{Q'} X \cdot (\s(q) - \s)\right| \le |X|^{\natural_{r}} \e/2.$$ Choose a smooth form $\a$ in $\R^n$ so that $\int \b -\a< \e/2.$

\end{proof}

\begin{proposition}\label{eta} If $\o$ and $\eta$ are $k$-forms of class $B^r$, then
$$\int_{Ch(\o)} \eta = (-1)^k\int_{Ch(\eta)} \o.$$
\end{proposition}

\begin{proof}
$$\int_{Ch(\o)} \eta = \int_M <\eta, \o> dV = (-1)^k \int_M<\o,\eta>dV = (-1)^k \int_{Ch(\eta)} \o.$$
\end{proof}

The pushforward of an exterior chainlet is not generally an exterior chainlet, unless $f$ is a diffeomorphism.  
 
\begin{theorem}  Suppose $M$ is compact.  $Ch: \cal{B}_k^r(M) \to \cal{N}_k^r(M)$  satisfies
\begin{enumerate}
\item[(i)] $\star Ch(\o) = Ch(\star \o)$;
\item[(ii)] $f_* Ch(\o) = Ch(f^{-1*} \o)$ if $f:M \to N$ is a diffeomorphism;
\item[(iii)] $\p Ch(\o) = Ch(\diamondsuit(\o))$;
\item[(iv)] $\d Ch(\o) = Ch(\p \o)$;
\item[(v)] $\Delta Ch(o) = Ch(\Box \o).$
\end{enumerate}

\end{theorem}

\begin{proof} (i) follows directly from the definitions of  $\star$ and $Ch$.  \\(ii)   Since $M = f^{-1}_*N$ we have
$$\begin{aligned} \int_{Ch(f^{-1*} \o)} \eta = \int_{f^{-1}_*N} <\eta,f^{-1*} \o> dV &= \int_M <f^* \eta, \o> dV \\&= \int_{Ch(\o)} f^* \eta \\&= \int_{f_* Ch(\o)} \eta.
\end{aligned}$$

 We give a proof to (v).  The remaining (i) and (iii) are similar.  
$$\begin{aligned} \int_{Ch(\Box \o)} \eta = \int <\eta, \Box \o> dV &= \int <\Box \eta, \o> dV \\&= \int_{Ch(\o)} \Box \eta \\&= \int_{\Delta Ch(\o)} \eta.
\end{aligned}
$$ The result follows.  
\end{proof}
\begin{conjecture} A chainlet $J$ of class $N^r$ is harmonic iff $J = Ch(\o)$ for some  harmonic $\o$ of class $B^r$.
\end{conjecture}

\subsection*{Covec(X)}   Define $$Covec: (\cal{N}_k^{\natural_{r}})' \to \cal{C}_k^0$$ as follows.  Recall the isomorphism $ (\cal{N}_k^r)'\buildrel {\phi} \over \cong \cal{B}_{k}^{r} $ of \S\ref{isosection}.     Let $\psi:\cal{A}_k^0 \to \cal{C}_k^0$ be the canonical isomorphism determined by the inner product of $\R^n$.    $$Covec(X) :=  \psi(Vec(Ch(\phi(X)))).$$

This maps cochainlets to the algebra $\cal{C}.$

\section{Cap product}

We have seen that $k$-vectors are in one-one correspondence with $k$-elements.  A $k$-vector $\a$ is also a  $k$-coelement via the inner product.  If we wish to consider $\a$ as a coelement, we will denote it by $\widetilde{\a}.$
 If $\a$ is a $k$-element and $\b$ is a $j$-element, we can define several products.  If $j = k$ we have the {\itshape \bfseries scalar product}
   $$\widetilde{\a} \cdot \b := <\a,\b>.$$  
   
\begin{lemma}
$\star \widetilde{ \a}   \cdot \b = \widetilde{\a} \cdot \star \b.$
\end{lemma}   
   

    If  $ j < k$, then $\widetilde{\a} \lfloor \b$ denotes the $(k-j)$-coelement
  $$(\widetilde{\a} \lfloor \b) \cdot \g :=  <\a ,  \b \wedge \g>,$$
   for all  $(k-j)$-elements  $\g$, and is called the {\itshape \bfseries interior product} of $\widetilde{\a}$ with $\b$.  This can be seen geometrically as the orthogonal complement of $\b$ in $\a$, i.e., $\b \wedge \nu = \a$ and $\widetilde{\nu} = \widetilde{\a} \lfloor \b.$
   
     If $j < k$, then $\widetilde{\a} \cap \b$ is the unique $(k-j)$-element given by the Riesz representation theorem
$$\eta \cdot  (\widetilde{\a} \cap \b) := < \eta \wedge \a,  \b>.$$  This is called the {\itshape \bfseries cap product} of  $\widetilde{\a}$ with $\b$.  A geometrical definition can be given as the $(k-j)$-element   $\nu$ orthogonal to $\a$ so that $\nu \wedge \a = \b.$

 A differential  $(j+s)$-form $\o$ of class $B^r$ and a $j$-cochainlet $X$ of class $N^r$ determine a product, also called {\itshape \bfseries cap product},
 $$X \cap Ch(\o): = Ch(\eta)$$ where $\eta(p) := \phi(X)(p) \cap \o(p).$

\begin{lemma}\label{XCH}
$$ \int_{X \cap Ch(\o)} \phi(Y) =  \int_M <\phi (Y) \wedge \phi (X), \o> dV.$$
\end{lemma}

\begin{proof}
Defining $\eta$ as above it follows that  
$$
\begin{aligned} \int_{X \cap Ch(\o)} \phi(Y) &= \int_{Ch(\eta)} \phi(Y) \\&= \int_M <\phi(Y), \eta> dV \\&= \int_M <\phi (Y) \wedge \phi (X), \o> dV.
\end{aligned}
$$

\end{proof}

\begin{lemma}\label{lemdot}  Suppose  $X$ is a $j$-cochainlet and $Y$ is a $k$-cochainlet of class $N^r$ with $j < k$.  Then 
$$ (Y \cup X) \cdot Ch(\o)=  Y \cdot (X \cap Ch(\o)).$$
\end{lemma}   

\begin{proof}  It follows from the definition of cap product that 
$$<\phi(Y)(p) , \phi(X)(p) \cap \o(p)> = <\phi (Y)(p) \wedge \phi( X)(p), \o(p)>.$$ Taking integrals we have
$$\int_M <\phi(Y)(p) , \phi(X))(p) \cap \o(p)> dV = \int_M <\phi Y(p) \wedge \phi X(p), \o(p)>dV.$$

  By  Theorem \ref{Chtheorem} and Lemma \ref{XCH} it follows that
$$\begin{aligned} (Y \cup X) \cdot Ch(\o)  &= \int_{Ch(\o)} \phi(Y) \wedge \phi(X) \\&=  \int_M <\phi (Y) \wedge \phi (X), \o> dV \\&=  \int_{X \cap Ch(\o)} \phi(Y) \\& = Y \cdot (X \cap Ch(\o)).\end{aligned}$$

\end{proof}

\begin{lemma}\label{Choineq}  Suppose $X$ is a $j$-cochainlet and $\o$ is a $k$-form, $j < k$.  Then 
$$|X \cap Ch(\o)|^{\natural_{r}} \le C|X|^{\natural_{r}}|Ch(\o)|^{\natural_{r}}.$$
\end{lemma}  

\begin{proof}  Suppose $Y$ is a $(k-j)$-cochainlet.  Then by the Fundamental Integral Inequality of chainlet geometry 
 $$
 \begin{aligned}|Y \cdot (X \cap Ch(\o))| &= |(Y \cup X) \cdot Ch(\o)| \\&\le |Y \cup X|^{\natural_{r}}|Ch(\o)|^{\natural_{r}} \\&\le C|Y|^{\natural_{r}}|X|^{\natural_{r}}|Ch(\o)|^{\natural_{r}}. \end{aligned}$$    It follows that $$|X\cap Ch(\o)|^{\natural_{r}} \le C|X|^{\natural_{r}}|Ch(\o)|^{\natural_{r}}.$$

\end{proof}

Let $J$ be a chainlet of class $N^r$.  By \ref{dense} $J = \lim Ch(\o_i).$   Define 
$$X \cap J := \lim X \cap Ch(\o_i).$$ This limit exists as a consequence of Lemma \ref{Choineq} and yields
 
\begin{theorem}\label{capcont} For each $r \ge 1$ there exists a constant $C>0$ such that
 if $X$ is a $p$-cochainlet and $J$ is a  $(p+q)$-chainlet of class $N^r$,  then $$|X \cap J|^{\natural_{r}} \le C|X|^{\natural_{r}}|J|^{\natural_{r}}.$$
\end{theorem}

\begin{theorem} If $X$ is a $p$-cochainlet $Y$ is a $q$ cochainlet and $J$ is a $(p+q)$-chainlet of class $N^r$, then \quad \\ \quad
\vspace{-.2in}
\begin{enumerate}
\item[(i)] $(X \cup Y) \cdot J = X \cdot (Y \cap J).$
\item[(ii)] $X \cap (Y \cap J) = (X \cup Y) \cap J$;
\item[(iii)] $\p (X \cap J) = (-1)^{p+1} dX \cap J + X \cap \p J$;
\item[(iv)] $I^0 \cap J = J;  I^0 \cdot (X \cap J) = X \cdot J$ where $I^0$ denotes the unit $0$-form $I^0(p) \equiv 1$.
\item[(v)] $f_*(f^*X \cap J)= X \cap f_* J.$
\end{enumerate}
\end{theorem}
 
\begin{proof}

\begin{enumerate}
   \item[(i)] First approximate $J$ with exterior chainlets $J = \lim Ch(\o_i).$  Then apply    Lemma \ref{lemdot} and Theorem \ref{capcont}.
 \item[(ii)]By (i)  if $Z$ is a cochainlet of class $N^r$, then
 $$
 \begin{aligned} Z \cdot (X \cap (Y \cap J)) &= (Z \cup X) \cdot (Y \cap J) \\&= ((Z \cup X) \cup Y) \cdot J \\&= (Z \cup (X \cup Y)) \cdot J \\&= Z \cdot ((X \cup Y) \cap J).
 \end{aligned}
 $$  It follows that $X \cap (Y \cap J) =  (X \cup Y) \cap J.$
 \item[(iii)] By (i) and Leibniz' rule for cochainlets
 $$
 \begin{aligned} Y \cdot (\p (X \cap J)) &= dY \cdot (X \cap J)
 \\&= (dY
 \cup X) \cdot J  \\&= (d(Y \cup X) + (-1)^{p+1} (Y \cup dX))
 \cdot J   \\&= (Y \cup X) \cdot \p J + (-1)^{p+1} (Y
 \cup dX) \cdot J \\&= Y\cdot (X \cap \p J + (-1)^{p+1} dX
 \cap J).
 \end{aligned}$$
 \item[(iv)] $Y \cdot (I^0 \cap J) = (Y \cup I^0) \cdot J =  Y \cdot J; \mbox{ and }   I^0\cdot (X \cap J) = (I^0 \cup X) \cdot J = X \cdot J.$
 \item[(v)] 
 $$
\begin{aligned}
Y \cdot f_*(f^* X \cap J) &= f^*Y \cdot f^*X \cap J \\&= (f^*Y \cup f^* X) \cdot J \\&= f^* (Y \cup X) \cdot J \\&= Y \cup X \cdot f_* J \\&= Y \cdot (X \cap f_* J)
\end{aligned}
$$ 
 
  \end{enumerate}

\end{proof}

 \section{Poincar\'e duality of chainlets and cochainlets}  An
 oriented compact $n$-manifold $M$ corresponds to a unique $n$-chainlet 
 $J_M$ in the sense that integrals of forms coincide
 $\int_M
 \o = \int_{J_M}\o.$  The cap product determines a
 Poincar\'e duality homomorphism
 $$PD: ({\cal N}_p^{r})^{\prime}
 \to  {\cal N}_{n-p}^{r}$$ by $$PD(X) := X \cap
 J_M.$$  
 
\begin{theorem}  If $\o$ is a $k$-form of class $B^r$, then
$$\phi(\o) \cap M = Ch(\star \o).$$
\end{theorem}

\begin{proof} Suppose $\eta$ is a $(n-k)$-form of class $B^r$.  By \ref{XCH} $$\int_{\phi(\o) \cap M} \eta = \int_M <\eta, \o> dV = \int_M \eta \wedge \star \o = \int_{Ch(\star \o)} \eta.$$

\end{proof}
It follows from \ref{eta} that
 $$X \cdot Ch(\o) = (-1)^k\int_{X \cap M} \o.$$
 
As another consequence, we have $$PD(X) = Ch(\phi(\star X)).$$
 
 The Poincar\'e duality homomorphism is not an
 isomorphism, although its image is dense in the space of chainlets. 
 Consider a $1$-cell
 $Q$ that is a straight line segment.  From the
 isomorphism theorem we know that if $\o \in {\cal
 B}_p^{r,\a}$,  then
 $$PD(\Psi(\o))(\nu) = \int_M \o \wedge\nu $$ for all $\nu \in
 {\cal B}_{n-p}^{r}$. Therefore the inverse would have
 to an $r$-smooth differential form $\o$ satisfying 
 $\int_Q \nu = \int_M \o \wedge \nu$ for all
 $r$-smooth $\nu$.   However, the only possibility
 would be a form defined only on $J$.        
 
 A cochainlet $X$ is {\itshape \bfseries harmonic} if
 $\Box X = 0.$   If $X$ is a cocyle, then it is harmonic if $d \d X = d \star d \star X = 0.$  A chainlet $J$ is {\itshape \bfseries harmonic} if $\Delta J = 0.$     If  $J$ is a cycle it is harmonic 
 if $ \p \diamondsuit J =  \p\star \p \star  J  = 0.$
  This extends the previous definition
 of $PD$ defined on cochainlets.
 \begin{theorem}\label{PD} $$PD: ({\cal N}_p^{r})^{\prime}
 \to  {\cal N}_{n-p}^{r}$$ is a homomorphism satisfying
 $$PD(dX) = (-1)^{p+1}\p PD(X)$$
 $$PD(\star X) = \star PD(X)$$
 $$PD(\d X) = \pm \diamondsuit PD(X);$$
 $$PD(H) \mbox{ is harmonic if }  H \mbox{ is harmonic and $M$ is closed.}$$
 \end{theorem}

 If $X$ and $Y$ are cocycles with $X - Y = dZ$, then $PD(X)$ and $PD(Y)$ are cycles and $PD(X) - PD(Y) = PD(dZ) = \p PD( (-1)^{p+1}Z). $ It follows that $PD$ passes to singular cohomology and homology classes.   Since classic Poincar\'e duality is an isomorphism of singular cohomology and homology classes, then $PD$ is also an isomorphism at this level since the definitions coincide.   
  
   This theorem also implies that PD preserves Hodge decompositions since coboundaries are sent to coboundaries, boundaries are sent to boundaries and harmonic chainlets are sent to harmonic chainlets.    Let $\cal{M}$ denote the dense subspace of chainlets of the form $Ch(\o).$  We know that $\cal{M}$ is closed under the operators of $\p$ and $\star.$
\begin{corollary}
 If $J \in \cal{M}$, then there exist $B, C, H  \in \cal{M}$ such that
  $$J = \p B + \diamondsuit C + H$$ where  $H$  is  a harmonic chainlet.
\end{corollary}   

Observe that this result is not possible via either the sharp or flat topologies of Whitney where harmonic chains are not defined.   Either the star operator or boundary operator is missing.  (See Table 1 of the preface.)   
   It should be of considerable interest to 
study the spectrum of the   geometric Laplace operator $\Delta$ on chainlets.  This is also important for the development of chainlet Hodge theory.    With the extension of chainlet geometry to Riemannian manifolds developed in these notes, and the isomorphism theorem of \cite{currents} between summable currents and chainlets extended to compact Riemannian manifolds $M$, a  Hodge decomposition for chainlets is immediate. \symbolfootnote[2]{The unpublished  1998 Berkeley thesis of J. Mitchell, initially drafted under the supervision of the author and completed with Morris Hirsch, attempted to do this, but not all tools were available at the time.  In particular, chainlets were not defined on Riemannian manifolds and the isomorphism between currents and chainlets was not well understood \cite{currents}. The geometric Hodge and Laplace operators were originally defined by the author for the purpose of developing a geometric Hodge theory for chainlets.}
\begin{theorem} If $C$ is a chainlet in a compact Riemannian manifold $M$ there exist unique chainlets $\p A, \d B$ and $H$ in $M$, where $H$ is harmonic, such that $$C = \p A + \delta B + H.$$
\end{theorem}  
\begin{proof} (sketch)
The chainlet  $C$ determines a unique summable current $c$ via $c(\o) = \int_C \o.$  The current has a unique Hodge decomposition 
$c = \p a + \d b + h$ where $a, b, h$ are currents of the appropriate dimension.    Each is associated to a unique chainlet $A, B, H$. The operators are respected under the operations of $\star$ and $d$.  The result follows. 
\end{proof}

 This is all well and good, but we would like to see a Hodge decomposition of chainlets for all Riemannian manifolds and constructed  geometrically.  We would also like to see   methods for calculations and specific examples. There is much remaining to do.  Some of this will be treated in more detail in the second part of these lecture notes.     
 \section{Inner products for subspaces of chainlets}

Since every Hilbert space is reflexive, we know that chainlet spaces are not  Hilbert spaces.  However, we can identify a dense subspace  of chainlets for which there is an inner product defined.
%

\subsection*{Effective Hilbert spaces} 
We call a Banach space $X$ with norm $|\cdot|$  an {\itshape \bfseries effective Hilbert space} if there exists a dense vector subspace $Y \subset X$ and an operator $<\cdot, \cdot>: X \times Y \to \R$ such that 
\begin{enumerate}
\item[(i)] $<\cdot, \cdot>$ is an inner product on $Y \times Y$;
\item[(ii)] $\sqrt{<B, B>} = |B|$ for all $B \in Y$;
\item[(iii)] $<u +v,w> = <u,w> + <v,w>$ for all $u,v \in X, w \in Y$, 
$<w, u+v> = <w,u> + <w,v>$ for all $w \in X, u,v \in Y$; 
\item[(iv)] $<au,v> = a<u,v> = <u, av>$ for all $u \in X, v \in Y, a \in \mathbb{F}.$ 
\end{enumerate}

We show that the Banach space $\cal{N}^{\i}$ of chainlets is an effective Hilbert space and find subspaces $Y$ and  inner products that satisfy the conditions of the definition.  
First set  $Y = \cal{M}$,  the subspace of exterior chailnets of $\cal{N}^{\i}$ consisting of chainlets of the form $X \cap M$ where $X \in \cal{N}^{\i \prime}$ is a cochainlet.  

\begin{lemma} If $M$ is compact, $$|X \cap M|^{\natural_{r}} \le vol(M) |X|^{\natural_{r}}.$$
\end{lemma}

\begin{proof}
$$|X \cap M|^{\natural_{r}} = |Ch(\phi(X)|^{\natural_{r}} \le vol(M) |X|^{\natural_{r}}.$$ 
 
\end{proof}

\setcounter{definition}{0}
\begin{definition}{Inner product on $\cal{N} \times \cal{M}$}
$$<J, X \cap M>:= |\star X \cdot    J|.$$
\end{definition}
\begin{proposition}
$<\p J,   Ch(\o)> = (-1)^k<J , \diamondsuit Ch(\o)>.$
\end{proposition}

Define $\|E\| = \sqrt{<E,E>}$
\begin{theorem}
$\|E\| \sim |E|^{\natural_{1}}.$  
\end{theorem}

Remarks and questions.

1.  Show that 
$$<Ch(\a), Ch(\b)> = <\a,\b>.$$

\subsection*{Orthonormal basis of $\cal{M}.$}     A Gram Schmidt process produces an orthonormal basis for $\cal{M}.$(material law)     This leads to  orthonormal basis of coelements which are dense in spaces of  differential forms.   The question is, how to find coefficients?    We can do this in the usual way for approximators to a chainlet.  (See below.) This should have broad application.

 \section{Locally compact abelian groups}  (Draft, under construction)  We recall that a topological group is {\itshape \bfseries locally compact} if and only if the identity $e$ of the group has a compact neighborhood.   Let $G$ by a locally compact abelian group.   Translation $T_v$ through a vector $v$ is replaced by translation through a group element $g$. That is, $T_g(U) = U + g$ where $U \subset G.$   (We write our group action as addition rather than the more standard multiplication simply to fit with our previous exposition.  This is strictly notational.)
 For our norms to carry through, we assume $k=0$ so that cells are all $0$-dimensional.  
 
 The cells $\g$ of $G$ will be sufficiently regular elements of the $\s$-algebra ${\cal A}(G)$ generated by the compact subsets.  It is remarkable that there exists an essentially unique natural measure, the {\itshape \bfseries Haar measure}, that allows us to measure the elements of ${\cal A}(G).$  Haar measures are unique up to positive scale factors.  
 More precisely, a right Haar measure $\mu$ on a locally compact group $G$ is a countably additive measure   defined on the Borel sets of $G$ which is right invariant in the sense that $\mu(A + x) = \mu(A)$ for $x$ an element of $G$ and $A$ a Borel subset of $G$ and also satisfies some regularity conditions.   For $x \in G$, denote $|x| = \mu(x).$  (We distinguish single elements $x \in G$ and elements of ${\cal A}(G)$ to fit with our previous exposition. Certainly $x \in {\cal A}(G)$.)  
   
\subsubsection*{Examples of locally compact abelian groups}  

\begin{itemize}
\item $\R^n$ with vector addition as group operation.

\item The circle group $T$. This is the group of complex complex numbers of modulus 1. $T$ is isomorphic as a topological group to the quotient group $\R/\Z.$ 
\end{itemize}

Haar measure allows to define the notion of integral for (complex-valued) Borel functions defined on the group. In particular, one may consider various $L^p$ spaces associated to Haar measure. Specifically, $$L_{\mu}^p(G) =\left\{f:G \to \C: \int_{G}|f(x)|^p d\mu(x)\right\}.$$

The Banach space $L^1(G)$ of all $\mu$-integrable functions on $G$ becomes a Banach algebra under the convolution $xy(g) = \int x(h) y(h^{-1}g) d\mu(h)$ for $x, y \in L^1(G)$.

 If $G$ is a locally compact abelian group, a {\itshape \bfseries character} of $G$ is a continuous group homomorphism from $G$ with values in the circle group $T$. It is well known that the set of all characters on $G$ is itself a locally compact abelian group, called the {\itshape \bfseries dual group} of $G$, denoted $G^{\wedge}$. The group operation on $G^{\wedge}$  is given by pointwise multiplication of characters, the inverse of a character is its complex conjugate and the topology on the space of characters is that of uniform convergence on compact sets. This topology is not necessarily metrizable. However, if the group $G$ is a 
  locally compact abelian group, then the dual group is metrizable.

\begin{theorem}
$(G^{\wedge})^{\wedge}$ is canonically isomorphic to $G$.
\end{theorem} 

The isomorphism is given by $x \mapsto \{\chi \mapsto \chi(x)\}.$
We define a {\itshape \bfseries distribution} to be an element of the dual space of $G.$  

Our next goal is to extend the definitions of the natural norm to  $G$ and obtain a theory of distributions on $G$ that   does not require the notion of differentiation.  (See \cite{feich} for a related result.)

For $g \in G$ let  $T_g$ denote translation through $g$
 $$T_g(A) = A + g.$$  Let   $\g^0 \in \cal{A}(G)$    and $g_1, \cdots, g_r \in G$.    Define  $$\g^1 = \g^0 - T_{g_1}\g^0$$  and   $$\g^{j+1} = \g^j  - T_{g_{j+1}} \g^j.$$
 
  Let
$$D^j =
\sum_{i=1}^m a_i \g_i^j$$
with coefficients $a_i \in \Z.$ This is called a $G$-chain of order $j$. The vector space of all such $G$-chains $D^j$ is
denoted  $\cal{D}^j.$      A $G$-chain $D^0$  of order 0 is sometimes simply called a $G$-chain.

\subsubsection*{$G$-chain mass} Given  $\g^j$   generated by $\g^0 \in G$ and $g_1,  \cdots , g_j \in G$, define $M(\g^0) = \|\g^0\|_0 = \mu(\g^0)$ and for $j \ge 1$, 
$$\|\g^j\|_j = \mu(\g^0)|g_1| |g_2| \cdots |g_j |.$$
 For $D^j = \sum_{i=1}^m a_i \g_i^j$, possibly overlapping, define $$\|D^j\|_j = \sum_{i=1}^m |a_i|\| \g_i^j\|_j.$$

\subsubsection*{Natural $G$-norms} 
 
For $r \ge 0$ define the {\itshape \bfseries r-natural} $G$-norm
$$|P|^{\natural_r}  = \inf\left\{\sum_{j=0}^r\|D^j\|_j  \right\}$$  
where the infimum is taken over all decompositions
$$P = \sum_{j=0}^r D^j  $$
where $D^j \in \cal{D}^j.$     It is clear $|\cdot|^{\natural_r}$ is a seminorm. 
 We shortly prove it is a norm.
 
  Define $$\int_{\g}f := \int_G \chi_{\g} f d\mu.$$

  Suppose $f:G \to T.$  Define $$\|f\|_0 := \sup \left\{\frac{|\int_{\g} f |}{\mu(\g)}: \g \in \cal{A}(G) \right\}.$$ 
  Inductively define $$\|f\|_r := \sup\left\{\frac{\|f -T_vf\|_{r-1}}{|v|}\right\}.$$  
 
Define $$|f|_0 := \|f\|_0$$ and  for $r \ge 1$,
  $$|f|_r := \max\{\|f\|_o, \cdots, \|f\|_r \}.$$
   We say that $f$ is of class $B^r$ if $|f|_r < \i.$  Let $ \cal{B}_G^r$ denote the space of functions on $G$ of class $B^r.$

Given a compact subset $K \subset M$, there exists a smooth  function $f_K$ which vanishes outside an $\e$-neighborhood of $K$ and which is nonzero on $K$.    If $M$ is infinitely smooth, we can also find a function $\phi_K:M \to \R$ which is nonzero on $K$  and such that all the derivative of $\phi_K$ are uniformly bounded by a constant $C$.

\begin{theorem}\label{oldintegral2} Let $P \in \cal{D}^0$, $r \in \Z^+,$ and $f \in \cal{B}_G^r$ defined in a neighborhood of $supp(P).$ Then
$$\left|\int_P f \right| \le |P|^{\natural_r}|f|_r.$$
\end{theorem}
 
\begin{proof}    We first prove $ \left|\int_{\g^{j}} f \right| \le \|\g^j\|_{j}\|f\|_{j}.$
By the definition of $\|f\|_0$ we know
$$\left|\int_{\g^0} f \right| \le   \|\g^0\|_0\|f\|_0. $$    
 
 Apply induction to deduce
$$\begin{array}{rll} \left|\int_{\g^{j}} f \right| =  \left|\int_{\g^{j-1} - T_{v_j}\g^{j-1}} f \right| 
&=  \left|\int_{\g^{j-1}} f - T_{v_j}^* f \right|  \\&\le  \|\g^{j-1}\|_{j-1}\|f - T_{v_j}^*f\|_{j-1} \\&\le  \|\g^{j-1}\|_{j-1}\|f\|_{j}|v_{j}| \\&= \|\g^j\|_{j}\|f\|_{j}\end{array}$$
 
  By linearity $$\left|\int_{D^j} f\right| \le \|D^j\|_j \|f\|_j$$
for all  $D^j \in \cal{P}_G^j$. 
 
  We again use induction to prove $\left|\int_P f\right| \le  |P|^{\natural_r}|f|_r.$ 
As before,   $\left|\int_P f\right| \le |P|^{\natural_0}|f|_0.$  
Assume the estimate holds for $r-1.$ 

  Let $\e > 0$. There exists $P = \sum_{j=0}^r  D^j   $ such that $|P|^{r} >
\sum_{j=0}^r \|D^j\|_j   - \e$. By induction
 $$\begin{array}{rll}  \left|\int_P f\right| &\le \sum_{j=0}^r \left|\int_{D^j} f \right|   \\& \le \sum_{j=0}^r \|D^j\|_j\|f\|_j  \\& \le (\sum_{j=0}^r
\|D^j\|_j  ) |f|_r\\&\le  (|P|^{\natural_r} +
\e) |f|_r.

\end{array} $$
Since the inequality holds for all $\e > 0$ the result follows.
\end{proof} 
\begin{corollary} $ |P|^{\natural_r}$ is a norm on the space of $G$-chains $\cal{P}_k$. \end{corollary}
\begin{proof} Suppose $P \ne 0$ is a $G$-chain. There exists a function $f$   such that $\int_P  f \ne 0$ and $|f|_r < \i.$  Then
$0 <  \left|\int_P f\right| \le |P|^{\natural_r}|f|_r$ implies $|P|^{\natural_r} > 0.$ \end{proof}
The Banach space of     $G$-chains $\cal{P}_k$ completed with the
norm $|\cdot |^{\natural_r}$ is denoted $\cal{G}^r $. The elements of $\cal{G}^r $ are called
{\itshape \bfseries $G$-chainlets of class} $N^r$.
 The characterization of the Banach space is essentially the same as for chainlets, except there is no boundary operator. 
 
\begin{theorem}
The Banach space $\cal{G}^{\i}$  is the smallest Banach space containing $X^0$ and which has  Lipschitz bounded translation operators.

\end{theorem}

Since the space is reflexive, these $G$-chainlets correspond to distributions over $G$.  We obtain a theory of distributions on $G$ that  does not require the notion of differentiation.

  The first  theory uses discrete chains and assumes that we have at our disposal classically defined smooth differential forms.      The second approach is a full discrete theory where both chains and cochainlets are supported in the same countable set of points and will be fully presented in the second part of these lecture notes.  In the full discrete sections we quantize matter and energy with models that make their identification transparent, up to a constant \symbolfootnote[1]{$e = mc^2$}.  
 In the full discrete theory, cochainlets are ``measuring sticks'' that match chains perfectly.  We characterize the cochainlets by what the we call ``discrete forms''.    Indeed, we propose alternates to the venerable ``Whitney forms'' in the way of discrete forms because our class of discrete forms is closed under the operations of $\star$ and $d$.  Furthermore, cochainlets are associative and graded commutative.  The operators $\star, d$ and $\wedge$ work seamlessly with each other at the discrete level and converge to the smooth continuum.        

\section{Miscellaneous results}

 \subsection{An equivalent discrete norm}  The notion of a difference $k$-cell of order $i$ naturally extends to all $k$-chainlets.  Let $\dot{\s}_i$ denote a difference $k$-element of order $i$.  That is, there exists a $k$-element  $\dot{\s_0}$ and vectors $v_0, \cdots, v_i$ such that $\dot{\s}_i = (Id -T_{v_i}) \circ \cdots \circ (Id - T_{v_0}) \dot{\s}_0.$  Let $\dot{D}^i = \sum a_j \dot{\s}^i_j$ denote a difference $k$-element chain of order $i$ and $\|\dot{D}^i\|_i = \sum_j |a_j| \|\dot{\s}^i_j\|_i.$  Denote the vector space of all such difference $k$-element chains of order $i$ by $\dot{\cal{D}}_k^i$.

\begin{theorem}   Let $\dot{P} \in V_k^0.$  Then 
$$|\dot{P}|^{\natural_{r}} = \inf\left\{\sum_{i =1}^r \|\dot{D}^i\|_i: \dot{P} = \sum_{i = 1}^r \dot{D}^i ,  \dot{D}^i \in \dot{\cal{D}}_k^i\right\}.$$ 
\end{theorem}

\begin{proof} To establish $\le$ this reduces to showing that  
 $\|\dot{D}^i\|^{\natural_{r}} \le \|\dot{D}^i\|_i$. This follows from Corollary \ref{natnorms}.  For the other direction, we again use Corollary \ref{natnorms}.   
 $$|\dot{P}|^{\natural_{r}} = \inf\left\{\sum\|D^i\|_i + |C|^{\natural_{r}}: \dot{P} = \sum D^i + \p C\right\}$$ where $D^i $  is a difference $k$-chainlet chain of order $i$ and $C$ is a $(k+1)$-chainlet of class $N^{r-1}.$   Thus for $\e > 0$, there exists $\dot{P} = \sum D^i + \p C$ such that 
 $$|\dot{P}|^{\natural_{r}} >  \sum\|D^i\|_i + |C|^{\natural_{r-1}} - \e/2.$$  We may assume that the RHS consists of difference element chains since the LHS is an element chain.    Furthermore, $C$ must be a  difference element chain, $C = \sum \dot{E}_i. $
 By induction $$|C|^{\natural_{r-1}} \ge \sum \|\dot{E^i}\|_i - \e/2.$$     (This is immediate for $r =1.$ )
   Thus 
$\dot{P} = \sum \dot{D^i}  + \sum \dot{E}^i $ and $$|\dot{P}|^{\natural_{r}} > \sum\|\dot{D^i}\|_i + \sum\|\dot{E^i}\|_i - \e.$$
\end{proof} 
 
We remark that one cannot omit the boundary term in the polyhedral definition of the natural norms and obtain an equivalent norm.   Without it, we would not be able to prove that staircases converge to the diagonal.  However, with elements, shape is no longer important.  Discrete staircases do converge to discrete diagonals without a boundary term being  used.  

\subsection*{Example}(Dedicated to my Ravello friends)  A vortex, such as a hurricane, is approximated by time parametrized $k$-element chains $$\sum f_{t*}(p_i, \a_{p_i})$$ where each $f_{t*}$ is a linear rotation.  This is related to the vortex models of Chorin, but the full calculus is established for our models, including the boundary operator, star operator and the divergence theorem.  

Similar examples exist for arbitrary dynamical systems, including saddles, sinks, sources, chaotic horseshoes, and the like.

\subsection*{Winding numbers, degree of a mapping, linking numbers}\  The following extends the presentation given in \cite{flanders} for degree of smooth mappings of manifolds to chainlets.  
Let $M^{n-1}$ be a closed oriented manifold and $$f:M^{n-1} \to \R^n - \{0\}.$$ This is a hypersurface that winds around the origin, possibly with self intersections.  Its winding number is given by the Kronecker integral.  Let $g = \pi \circ f: M^{n-1} \to S^{n-1}$ where $\pi(x) = x/|x|.$  Let $\s$ denote the $(n-1)$-dimensional volume on $S^{n-1}$ and $A_{n-1} = \int_{S^{n-1}} \s.$  Then 
$$deg \, g = \frac{1}{A_{n-1}}\int_M g^*\s.$$
(Proof:  $$\int_M g^* \s = \int_{g_*M} \s = (deg \, g) \int_{S^{n-1}} \s = A_{n-1} \int_M g^* \s.)$$  

Now suppose $P_k \to J$ in the chainlet norm and $f$ is a mapping defined in a neighborhood of $supp(J)$.  Then 
$f_* P_k \to f_* J$ and $g_* P_k \to g_* J.$  Hence $\int_{P_k} g^*\s \to  \int_{J} g^*\s.$  The above argument applies to $P_k$ and we may define
$$deg\,g =  \frac{1}{J_{n-1}}\int_{J} g^*\s.$$   Continuity of the integral yields
\begin{theorem}
Degree of a smooth mapping of chainlets extends the definition degree defined for smooth mappings of smooth manifolds.
\end{theorem}

If $f:M^n \to N^n$ is smooth, let $\b$ denote the volume form on $N^n$ with $\int_{N^n} \b = 1.$  Then 
$$deg \, f =  \int_{M} f^* \b.$$  By Seifert-Threfall, we know 
$$f_*M = deg \, f N.$$  Hence 
$$\int_M f^* \b = \int_{f_*M} \b = (deg \, f) \int_N \b = deg \, f.$$  This again extends to chainlets $J$ by taking $M_i \to J$ in the chainlet norms.  We can define $$deg \, f = \int_J f^* \b.$$  
   
If $f:M \to N$ is a smooth mapping of closed, orientable manifolds, then $f_*M$ is an integer multiple of $N$ plus a boundary.  This integer is called $deg \, f.$  
Now suppose $P_i \to J$ in a manifold $M$ where $dim_TJ = n$ and $f: J \to N^n$ is smooth and defined in a neighborhood of $supp(J)$.  We may assume the $P_i$ are contained in this neighborhood.   
We have $f_* P_i = (deg \, f)N$ for each $i$ where $B_i = \p C_i.$  Each $deg \, f$ depends on $i$, but by continuity, it limits to an integer which is $deg \, f$.  $f_*J = (deg \, f)N$.      This is a homotopy invariant and agrees with degree defined for submanifolds.  

 
\subsection*{Cartesian product}   In this section we define   the Cartesian product $J \times K$ of two chainlets of arbitrary dimension. 

 The Cartesian product of a $k$-cell and a $j$-cell is a $(j+k)$-cell.
 Suppose $P = \sum a_i \s_i $ and $P = \sum b_j \t_j$ are polyhedral $k$- and $j$-chains, respectively.  Define $$P \times Q:= \sum_{i,j} a_i b_j \s_i \times \t_j.$$
\begin{lemma}\label{products}  If $D^s$ is an $s$-difference $j$-chain and $E^t$  is a $t$-difference $k$-chain, then
$$|D^s \times E^t|^{\natural_{s+t}} \le \|D^s\|_s \|E^t\|_t.$$
\end{lemma}

\begin{lemma}\label{boundaryP} If  If $D^s$ is an $s$-difference $j$-chain and $P$ is a polyhedral $k$-chain, then 
$$|D^s \times \p P|^{\natural_{s+r}} \le k\|D^s\|_s|P|^{\natural_{r-1}}.$$
\end{lemma}

 We need to break $D^s$   into cubical differences.  
 
\begin{proposition}\label{Cartesian}  Suppose $P \in \cal{P}_j$ and $Q \in \cal{P}_k$ are polyhedral chains.  Then 
$$|P \times Q|^{\natural_{r +s}} \le  (j+1)(k+1)|P|^{\natural_{r}}|Q|^{\natural_{s}}.$$
\end{proposition}

\begin{proof}  The proof proceeds by induction.  The result holds for $r =s = 0.$   
Assume it holds for $r$ and $s$.  We show it holds for $r+1$ and $s+1$.  

Now $\p B \times \p C  = (-1)^k\p(\p B \times C).$    By induction, 
$$|\p B \times \p C|^{\natural_{r+1+s+1}} \le |\p B \times  C|^{\natural_{r+1+s}} \le |\p B|^{\natural_{r+1}}| C|^{\natural_{s}} \le |B|^{\natural_{r}}|C|^{\natural_{s}}.$$
Combine this with Lemmas \ref{products} and \ref{boundaryP} to complete the proof in the usual way.    

Let $\e > 0.$  There exist decompositions $P = 
\sum_{i=0}^r D^i + \p B$ and $Q = \sum_{j=0}^s E^j + \p C$ such that $$|P|^{\natural_{r}} > \sum_{i=0}^r \|D^i\|_i + |B|^{\natural_{r-1}} -\e $$ and
$$|Q|^{\natural_{s}} > \sum_{j=0}^s \|E^j\|_j + |C|^{\natural_{s-1}} \e. $$

Therefore $$P \times Q = \sum_i\sum_j D^i \times E^j + \sum_i D^i \times \p C + \sum_j \p B \times E^j + \p B \times \p C.$$
 
 Hence
 $$\begin{aligned}  |P\times Q|^{\natural_{r+s}} &\le \sum_i\sum_j |D^i \times E^j|^{\natural_{i+j}} + \sum_i |D^i \times \p C|^{\natural_{i+s}} \\&+ \sum_j |\p B \times E^j|^{\natural_{r+j}} + |\p B \times \p C|^{\natural_{r+s}} \\ &\le \sum_i\sum_j \|D^i\|_i\|E^j\|_j + \sum_i \|D^i\|_i |C|^{\natural_{s-1}} \\&+ \sum_j | B|^{\natural_{r-1}}\|E^j\|_j + |B|^{\natural_{r-1}}| C|^{\natural_{s-1}} \\ &\le  |P|^{\natural_{r}}|Q|^{\natural_{s}} + \e.
 \end{aligned}$$
\end{proof}

 Let $J \in \cal{N}_j^r$ and $K \in \cal{N}_k^s$ be chainlets.   Choose $P_i \buildrel {\natural_{r}} \over \to J$ and $Q_i \buildrel {\natural_{s}} \over \to K$.  Define $$J \times K := \lim P_i \times Q_i.$$    By the lemma, we deduce
\begin{theorem}
 Let $J \in \cal{N}_j^r$ and $K \in \cal{N}_k^s$ be chainlets.  Then $J \times K \in \cal{N}_{j+k}^{r+s}$ is a well defined chainlet  satisfying
 $$|J\times K|^{\natural_{r+s}} \le  (j+1)(k+1)|J|^{\natural_{r}}|K|^{\natural_{s}}.$$
\end{theorem}

\begin{proof}

\end{proof}

Suppose $J_1 = \lim A_{1,i}$ and $J_2 = \lim A_{2,i}.$ 
\newpage 
\subsection*{Linking numbers}   Suppose $M^j, M^k$ are two disjoint, orientable, closed submanifolds in $\R^n$ where $j+k = n-1$.  Their {\itshape \bfseries linking number} is defined as follows:  Form the product space $M \times N$ which is a manifold of dimension $j+k = n-1.$  Let $f: M \times N \to \R^n - \{0\}$ given by $$f(x,y) = y-x.$$  Define $$link(M,N) = deg \, f.$$  Cartesian product of is continuous operator on chainlet spaces.  Thus linking number is well defined, by taking limits, of nonintersecting, closed chainlets $J, J'$ supported in $\R^n.$ 

An interesting example is the toral solenoid obtained by wrapping the torus around itself twice and taking a  limit.
 This set supports many chainlets. We consider the canonical one which has unit mass and dimension one.   We will call this {\itshape \bfseries the unit chainlet solenoid}.  The linking number of a meridian circle about the torus with the unit chainlet solenoid is one, but we may obtain real linking numbers if the second curve cuts through the solenoid.  

Imagine a unit electric current flowing around the chainlet soleonid $N$.  By Amp\`ere's law for smooth $M$ and $N$
$link(M,N)$ is the work done by this field on a unit magnetic pole which makes one circuit of $M$.    (See \cite{flanders}, p. 81.)   For chainlets, we take limits and make the same deduction.  ($\int_N f^* \t$ is well defined and continuous. )

\subsection*{Cup product of chainlets}
It is not possible to define  cup product as a continuous operator
$$\cal{N}^{\natural_{r}}_j \times \cal{N}^{\natural_{r}}_k \to
\cal{N}^{\natural_{r}}_{j+k}$$  that extends cup product of  $k$-elements and coincides with  cup product of differential forms.  Cup product and Hodge star leads to an inner product but     the space of $L^1$ functions, a subspace of chainlets, is not a Hilbert space.   However, cup product is defined on pairs of chainlets where one element of the pair is contained in a dense subspace of chainlets,  namely the exterior chainlets defined earlier.
  $$\cal{N}^{\natural_{r}}_k \times \cal{M}^{\natural_{r}}_j \to
\cal{N}^{\natural_{r}}_{j+k}.$$  
     
     Cup product converges in the natural norm, to define $J \cup Ch(\o)$ where $J$ is an arbitrary chainlet and $Ch(\o)$ is an exterior chainlet.    This extends the standard wedge product on forms  $Ch(\eta \wedge \o) = Ch(\eta) \cup Ch(\o)$.   This leads to an inner product  $$<J,Ch(\o)> = \int_{J \cup *Ch(\o)} dv.$$

We first define cup product for pairs of element chains and exterior chainlets, and then prove continuity in the first variable.  

Let $Ch(\o)$ be the exterior $j$-chainlet associated to a differential $j$-form $\o$.  In local coordinates,  $\o(p) = \sum a^H(p) dx^H.$   The differential $j$-form $dx^H$  at $p$ determines a unique unit $j$-element $Ch(e^H)$ supported in $p$ satisfying $dx^H(Ch(e^H)) = 1.$ 
Define $$Ch(\o)(p) := \sum a^H(p) Ch(e^H).$$    
\begin{lemma}
Let $p,q \in \R^n.$   If $\o$ is smooth, then 
 $$|Ch(\o)(p) - Ch(\o)(q)|^{\natural_{r}} \le |p-q||\o|^{\natural_{r}}.$$
\end{lemma} 

Let $\dot{\s}$ be a  $k$-element supported in $p$.   
Cup product of $k$-vectors and $j$-vectors is defined for $k$-elements and $j$-cells supported in a point $p$, of course, since it is defined for all vector spaces.  This can be used to define   $$\dot{\s} \cup Ch(\o)(p).$$   If $A = \sum a_i \a_i$ is a  $k$-element  define 
$$A \cup Ch(\o) = \sum a_i \a_i \cup Ch(\o).$$ 
\begin{theorem}
$$|A \cup Ch(\o)|^{\natural_{r}} \le |A|^{\natural_{r}} |\o|^{\natural_{r}}.$$
\end{theorem}

\begin{proof}

Suppose $A$ is a $k$-element  chain  and $\e > 0$.  There exists a decomposition $A = \sum \dot{D}^i $ with $$|A|^{\natural_{r}} > \sum \|\dot{D}^i\|_i   - \e.$$   
$$|A \cup Ch(\o)|^{\natural_{r}} \le \sum 
|a_i||\dot{D}^i \cup Ch(\o)|^{\natural_{r}}.    $$

This reduces to showing $\|\dot{D}^i \cup Ch(\o)\|_i \le \|\dot{D}^i\|_i |\o|^{\natural_{i}}$
 
Now $\a^i \cup Ch(\o)$ can be written as the sum of two chainlets, the first is an $i^{th}$ order dipole based on $\a \cup Ch(\o)$ and vectors $v_1, \dots, v_i.$  Its dipole norm is bounded by $$\|\a \cup Ch(\o)\|_0|v_1| \cdots |v_i| \le \|\a\|_0|v_1| \cdots |v_i| |\o|^{\natural_{0}} =  \|\a^i\|_i  |\o|^{\natural_{0}}.$$  The second term is bounded by $$\begin{aligned} M(\s)(|\o(p) - \o(p+v_1)  &+ \o(p+v_1+v_2) -\o(p+v_2)| + \cdots) \\&\le M(\a)|v_1||v_2| \cdots |v_i| |\o|^{\natural_{i}} \\&= \|\a^i\|_i|\o|^{\natural_{i}}.\end{aligned}$$  Hence 
$$\|\a^i \cup Ch(\o)\|_i \le \|\a^i\|_i |\o|^{\natural_{i}}.$$  

In the top dimensional case, $k = n$, the result follows since $A \cup Ch(\o) = 0.$ 
By induction, 
$$
\begin{aligned} \sum 
|a_i||\dot{D}^i \cup Ch(\o)|^{\natural_{r}}   &\le \sum |a_i| \|\dot{D}^i \cup Ch(\o)\|_i    \\&\le 2|\o|^{\natural_{r}}\sum |a_i|\|\dot{D}^i\|_i   \\& \le 2|\o|^{\natural_{r}}\left(\sum |a_i|\|\dot{D}^i\|_i  \right) 
\end{aligned}
$$
\end{proof} 
\subsection*{Properties of cup product} 
\begin{enumerate}
\item Associative
\item Bilinear   $$(J+J') \cup Ch(\o) = (J \cup Ch(\o)) + (J' \cup Ch(\o)).$$
  $$J \cup (Ch(\o) + Ch(\eta)) = (J \cup Ch(\o)) + (J \cup Ch(\eta)).$$
\item Operators:  $$\p ( J \cup Ch(\o) ) = (\p J) \cup Ch(\o) + (J \cup \p Ch(\o)).$$
\item $$\perp  (J \cup Ch(\o) ) = \perp J \cup Ch(\star \o)$$
\item $$f_*( J \cup Ch(\o) ) =   f_* J \cup f_*(Ch(\o))$$
\item $$(X \cap M) \cup (Y \cap M) = \int_M \phi(X) \cup \star \phi(Y) = (X \cup Y) \cap M$$
\item anti commutative  $$J \cup \perp J' = (-1)^k J' \cup \perp J$$   
\item $J \cup J = 0$ if $J$ is simple.  
\end{enumerate}

\begin{theorem} If $J$ is a   $k$-chainlet of class $N^r$ and $\o$ is a $k$-form of class $B^r$, then
$$\int_J \o = \int_{J \cup \star Ch(\o)} dV.$$ 
\end{theorem}
  
 \begin{corollary} Assume $M$ has no boundary.  Then  $\d  \o = 0 \iff  \p Ch(\o) = 0.$ Also, $d \o = 0 \iff  \diamondsuit Ch(\o) = 0.$
$\o$ is harmonic $\iff$ if $Ch(\o)$ is harmonic.
\end{corollary}  

\begin{proof} By the Leibniz rule for wedge product, if $\a$ is a $p$-form and $\b$ a $q$-form, then, $d(\a \wedge \b) =  (-1)^p \a \wedge d \b + d \a \wedge \b.$  

Assume  $\d \o = 0,$  Then $d \star \o = 0$.   Then $$X \cdot \p Ch(\o) = dX \cdot Ch(\o) = \int d\phi(X) \wedge \star \o =(-1)^{k+1}\int \phi(X) \wedge d \star \o = 0.$$  Since this holds for all $X$ we deduce $ \p Ch(\o) = 0.$
Similarly, $$\d(\a \wedge \b) = \star d \star (\a \wedge \b) = \star d(\star \a \wedge \star \b) = \star (d \star \a \wedge \star \b + \star \a \wedge d \star \b) = \d \a \wedge \b + \a \wedge \d \b.$$    Note that $\diamondsuit M = 0$.  Hence $\int_M \d(\phi(X) \wedge \star \o) = 0.$   

Assume $d \o = 0$.  Then $\d \star \o = 0.$    Let $\o$ be a $(k-1)$-form and $X$ a $k$-cochainlet.  Then $\d(\phi(X) \wedge \star \o) = 0.$  Then $$X \cdot \diamondsuit Ch(\o) = \d X  \cdot Ch(\o) = \int \d \phi(X) \wedge \star \o = \int \phi(X) \wedge \d \star \o = 0.$$  It follows that $ \diamondsuit Ch(\o) = 0.$

Conversely, suppose $\p Ch(\o) = 0. $ Then $$\d \o \cdot P = \o \cdot \diamondsuit P = \int_{\diamondsuit P \cup \star Ch(\o)} dV = \int_{P \cup \diamondsuit \star Ch(\o)} dV = 0.$$   Here, we use $\d dV = 0.$    Now suppose $  \diamondsuit Ch(\o) = 0.$  Then 
$$d \o \cdot P = \o \cdot \p P = \int_{\p P \cup Ch(\o)} dV = \int_{P \cup \p Ch(\o)} dV  = 0.$$ Here we use  $\int_{\p (P \cup Ch(\o))}dV = 0.$
\end{proof}

\begin{theorem}[Cartan's magic formula for differential elements] \label{Cartan}
$$\cal{L}_X(\a) = (Ch(\phi(X)) \cup \p \a) + \p (Ch(\phi( X)) \cup \a).$$
\end{theorem}
This extends to discrete chains and then to chainlets $J$ by taking limits.  We deduce the classical result for forms.
Observe $i_X \a := Ch(\phi(X)) \cup \a.$  Then 
$i_X \o(p; \a) = \o(p; i_X \a).$
\begin{corollary}[Cartan's magic formula for forms]
$$\cal{L}_X \o =d(i_X \o) + i_X d \o .$$
\end{corollary}

\begin{proof} By Theorem \ref{Cartan}
$$\begin{aligned}\cal{L}_X \o(p; \a) &= \o(p; \cal{L}_X(\a)) \\&= \o(p; Ch(\phi(X)) \cup \p \a) + \o(p; \p (Ch(\phi(X)) \cup \a)) \\&= d(i_X \o)(p; \a) + i_X d \o(p;\a).\end{aligned}$$
\end{proof}
  
There is an orthonormal basis to $k$-element chains supported in a finite set of points $J$. An inner product can be defined on these chains leading to  new numerical methods.

  We can use Leibniz notation for these $\frac{\p dx^i}{\p x^j}.$

 In the case of $V = \R^n$ we use Euclidean dot product to represent the form as an element of the direct sum of the dipole versions of the differentials $dx^1, dx^2, \dots $  
 
\begin{theorem}
The space of differential forms (rel to K) is a DGA.
\end{theorem}

To obtain continuity to the smooth continuum, we   let the distance between points of $K$ to tend to zero and require that both chains and cochains be defined over the same set of points.       The rate of convergence is on the order of the distance between the points of $K$.  If we desires a faster rate of convergence we   may add   higher order elements much as in Taylor's expansion.

 The convergence will be faster than with the linear approximations.  A combination of higher order curvatures and decreased distance between the supporting points can be maximized, much as with the Galerkin method to find faster rates.

 
   \subsection*{Cup product of discrete cochains}  
\begin{theorem}
Wedge product of discrete forms is continuous and converges to wedge product of forms.
\end{theorem}

     \section{Applications to the applied sciences}  (Draft, under development)  The applications that follow are, more often that not,  the first, simple ideas that arise from blending chainlets with classical techniques.     There is room for significant expansion of these rudimentary ideas.   However, extending classical methods to chainlets is a worthy effort since chainlets include a wide variety of domains beyond smooth manifolds, including     discrete (e.g., points), rough (e.g., fractals) and bilayer (e.g., soap films).     The reader should keep in mind that this section covering applications is the least developed and organized of the entire paper, but should be improved shortly, some with the help of collaborators in the relevant areas.

\subsection*{Calculus of variations over chainlet domains}

This is a rich area of development that is a natural extension of the classical theory of calculus of variations for smooth domains.  The author has shown how calculus of variations methods produce solutions permitting soap film structures.  Many of these methods may easily be generalized to include examples such as electrostatic discharge, the lungs of a dog, crystal growth, and so forth.  The ingredients are within this paper and in \cite{soap} and \cite{plateau}.

For example, suppose we have a compact subspace $X$ of chainlets and we wish to minimize some function $f$ of elements of $X$, such as mass.  As long as the function is lower semicontinuous, then we may proceed as follows.   Let $m_0 = \inf\{f(J): J \in X\}.$  Then there exists $J_i \in X$ such that $f(J_i) \to m_0.$  Since $X$ is compact, there exists a subsequential limit $J \in X$ of the $J_i$.  By lower semicontinuity of $f$ we deduce that $f(J) \le m_0.$  Since  $J \in X$, we conclude $f(J) = m_0.$  The common difficulty in this approach is to establish compactness.    This was applied to $f(J) = m(J)$ in \cite{plateau} where $X$ was a certain space of $2$-chainlets spanning a given wire loop.   
Local extrema of functions of chainlets may be studied in chainlet space-time, as well.  

This topic is under investigation with Robert Kotiuga.

\subsection*{Quantum electrodynamics} 

Consider QED is dimension four.   The gauge potential $A$ is a $1$-form and the EM field $F = dA$ is a $2$-form.    In the classical theory, the action for the gauge potential in QED is given by 
$$S(A) = <dA, dA> = \int_M dA \wedge  * dA.$$  Since $d, \wedge$ and $*$ are well defined over chainlet domains, this extends to chainlets $J$:
$$S_J(A) = <dA, dA> = \int_J dA \wedge  * dA.$$ 

Simiarly, for the Abelian Chern-Simons theory the gauge field $A$ and EM field $F = dA$ are $1$- and $2$-forms as before.  The action for the theory is given by 
$$S(A) =  <A, *dA> = \int_M A \wedge dA$$ where the spacetime $M$ is three dimensional.   For chainlet domains we have
$$S_J(A) =  <A, *dA> = \int_J A \wedge dA.$$
The {\itshape \bfseries curvature form} is $dA + A \wedge A$.

Using our algebras $\cal{A}$ and $\cal{C}$, we have seen how to discretize the forms, the inner product, the integral, and the operators $(\wedge, d, *)$ so that discrete analogues of the continuum relationships hold and there is convergence to the continuum limit.  In particular, the gauge potential is approximated by a $1$-element chain and the EM field $F$ by a $2$-element chain, both supported in finitely many points.

\subsection*{Classical mechanics, a dictionary}(Draft, under construction)
A {\itshape \bfseries velocity vector} $v$ corresponds to a $1$-element $\a$.  Its speed is the quantity $M(\a)$ and its direction is the direction of $\a$.  Acceleration corresponds to a $1$-element of order one.  

An element $m \a(x)$ is called a {\itshape \bfseries particle} where the mass of $\a$ is assumed to be one.  The mass of the particle is $m$, its support, or location, is the point $x$.

It is customary to model momentum as $p = mv$.   
In this model we convert $p$ to a $1$-element $m\a.$  
Position of the particle  is its support $x$.  If considered to be a vector $v$, this corresponds to a $1$-element $Vec(v)$, supported in $x$.  

We propose another model for momentum.   Give a velocity vector $v$ and a mass $m$, we use the $3$-element of order $1$  to model $p$.   Let $V^3 = (e_1, e_2, e_3)$ and $\a = \a(V^3).$  Define $$\widetilde{p} := m\a^1 = m \frac{\p \a}{\p v}.$$  This takes account the particle $\a$, not just its mass and velocity.  

{\itshape \bfseries Angular momentum} corresponds to the cross product which takes the form  $r \times p = \perp (Vec(r) \wedge m\a).$

In our alternate view, suppose we have a particle $\alpha$ with mass $m$ moving with velocity $v$ at a position $r$, relative to the origin.  We propose  using the cross product vector $w = r \times v$ and defining angular momentum as $m \frac{\p \a}{\p w}.$

 If each particle  is a distance $d_i$ from a particular axis of rotation, then the {\itshape \bfseries moment of inertia} of the rigid body $A = \sum m_i \a_i(r_i)$ about that axis is given by:

$$I = \sum m_i r_i^2.$$  This limits to chainlets to become

$$I = \int_J r^2 dm$$ where $dm =  \rho dV$ and $\rho$ is the {\itshape \bfseries density of mass} where $J$ is any $3$-chainlet with bounded, measurable distribution function $\rho(r).$  

This corresponds to the integral version of   moment of inertia  over a solid body in classical physics  $$I =  \int \rho(r)r^2 dV.$$ where $r$ is the distance from the axis of rotation.

The {\itshape \bfseries angular velocity vector} $\vec{w}$ is given by 
$$\vec{w} = \frac{r \times v}{|r|^2}$$  where $v = r'(t)$ is the velocity vector.   With chainlet terminology, the vectors become elements.   Replace $r(t)$ with the $1$-element $\a(t)$   and $\a(t)$ is the $1$-element with the same direction as $r(t) $ and mass $|r(t)|.$   The velocity vector $r'(t) = \lim_{h \to 0}\frac{r(t+h) -r(t)}{h}$ transforms to the directional derivative
$$\frac{\p r(t)}{\p u} = \lim_{h \to 0} \frac{\Delta r(t)}{\Delta h u}$$ where $u$ is the unit vector of time.  We also denote this as $\a'(t).$ 
 We obtain   
$$\vec{w} =  \frac{\a(t) \wedge \perp \a'(t)}{M(\a(t))^2}.$$

{\itshape \bfseries Density of a chainlet}  
Let $J$ be a $k$-chainlet  with finite mass, $k > 0$.  Let $S(x,r) = \{y: |x-y| \le r\}.$  Recall the Borel measure $\mu_J$ of $J$.  For brevity, let $\mu(x,r) = \mu_J(S(x,r)).$  Let $\d(k)$ denote the $k$-mass of a disk of radius one.  The upper and lower densities of $J$ at a point $x$ are 
$$D_{*}(x) =  \liminf_{r \to 0^+}\frac{\mu(x,r)}{\d(k)r^k}; \quad D^{*}(x)  = \limsup_{r \to 0^+}\frac{\mu(x,r)}{\d(k)r^k},$$ respectively.
 
 {\itshape \bfseries Angular momentum}  $L = r \times p$ converts in a similar fashion, using wedge product and the operator $\perp$.  
 
{\itshape \bfseries  Angular momentum in relativistic mechanics } looks similar for a rigid body.  This form is called the $2$-form Noether charge.  

$$\sum_i r_i \wedge p_i.$$  

This easily translates  into chainlet terminology as above.  

{\itshape \bfseries Torque} of a force $F$ on a particle $x$ is defined as  $$T(x) = r(x) \times F(x).$$  In chainlets, we convert $T(x)$ to $\perp(Vec(r(x)) \wedge Vec(F(x)))$

The bras and kets of Dirac can be modeled as elements and coelements in the algebras $\cal{A}$ and $\cal{C}.$  Observables are the space $\cal{A}$, thus  quantum states are modeled.
(A pure quantum state is a ket vector, thus an element.)

The Riesz representation theorem implies $\cal{A}$ and $\cal{C}$ are isomorphic, w.r.t. metric.  Thus,   fields and particles become isomorphic.  Thus measures of energy and mass are related, up to a constant squared (coming from the inner product).    This appears to recover Einstein's equation, $e = mc^2$, with the addition that higher order elements and coelements are also found to be isomorphic.  (under development)

 \subsection*{Methods of solving ordinary differential equations}
The existence and uniqueness theorem of ordinary differential equations is valid for Banach spaces.  This lets us apply standard methods to our Banach spaces of chainlets.  

Suppose $J:[t_0, \i) \to \cal{N}_k^{\natural_{r}}$  
and the limit $J'(t) = \lim_{h \to 0} \frac{J(t+h) -J(t)}{h}$ exists for all $t \ge t_0.$  Suppose $f:[t_0, \i) \times \cal{N}_k^{\natural_{r}} \to \cal{N}_k^{\natural_{r}}$ and the initial condition $J_0 \in \cal{N}_k^{\natural_{r}}$ is given.  We want to approximate the solution of the differential equation
$$ J'(t) = f(t, J(t)), J(t_0) = J_0.$$ 

Two elementary methods are given to give the reader a feeling for the subject.  We give pointers to other methods that are more accurate and which are being developed for a separate publication.  

\subsection*{Euler method}  Starting with the differential equation, we replace the derivative $J'$ by the finite difference approximation 
$$J' \approx \frac{J(t + h) - J(t)}{h},$$ which yields
$$J(t+h) \approx J(t) + h f(t, J(t)).$$ 
 
 A simple way to apply this formula is to choose a step size $h$ and construct the sequence $t_0, t_1 = t_0 + h, t_2 = t_0 + 2 h, \dots.$  We denote by $J_n$ a numerical estimate of the exact solution $J(t_n).$  We compute these estimates by the following recursive scheme
 $$J_{n+1} = J_n + h f(t_n, J_n).$$
 
 This is the {\itshape \bfseries Euler method} applied to chainlets.  
 
 If we use the approximations $$J' \approx \frac{J(t ) - J(t-h)}{h},$$ we get the {\itshape \bfseries backward Euler method}:
 $$J_{n+1} = J_n + h f(t_{n+1}, J_{n+1}).$$

The backward Euler method is an implicit method.  Thus we have to solve an equation to find $J_{n+1}.$    A functional iteration or a modification of the Newton-Raphson method may be used to achieve this.  

To obtain higher order approximations, there are standard methods which may be extended to chainlets.  One possibility is the {\itshape \bfseries multstep method} which uses more past values than just the previous one.  Another possibility is to use more points in the time interval $[t_n, t_{n+1}]$.  This leads to an extension of the Runge-Kutta methods.    Both ideas can be combined resulting in {\itshape \bfseries general linear methods.}  There are variable step-size methods, variable order methods, extrapolation methods to be considered.   Methods involving $f$ and its derivatives include Hermite-Obreschkoff methods and Fehlberg methods.  

All higher order equations can be transformed to first order equations, but may not be the best way to proceed.  Nystr\"om methods appear to extend to chainlets, as well.   Geometric integration methods are defined for special classes, such as Hamiltonian equations and reversible equations.  The numerical solution will respect the underlying structure of these classes.

An example of special interest is $f(t, J(t)) := \Box(J(t))$ and $J_0 = u K$  where $u$ is a heat distribution and $K$ is a chainlet with local density one.  It appears that Euler's method leads to  a solution  of the heat equation.   $J = f K$.     
 $$J_{n+1} = J_n + h \Box(J_n).$$
 \subsection*{Heat equation for chainlet domains} 
  \subsection*{Heat conduction}  There are some immediate applications to modeling heat flow across a chainlet domain $J$.   Recall that a chainlet domain $J$ does not have to be smooth, nor does it need to be connected.   We remind the reader that this includes all chainlets, smooth, rough, discrete, bilayer (e.g., soap films).  For example,  $k$-element chains $J = \sum a_i \a_i$ are permitted which are supported in finitely many points. This may well be the first theory that can treat heat flow across branching surfaces.  (?)

We give two simple examples to demonstrate how certain classical methods extend readily to chainlet domains.  

\subsection*{Example 1}
 In the classical approach,  the rate of heat flow into a  non-homogeneous anisotropic region is given by the surface integral
 $$q_t(V) = - \int_{\p V} H(x) \cdot n(x) ds.$$   We may extend this naturally to $3$-chainlets $J$ of class $N^r$ and apply our  chainlet divergence theorem to obtain  the rate of heat flow into a  non-homogeneous anisotropic chainlet region.
 $$q_t(J) =    - \int_{\perp \p J} H(x) = -\int_J d \star H(x).$$

\subsection*{Example 2}  This   simple example generalizes one seen in \cite{flanders} to chainlet domains.  The extension is possible since Stokes' theorem is valid on chainlets.  
Uniqueness of solutions: Let $J$ be a chainlet in $\R^n.$
Let $R = J \times [0,b].$  Then $\p R = \p J \times [0,b] + (-1)^n J \times b + (-1)^{n-1} J \times 0.$  
Consider the heat equation 
$$\Delta u = \sum \frac{\p ^2 u}{\p x_i^2} = \frac{\p u}{\p t}.$$  Suppose $u$ vanishes on $\p J \times [0,b]$ and on $J \times 0.$  Now $dt = 0$ on $J \times b$ since $t = b.$ 
Hence 
$$\int_{\p R} u *dudt = 0.$$   

Set $$\b = 2u(*du)dt + (-1)^{n-1} u^2 dv$$ where $dv = dx_1dx_2 \cdots dx_n.$  Since $$du \wedge (*du) = (grad \, u)^2 dv = \sum \left(\frac{\p u}{\p x_i}\right)^2 dv$$ we have
$$d \b   = 2(grad \, u)^2 dvdt.$$  Therefore, by Stokes' theorem for chainlet domains $$\int_{\p R} \b = 2\int_R  (grad \, u)^2 dvdt.$$  

Hence 
$$(-1)^{n-1}\int_{\p R} u^2 dv= 2\int_R (grad \, u)^2 dv dt.$$ I.e., 
$$\int_{J \times b}u^2 dv + 2 \int_R(grad\, u)^2dv dt = 0$$ and thus 
$$(grad \, u)^2 = \sum \left(\frac{\p u}{\p x_i}\right)^2 = 0,$$ implying  $$\frac{\p u}{\p x_i} = 0$$ on $R$. Hence $u$ is identically $0$ on $R$.    We conclude that if two temperature distributions coincide initially at $t = 0$ and always on $\p R$, then they must be the same at each point of $R$ and for each $t$.

\subsection*{Numerical implementation of discrete calculus}

\subsection*{Matrix methods}   The space of discrete chainlets is finitely generated.  This leads to simple matrix methods.    There is an orthonormal basis for each $\cal{A}_k^j(K)$ and coefficients can easily be calculated.

As we have seen, one may approximate a chainlet $J$ by choosing some sample points.  Then use standard techniques to form a Delauney triangulation. Each simplex $\s_i$ of the the triangulation determines a unique $k$-element $\a_i = \a(\s_i).$  Choose a base point $p_i$ near each $\s_i$.  The element chain $A = \sum \a_i(p_i)$ approximates $J$ if $J$ is a smooth manifold.

An alternate method is to cover $\R^n$ by a binary cubical mesh.  At the midpoint of each cube $Q_i$, place an orthonormal basis of element chains of   dimension $k$:  $\a_{i,1}, \dots, \a_{i,s}$.  Let $\g_{i,1}, \dots, \g_{i,s}$ denote the $k$-coelements dual to these element chains.    Now let $J$ be a $k$-chainlet with finite mass.  We have defined $J\lfloor_{ Q}$ as a chainlet.  Define  $a_{i,m} = \int_{J\lfloor_{ Q_i}} \g_m$  Then $A = \sum_i \sum_m a_{i,m} \a_{i,m}$ approximates $J$ in the $r$-natural norm.     

One may then solve problems of calculus on this approximating element chain $A$.   For example, the boundary operator $\p A$ approximates $\p J$.  Flux across $\p J$ of a vector field may be estimated by summing over each of the boundary elements of $A$.  $$\int_{\perp \p J} \o \sim  \sum_i  \int_{\perp \p \a_i} \o = \int_{\a_i} d\o.$$  At the level of elements, the operators $\p$ and $\perp$ are linear and are therefore represented by matrices.  The dimensions of the matrices are small, on the order of the dimension $n$.  Orthornormal bases may be chosen for the algebras of $k$-elements of order $j$ which leads to orthogonal projection and coefficients in the usual way.     This will be developed and expanded, but the basic ideas are simple and well understood at this point.  All of our operators and products in this paper are defined initially at the level of the algebras and thus have matrix representations.  At the level of chainlets, we approximate the chainlet by choosing a finite set of base points $P$ and using element chains from $\cal{A}(P).$   The results converge to the classical continuum limit by our continuity theorems.

\section{Element-valued and coelement-valued integrals}\label{preint} So far, we have integrated $k$-forms over $j$-chainlets where $k = j$.  Now we consider how to treat the general case where $k$ and $j$ might be different.  We first examine the infinitesimal case. 

If $\a$ is a $k$-element and $\o$ is a $(k+j)$-coelement, define $\o(\a)$ to be the $j$-coelement satisfying $\o(\a)(\b) = \o(\a \wedge \b)$ for all $j$-elements $\b$.   Suppose $\a$ is a $(k+j)$-element and $\o$ is a $k$-coelement. Define $\o(\a)$ to be the $j$-element satisfying $\eta(\o(\a)) = \eta \wedge \o(\a)$ for all $j$-coelement $\eta.$

\begin{theorem}  Suppose $\a$ is a $(k+j)$-element and $\o$ is a $(k-1)$-form.  Then
$$\o(\p \a) = \p(\o(\a)) + (-1)^j d\o(\a).$$
\end{theorem}

\begin{proof}
By the product rule.
$$d(\eta \wedge \o)(\a) = d\eta \wedge \o(\a) + (-1)^j \eta \wedge d \o(\a)$$ for all $j$-   Then $$\eta( \o(\p \a)) = d\eta( \o(\a)) + (-1)^j \eta ( d \o(\a)).$$  The result follows since $d\eta(\o(\a)) = \eta(\p (\o(\a)).$ 
 \end{proof}
 
 \begin{theorem}  Suppose $\a$ is a $k+1$-element and $\o$ is a $(k+j)$-form.  Then
$$\o(\p \a) = d(\o(\a)) + (-1)^{k+1} (d\o)(\a).$$
\end{theorem}

\begin{proof}
By the product rule for elements, if $\b$ is a $j$-element, then 
$$\o(\p(\a \wedge \b)) = \o(\p\a \wedge \b) + (-1)^{k+1} \o(\a \wedge \p \b).$$ xxx check sign xxx Hence 
$$(d\o)(\a)( \b) =  \o(\p \a)(\b) +  (-1)^{k+1} d(\o(\a))( \b).$$
\end{proof}

One might be tempted to try and use these concepts to develop vector-valued and covector-valued integrals.   However, these cannot may not be defined on  a manifold as there is no way to add vectors based at different points, unless we have a Lie algebra.  To handle manifolds, we define a new concept of a {\itshape \bfseries preintegral} of a form over a chainlet $J$.   The value of this preintegral is a chainlet whose support is $J$ itself, even if the form has the same dimension as $J$.

For a $k$-element chain $A = \sum a_i \a_i$ and a $j$-form $\o$, define $$\cal{I}_A \o := \sum a_i \o(\a_i).$$  Take note that the RHS is supported in finitely many points, wherever the $\a_i$ are supported.  We are not taking the sum to be a single vector or covector.   It can be shown that this preintegral is continuous in the chainlet topology if $\o$ is of class $B^{\i}.$ It thus leads to the preintegral defined for chainlet domains:
$$\cal{I}_J \o  : = \lim_{A_i \to J} \cal{I}_{A_i} \o.$$
For example, if $\o = dx$ and $C$ is a solid cylinder in 3-space with its central core along the $x$-axis, then $\int_C dx$ is the foliation of $C$ by two dimensional slices, orthogonal to the $x$-axis.  This chainlet has a boundary which is represented by the foliation of $\p C$ by the bounding circles.   Of course, $\int_C d(dx) = 0.$

\begin{corollary}\label{preintegralb} Suppose  $\o$ is a $(k-1)$-form.  Then  for $(k+j)$-chainlets $J$
$$\cal{I}_{\p J} \o = \p \cal{I}_J \o + (-1)^j \cal{I}_J d\o.$$
\end{corollary}

\begin{proof} Apply continuity of the integral in this category (xxx You have proved this in your stolen laptop.  Redo it.  Check the sign. xxx).   
\end{proof}

A similar result holds if the dimension of the form exceeds that of the chainlet.  
\begin{corollary}\label{preintegrald} Suppose  $J$ is a $(k-1)$-chainlet.  Then  for $(k+j)$-forms $\o$
$$\cal{I}_{\p J} \o = d \cal{I}_J \o + (-1)^j \cal{I}_J d\o.$$
\end{corollary}
xxx check the sign xxx

  If the forms have values in a Lie algebra, we may use the operators $Vec$ or $Covec$ to project preintegrals to the standard integral:  $$Vec(\cal{I}_J \o) = \int_J \o$$ if $dim(\o) > dim(J)$  and $$Covec(\cal{I}_J \o) = \int_J \o$$ if $dim(\o) < dim(J)$.

This has immediate applications.

1.   If the dimensions are the same, then $j = 0.$  Instead of a scalar, our preintegral yields a function times $J$.  

2.    Apply $Vec$ in Corollary \ref{preintegralb} and the second integral, with the boundary, vanishes.  In case the dimension of the form and chainlet match, this retrieves Stokes' theorem. 

3.  Apply $Covec$ in Corollary \ref{preintegrald} and the second integral vanishes. 

4.  If $J$ is a cycle, then the integral on the left is zero in both Corollaries. 

5.  If $\o$ is closed, the last integral vanishes in both Corollaries.  

 \subsection*{Maxwell's equations} (under development)    Discrete, rough and bilayer versions of Maxwell's equations are available, including extensions of the constitutive relations. (Full research paper in preparation with Alain Bossavit and Robert Kotiuga.)

Remarks:
\begin{itemize}
\item The discrete theory of chainlets carries the basic operators $(\perp, \wedge, \p)$ and the dual operators $(*, \wedge, d)$, needed to state Maxwell's equations at a single point, and thus over a discrete chain.   The pushforward operator leads to time varying equations. The cochains are graded commutative and associative.   Convergence to the smooth continuum holds by the continuity theorems of chainlet geometry.
\item We may use the well established integral versions of Maxwell's equations for extending the theory to chainlet domains.  The integral is defined and the divergence theorem holds.  Domains  include discrete (e.g., energy points), rough (e.g., fractals), and bilayer (e.g., soap films, lightening).  (This may be the first theory to handle bilayer  domains in EM.   This is a further topic under development.)  
\item  We may model Maxwell's equations using the preintegral defined above.      Consider  $E$, the $1$-form representing the electric field over a chainlet domain.  The preintegral $\cal{I}_J E$ represents the magnetic $2$-form.  An interesting question arises: what do the three terms in   Theorem \ref{preintegralb} and \ref{preintegrald} correspond to in the known theory?  Do they tell us anything new?
\item We may use  the preintegral to model the energy momentum tensor.   The three terms in Theorem \ref{preintegralb} and \ref{preintegrald} may well predict physical phenomena.  (Under investigation with UC Berkeley physicists)
\end{itemize}

\section{Calculus on fractals}
\subsection*{Fractal norms on polyhedral chains}  
We recall the 1-parameter family of natural norms on polyhedral chains in $\R^n$, depending on a real parameter $r \ge 0$, as defined in  \cite{continuity}.   This duplicates what appeared earlier in these notes, except that a scalar $0 < \a \le 1$ is incorporated into the definitions.

 For consistency of notation, we sometimes denote the $k$-mass of a polyhedral   $k$-chain $P$  as  $M(P) = \|P\|_0.$  
 For $0 < \a \le 1$, the $\a$-weighted mass of a $k$-cell $\s$   is defined to be  its $(k-1+\a)$-Hausdorff measure and is denoted $|\s|_{0,\a}.$   For $\a = 1$ this is merely $k$-volume again.   
   For a polyhedral   $k$-chain $P$, define $$\|P\|_{0,\a} = \inf\left\{\sum |a_{i}||\s_{i}|_{0,\a}: P = \sum a_{i} \s_{i}\right\}$$    where the $\s_{i}$ are $k$-cells.    
Given a $k$-cell $\s^0$ and vectors $v_1, \cdots, v_i$, define the   $1$-difference $k$-cell as $\s^1 = \s^0 - T_{v_1}\s^0. $  For consistency of notation a $k$-cell is sometimes referred to as a $0$-difference $k$-cell. Inductively define the  $i$-difference $k$-cell by $\s^i = \s^{i-1} - T_{v_i} \s^{i-1}.$ Define
 $\|\s^{1}\|_{\a} = |\s^0|_{0}|v_1|^{\a}$ and  $\|\s^i\|_{i-1+\a} = |\s^0|_{0}|v_1| \cdots |v_{i-1}||v_i|^{\a}.$  
 For  $D^i = \sum a_i \s^i$, $i \ge 1$,  define 
 $$\|D^i\|_{i-1+\a} = \sum |a_i|\|\s^i\|_{i-1+\a}.$$

Define $|P|^{\nat_0} = \|P\|_0$ and

$$|P|^{\nat_{\a}} = \inf\left\{ \|D^0\|_0 + \|D^1\|_{\a} + \|C\|_{0,\a}\right\}$$
where $P = D^0 + D^1 + \p C.$

For $r>1$, set $r=r'+\a$ where $r' \in \Z$ and $0 < \a \le 1.$  Define
$$|P|^{\nat_r} = \inf\left \{ \sum_{i=0}^{{r'}} \|D^i\|_i  +\|D^{r'+1}\|_r  + |C|^{\nat_{r-1}}\right\}$$ where $P = \sum_{i=0}^{r'+1} D^i + \p C.$

If $\o$ is a differential form $\|\o\|_0$ denotes its essential supremum.   For $i \ge 0$ and $0 < \a \le 1$, define
$$\|\o\|_{i+\a} = \sup\frac{\int_{Q^{i+1}}\o}{\|Q^{i+1}\|_{1+\a}}.$$
For $0 < \a < 1$ define
$$|\o|^{\natural_{\a}} = (k+1)\sup\{\|\o\|_0, \|\o\|_{\a}\},$$  and for
$r \ge1$
$$|\o|^{\natural_{r}} = \sup\{\|\o\|_0, \cdots, \|\o\|_{{r'}}, \|\o\|_r,  \|d\o\|_0, \cdots, \|d\o\|_{r'-1}, \|d\o\|_{r-1}  \}.$$
Clearly, $|d\o|^{\natural_{r-1}} \le |\o|^{\natural_{r}}$ if $r \ge 2.$ 
 \begin{lemma} $$\|\o\|_{i+\a} = \sup_v \left\{\frac{\|\o -T_v\o\|_{i}}{|v|^{\a}}\right\}.$$
\end{lemma}

\begin{proof} The proof proceeds by induction.   Suppose $i = 0.$  Now 
$$\begin{aligned} \frac{|\int_{\s^1}\o|}{\|\s^1\|_{\a}} &=  \frac{|\int_{\s^0-T_v\s^0}\o|}{\|\s^1\|_{\a}} \\&=  \frac{|\int_{\s^0}\o -T_v^*\o|}{\|\s^1\|_{\a}} \\&\le \frac{\|\o-T_v^*\o\|_0 \|\s^0\|_0}{\|\s^0\|_0|v|^{\a}}
\end{aligned}$$
and 
$$\begin{aligned} \frac{\|\o-T_v^*\o\|_0}{|v|^{\a}} \le \sup_{\s^0}\frac{|\int_{\s^0}\o -T_v^*\o|}{\|\s^0\|_0|v|^{\a}} =  \frac{|\int_{\s^0-T_v\s^0}\o|}{\|\s^1\|_{\a}} =  \frac{|\int_{\s^1}\o|}{\|\s^1\|_{\a}}.
\end{aligned}$$

Now suppose the result holds for $i$.  Then
$$\begin{aligned} \|\o\|_{i+\a} = \sup\frac{|\int_{\s^{i+1}}\o|}{\|\s^{i+1}\|_{i+\a}}  &=\sup\frac{|\int_{\s^i -T_v\s^i}\o|}{\|\s^{i+1}\|_{i+\a}} \\&=
\sup\frac{|\int_{\s^{i}}\o -T_v^*\o|}{\|\s^{i+1}\|_{i+\a}} \\&\le 
\sup\frac{\|\o -T_v^*\o\|_i\|\s^i\|_i}{\|\s^i\|_i|v|^{\a}} \\&= 
\sup\frac{\|\o -T_v^*\o\|_i}{|v|^{\a}}
\end{aligned}$$
and 
$$\begin{aligned} \frac{\|\o -T_v^*\o\|_i}{|v|^{\a}} &=\sup\frac{\int_{\s^i}\o -T_v^*\o}{\|\s^i\|_i|v|^{\a}} \\&=
\sup\frac{\int_{\s^i -T_v\s^i}\o}{\|\s^i\|_i|v|^{\a}} \\&= 
\sup\frac{\int_{\s^{i+1}}\o}{\|\s^{i+1}\|_{i+\a}}
\end{aligned}$$
\end{proof}

 \begin{lemma} For integers $ i \ge 0$ and $\o \in \cal{B}^{i+\a}$ $$\left|\int_{D^{i+1}} \o\right| \le \|D^{i+1}\|_{i +\a}\|\o\|_{i+\a}$$ and

  \end{lemma}
  
  \begin{proof} Let $i = 1$.  Then 
  $$\begin{aligned} \left|\int_{\s^{1}} \o\right| = \left|\int_{\s^{0} -T_v\s^{0}} \o\right| &= \left|\int_{\s^{0}} \o - T_v \o\right| \\& \le  
  \|\s^0\|_0\|\o -T_v\o\|_0 \\&\le
  \|\s^0\|_0|v|^{\a}\|\o\|_{\a} = \|\s^1\|_{\a}\|\o\|_{\a}.\end{aligned}$$ By induction
   $$\begin{aligned} \left|\int_{\s^{i+1}} \o\right| = \left|\int_{\s^{i} -T_v\s^{i}} \o\right| &= \left|\int_{\s^{i}} \o - T_v \o\right|  
    \\&\le \|\s^i\|_i\|\o-T_v^*\o\|_i
   \\&\le \|\s^i\|_i|v|^{\a}\|\o\|_{i+\a} \\&= \|\s^{i+1}\|_{i+\a}\|\o\|_{i+\a} 
   \end{aligned}$$ 
   
   Hence if $D^{i+1} = \sum a_j \s_j^{i+1}$, then $$\left|\int_{D^{i+1}} \o\right|  \le \sum_j |a_j|\left|\int_{\s_j^{i+1}} \o\right|  \le \left( \sum_j |a_j|  \|\s^{i+1}_j\|_{i-1+\a}\right) \|\o\|_{i+\a }.$$
It follows that  $$\left|\int_{D^{i+1}} \o\right| \le \|D^{i+1}\|_{i+\a }\|\o\|_{i +\a}.$$  \end{proof}

\begin{lemma} \label{derivative} If $\o \in \cal{B}^{1+\a}$ and $P$ is a polyhedral chain, then $$\left|\int_{\p P} \o\right| \le|P|^{\nat_{\a}}|\o|^{\natural_{1+\a}}.$$
\end{lemma}

\begin{proof} Let $\e > 0.$  There exists $P = D^0 + D^1 + \p C$ such that $|P|^{\nat_{\a}} > \|D^0\|_0 + \|D^1\|_{\a} + |C|_{0,\a} -\e.$
Since $\p P = \p D^0 + \p D^1$  we have 
$$\begin{aligned} \left| \int_{\p P} \o \right| &\le \left| \int_{ D^0}d \o \right|  + \left| \int_{ D^1} d\o \right|  \\&\le \|D^0\|_0\|d \o\|_0 + \|D^1\|_1\|d \o\|_{\a} \\&< 
(|P|^{\nat_{\a}} +\e)| \o|^{\natural_{{1+\a}}}.\end{aligned}$$

\end{proof} 

  \begin{theorem} If $\o \in \cal{B}^r$, $r \in \Z, r \ge0$ and $0 < \a \le 1$,  then $$\left| \int_{P} \o\right| \le |P|^{\nat_r }|\o|^{\natural_{r}}.$$
\end{theorem}

\begin{proof} This is the standard inequality of integration  if $P$ is a polyhedral chain and $r = 0$, i.e., $\o$ is bounded, measurable.    Suppose $r > 0.$ For $\e >0$ there exists $P = D^0 + D^1 + \p C$ such that $$|P|^{\nat_{\a} }> \|D^0\|_0 + \|D^1\|_{\a} + |C|_{0,\a} - \e.$$

  First observe  that  if $C = \sum a_j Q_j$ and $Q_j$ has edge length $\e_j$, then  $$\left|\int_{\p C} \o\right| \le \sum |a_j|\left| \int_{\p Q_j} \o\right| \le  (k+1)\|\o\|_{\a} \sum|a_j| |\e_j|^{k-1+\a }.$$  Hence $$\left|\int_{\p C} \o\right| \le |C|_{0,\a} |\o|^{\natural_{{\a}}}.$$
  
  By the previous lemma
 $$\begin{aligned} \left|\int_{P} \o\right|  &\le\|D^0\|_0\|\o\|_0 + \|D^1\|_{\a}\|\o\|_{\a} + (k+1)|C|_{0,\a}\|\o\|_{\a} \\&\le (\|D^0\|_0 + \|D^1\|_1 + |C|_{0,\a}) |\o|^{\natural_{\a}}
\\&< ( |P|^{\nat_{\a}} +\e)|\o|^{\natural_{\a}} .\end{aligned}$$
We conclude $$\left| \int_{P} \o\right| \le |P|^{\nat_{\a}} |\o|^{\natural_{\a}}.$$

  Let $\e > 0.$  There exists $P = \sum_{i=1}^{r'+1} D^i + \p C $ such that $$|P|^{\nat_r} > \sum_{i=1}^{{r'+1}} \|D^i\|_i +\|D^{r'+1}\|_r  + |C|^{\nat_{r-1}} -\e.$$  Suppose $r \le 2.$  By \ref{derivative} 
    $$
  \begin{aligned} \left|\int_P \o \right| &\le \left|\int_{D^0} \o \right| +
  \left|\int_{D^1} \o \right| + \left|\int_{\p C} \o \right| \\&\le \|D^0\|_0\|\o\|_0 + \|D^1\|_{1+\a}\|\o\|_{1+\a} + |C|^{\nat_{\a}}|\o|^{\natural_{1+\a}} 
  \\&\le (\|D^0\|_0 + \|D^1\|_1 + |C|^{\nat_{\a}})|\o|^{\natural_{1+\a}}  \\&< (|P|^{\nat_{1+\a}} +\e) |\o|^{\natural_{1+\a}}.\end{aligned}$$
  
 For $r > 2$ we use induction.  Now $d\o \in B^{r-1}$ with $|d\o|^{\natural_{r-1}} \le |\o|^{\natural_{r}}.$   By induction, 
  $|\int_{\p C} \o|= |\int_{C} d\o| \le |C|^{\nat{r-1}}|\o|^{\natural_{r}}.$  Thus $$\begin{aligned} 
\left|\int_P \o \right|  =\left| \int_{\Sigma D^i + \p C}\o \right|&\le \sum_{i = 1}^{{r'+1}} \left| \int_{D^i} \o \right|  + \left| \int_{C} d\o \right| \\&\le \left( \sum_{i = 1}^{r} \|D^i\|_i  +\|D^{r'+1}\|_r + |C|^{\nat_{r-1}}\right) |\o|^{\natural_{r}} \\&\le(|P|^{\nat_r} +\e)|\o|^{\natural_{r}}
\end{aligned} $$

\end{proof}

It follows that $ |P|^{\nat_r }$ is a norm on polyhedral chains since there exists a smooth form $\o$ with $\int_P \o \ne 0.$  

The dual space of cochains has norm 
$|X|^{\nat_r} = \sup_P \frac{X \cdot P}{|P|^{\nat_r}}$ where $P$ is a nonzero polyhedral chain.  

Define $\|X\|_s = \sup \frac{X \cdot D^{s'+1}}{\|D^{s'+1}\|_s}.$
\begin{lemma}\label{cochain} $$|X|^{\nat_{\a}} = (k+1) \sup\{\|X\|_0, \|X\|_{\a} \}$$ and $$|X|^{\nat_r} =  \sup\{\|X\|_0, \cdots, \|X\|_{r'}, \|X\|_r, \|dX\|_0,\cdots, \|dX\|_{r'-1}, \|dX\|_{r-1}\}.$$\end{lemma} 

We defer the proof until the next section where we present a more general version.

 \section{Additional results}
\medskip
 \begin{lemma} If $J$ is a nonzero $r$-natural chain, then $supp(J) \ne \varnothing.$
\end{lemma}

\begin{proof} It suffices to show that $X \cdot J = 0$ for any compact $X$.  Each
point $p
\in Q = supp(X)$ is in some neighborhood $U(p)$ such that $Y \cdot J = 0$ for any
$r$-natural $Y$ with $\phi(Y) = 0$ outside $U(p)$.  Finitely many of these neighborhoods,
$U_1,U_2, \dots, U_m$ cover $Q$.  Let $\phi_0, \phi_1, \dots, \phi_m$ be a partition
of unity subordinate to this cover.  Then $\Phi_0(p)\phi(X)(p) = 0$ for all $p$, so that
$\phi_0X = 0.$  Note that $X = \phi_1X +  \dots + \phi_mX$ and $D_{\phi_iX} = \phi_i
\phi(X) = 0$ outside $U_i$.  Hence
$$X \cdot J = \sum (\phi_i X \cdot J) = 0.$$ 
\end{proof} 

The support of a chainlet $J$ is a piece of information about $J$ but far
from all of it. For example, consider a Jordan curve $\g$ in the plane with positive
$2$-dimensional Lebesgue measure.  Approximate
$\g$ with polyhedral  approximators $P_k$ inside $\g$ and $Q_k$ outside $\g.$  Then
both
$P_k$ and $Q_k$ are Cauchy sequences in the $1$-norm but they converge to distinct
chains $J \ne K$ because $|P_k - Q_k|_1$ does not tend to 0.

\begin{theorem} Let $K \subset \R^n$.  Then $S = \left\{J \ne 0 \in {\cal N}_k^r :
supp(J)
\subset K \right\}$ is convex.
\end{theorem}

\begin{proof}  Let $J_1, J_2 \in S$ and $T = tJ_1 + (1-t)J_2$.  We show $supp(T)
\subset P .$ Since $supp(T) \ne \varnothing$  there exists $p \in supp(T).$  Then for
every
$\e > 0$, there exists
$X$ such that $X \cdot T  \ne 0$ and
$\phi(X)(q) = 0$ outside the neighborhood $U_{\e}(p)$.  We show $p \in P.$  Since $X
\cdot T
\ne 0$, either
$X
\cdot J_1$ or $X
\cdot J_2 \ne 0.$  Say $X
\cdot J_1 \ne 0$.  The cochain $ X$     may be used to show that $p \in supp(J_1)
\subset P.$  
\end{proof} Recall the previous example of a Jordan curve with positive Lebesgue area
bounding an open subset
$U$ of
$\R^2$.  We have seen that the chain $\p U$ may be approximated by polyhedral  curves
$P_k$ on the inside of $U$ and by by polyhedral  curves
$Q_k$ on the outside of $U$.  Both $P_k$ and $Q_k$ have unique limits $P$ and $Q$ in
the
$1$-norm but the limits are distinct.  Our theorem shows that any convex combination
of
$P$ and
$Q$ also has support in $\p U$.  

\begin{proposition}\label{prop.open} If $U \subset \R^n$ is an oriented, bounded open
subset with finite volume
$m(U)$, then $U$ supports  a
$1$-natural chain $J$.  If the $n$-dimensional Lebesgue measure of $\p U$ is $0$,
$m(U) = |J|_0$, and  $U$ and $J$ are similarly oriented, then $J$ is unique.
\end{proposition}

\begin{proof} $U$ is the union of positively oriented Whitney cubes $\s_i$ with
disjoint interiors contained in $U$.  The series $\sum \s_i$  converges in the mass
norm to a 
$0$-natural chain $J$ since $|U|_0 <  \i$ and the support of $J$ is the pointset
$U$.  It is immediate that  $\p J$ is $1$-natural. 

Suppose $J_1 = lim P_k$ and $J_2 = limQ_k$ where the $P_k$ and $Q_k$ are polyhedral ,
$supp(J_i) = U$,  the $n$-dimensional Lebesgue measure of $\p U$ is
$0$, $m(U) = |J_i|_0$, and  $U$ and $J_i$ are similarly oriented for $i = 1,2.$ 
Since  
$\p U$ has 0 volume,  it has tubular neighborhoods with volume tending to 0. 
$|J_1-J_2|_0 \le lim|P_k - Q_k|_0.$  But $P_k - Q_k = B_k + C_k$ where $B_k \subset
\e-$tubular neighborhood of $\p U$ and $|C_k|_0 \to 0$ as $k \to \i.$  If $|C_k|_0$
does not tend to $0$, then the supports of $J_i$ are not the same.
\end{proof}
 
\subsection*{Compact chains}  

We now show that most of any chainlet lies in a compact set. We say a chainlet is {\itshape \bfseries compact} if its support is compact.
\begin{theorem} For any $J \in \cal{N}_k^{\nat_r}$ and $\e > 0$ there is a compact chainlet $J' \in \cal{N}_k^r$ such that 
\begin{enumerate}
\item $|J'-J|^{\nat_r} < \e$,
\item $supp(J') \subset supp(J)$,
\item $|J'-J|_{s,r} < \e$ if $|J|_{s,r} < \i$,
\item $|\p(J'-J)|_{s+1,r+1} < \e$ if $|\p J|_{s'+1, r'+1} < \i.$
\item Moreover, given $N > 1$, there exists a compact set $Q$ such that for any compact Lipschitz function $\phi$ with $0 \le \phi(p) \le 1$, $\phi(p) = 1$ in $Q$, $\|\phi\|_1 \le N$, we may use $J' =\phi J.$

\end{enumerate}
\end{theorem}

\subsection*{Polyhedral  approximation}
\begin{theorem} For any $J \in \cal{N}_k^r$, any neighborhood $U$ of $supp(J)$ and any $\e > 0$, there is a polyhedral  chain $P$ such that
\begin{enumerate}
\item $|P-J|^{\nat_r} < \e$,
\item $supp(P) \subset U$,
\item $|P|_{s,r} < |J|_{s,r} + \e$ if $ |J|_{s,r} < \i$,
\item $|\p P|_{s+1,r+1} < |\p J|_{s'+1, r'+1} + \e$ if  $|\p J|_{s'+1, r'+1} < \i.$
\item $|*P|_{s,r} \le |*J|_{s,r} + \e.$
\end{enumerate}
\end{theorem}

\medskip \noindent {\bf Multifolds}  Introduced in these lectures are what we call a {\itshape \bfseries multifold} $M^m$ which is defined to be a chainlet in $\R^n$ whose support is locally the graph of an $L^1$ function of $\R^m$, where $m \le n$.   That is, if $x \in M$ there exists a function $f: U \subset \R^m \to M^m$ such that $x = f(p)$ for some $p \in U.$  Furthermore, the overlap mappings are smooth.  This differs from a manifold in that we do not require that the local coordinate charts be homeomorphic images of open subsets of $\R^m.$   To distinguish this feature, we call the pair $(U,f)$ a {\itshape \bfseries local function chart} and $f$ is a {\itshape \bfseries local coordinate function}.   We will return to this class of domains later and show that a multifold $M$ carries with it a full theory of calculus, including a boundary, a way to measure flux of a vector field across the boundary, measurements of mass,  vector bundles, characteristic classes, homology, cohomology, homotopy, etc..  Of course, smooth manifolds embedded in $\R^n$ are examples of multifolds.    Compact, topological manifolds that are locally the graph of a continuous function are also multifolds since the local coordinate functions will necessarily be integrable. 

  Later we will add to this notion of multifold and define higher order multifolds.  We will consider  $(k,s)$-multipole versions of $L^1$ functions and create  $(k,s)$-multifolds.   (Replace the graph of step functions approximating $f$ with dipole steps, etc.  Sheer can be added by using a translation vector not orthogonal to the steps. If there is not too much variation, then the limit exists in the $s$-natural norm.)    An example would be a dipole circle.  Sheafs can be used to make this rigorous.

 \subsection*{Multiplication of distributions and functions} Let
 $p = q = 0.$  Then the cap product $X \cap A$ is defined for  a
 $0$-cochain $X$ and a   $0$-chainlet $A$.  On a compact
 manifold,    $0$-chainlets  are
 isomorphic to distributions
  and   $0$-cochains are isomorphic to functions.  Thus cap
 product extends multiplication
  of  distributions  and functions.

\subsection*{Chainlet homology}(Draft, under development)

\begin{theorem}
If $J_i \to J$ and $J_i$  are chainlet $k$-cycles, then  $J$ is a chainlet $k$-cycle.  $H_k(J_i) \to  H_k(J).$   If $J_i = \p C_i$, then $J = \p C.$   
\end{theorem}

Norms on homology classes.  Define $$|A|^{\natural_{r}} = \inf \{|J|^{\natural_{r}} : A = [J]\}.$$

\begin{proposition}
Let $M$ be compact manifold.  There exists $\e > 0$ such that if $A$ is a homology  cycle with $|A|^{\natural_{r}} < \e$, then $A$ is homologous to zero.
\end{proposition}

  \subsection*{Tensor and convolution products} Let $M$ and $N$ be manifolds and $M \times N$ their Cartesian product.  Then each space  $\cal{N}_k(M) \otimes \cal{N}_j(N) \subset \cal{N}_{j+k}(M \times N).$  The {\itshape \bfseries tensor product} $\otimes$ is a bounded operation in each of the $r$-natural spaces. 
  
  Theorem.   $\p(P \otimes Q) = \p P  \otimes Q + P \otimes \p Q.$  $\p(\p P \otimes Q) = \p(P \otimes \p Q).$ xxx check the sign xxx
 
  In $\R^n$ define the {\itshape \bfseries convolution product} of $A \in \cal{N}_k^{\natural_{r}}$ and $B \in    \cal{N}_j^{\natural_{r}}$ by  $$X \cdot A *B := \int_A \left( \int_{T_y B} \o \right)$$  for all  $(j+k)$-cochains $X$ of class $B^r$.  
  

\begin{theorem} If $f$ is a function of class $B^r$ with compact support, then $Ch(f) * A$ is a $k$-vector field of class $B^r$ for any $A \in \cal{N}_k^{\natural_{r}}.$
\end{theorem}

If $M$ is a Lie group  convolution product is determined by 
 $$X \cdot A *B = \int_A \int_B \phi(X)(xy)$$ for all cochains $X \in (\cal{N}_{j+k})^{\prime}.$  
 
 Convolution product is bounded in the $r$-natural norms.  It is associative.

\begin{theorem} If $A$ and $B$ are summable $j$- and $k$-vector fields on $M$, then $A *B$ corresponds to the exterior product $A \wedge B$ as follows: 
$$X \cdot (A*B) = \int_{M\times M} \phi(X)(xy) \cdot[A(x) \wedge B(y)]dxdy$$ for all $X \in \cal{N}_{j+k}(M)^{\prime}.$ 
\end{theorem}

 This relates the convolution product in the homology ring of a Lie group to the cup product in cohomology.  The convolution product coincides with the Pontrjagin product on the homology classes. \cite{hopf}
 
 \subsection*{Degree of a map}
 \begin{example} Let $M$ be a compact connected
 orientable Riemannian manifold and
 $\o$ an
 $n$-form on $M$.   Then $$<M, \o> = 0 \iff \o = d\eta.$$
 \end{example}

 \begin{proof}A compact connected orientable Riemannian
 manifold can be triangulated and so supports a chain, also
 denoted
 $M$. The dimension of $H_n(M)$ is one so any element of
 $Z_n(M)$ can be written as $$cM + \p B$$ where $c$ is a real
 number. By de Rham's theorem and the fact that all
 $n$-forms on $M$ are cocycles we have
 $$
 \begin{array}{rll}
 \o = d\eta &\iff <C,\o> = 0 \mbox{ for all } C \in Z_k
 \\&\iff <cM +
 \p B, \o> = 0 \mbox{ for all } c \in \mathbb{F} \\
 &\iff c<M,\o> = 0 \mbox{ for all } c \in \mathbb{F} \\&\iff <M,\o> = 0.

 \end{array}
 $$
  \end{proof}

 This implies that if $\o$ is an $n$-form with compact
 support, then $$\o = c\phi + d\theta$$ where $c$ is a real
 number, $\phi$ is some fixed $n$-form with compact
 support such that $$<M, \phi> \ne 0$$ and $\theta$ is an
 $(n-1)$-form with compact support.

 Then if $f:M \to N$ is a proper differentiable mapping and if
 $N$ is also connected and triangulable, then there exists a
 number $k$ such that $$\int_M f^*\o = k \int_N \o.$$ for any
 $\o$ with compact support.

There are several natural settings for this study.  (Draft mode, paragraph under development)     We say that a chainlet $B$ is {\itshape \bfseries in a chainlet $J$} if there exists algebraic chains $P_k \to J$ in the natural norm and algebraic $Q_k \to B$ with $Q_k$ supported in $P_k$.
\begin{enumerate}
\item{Chainlets in smooth manifolds}
We can study chainlets in smooth abstract manifolds by mapping local chainlets into M via the coordinate mappings.  The restricted overlap maps should be smooth.   We can use these to compute homology.  The homology classes should coincide with the singular classes. (See theorem below.)
\item{Chainlets in a fixed chainlet $J$ in a smooth manifold} We can compute homology classes of $J$ by studying   $k$-chainlets $B$ in $J$.  Cycles are well defined since we have the boundary operator.  We can equate two cycles iff their difference is a $(k+1)$-chainlet.  We obtain a Stokes' theorem.  Do we have the Poincare duality  homomorphism?
\item{Chainlets in a fixed chainlet in a smooth Riemannian  manifold}
Here we obtain divergence
\item{Element chainlets}  
  We need to study generalized homology, giving up the dimension axiom.  We only obtain approximate homology groups, depending on a small parameter $\l$.  These are increasing and converge to the homology group of the limit chainlet.  Say that $J \sim_{\l} B$ if $|J-B|^{\natural_{r}} \le \l.$  We have   $\l$-cycles if $\p J \sim_{\l} 0$ and $\l$-homology classes if $J - B \sim_{\l} \p C.$   

\end{enumerate}

\begin{lemma} Let $F$ be the fixed point set of a cochain $X$. Then $supp(\p \perp (X \cap M)) \subset
F.$   
\end{lemma}

\begin{theorem} Suppose $H_p(M^n) = 0 p < n.$ Every closed $p$-form $\omega$
has a zero.
\end{theorem}

\begin{proof} Let $X$ be the  cochain of $\o$.  If $X \ne 0$, then $\p *(X \cap M) = 0.$  Thus
$*(X \cap M)$ is a cycle.  Since $H_p(M) = 0$ we know $*(X \cap M) = \p C.$   Now $X \cdot *(X
\cap M) = 1.$  Thus $dX \cdot C = 1.$  Hence $dX \ne 0.$
\end{proof}

The converse of this is false if there are manifolds with $H_p = 0$ and $H^p \ne 0.$


\begin{theorem} If $\o$ is any $(n-1)$-form on orientable M without boundary, then there is a $p$
st
$d\o(p) = 0.$
\end{theorem}  

\begin{proof} Suppose $Y$ is an $(n-1)$-cochain.  Associated to $dY$ is the chainlet $A = *(X
\cap M).$  If $dY \ne 0$, then $A$ is a cycle.   But since $M$ is orientable, the only $n$-cycles
are multiples of $A$.  So $1  = dY \cdot A =  dy \cdot KM = k y \cdot  \p M = 0.$
\end{proof}

Potential theory can be extended to chainlets by way of Green's identities on chainlets.  (See   pp 72-74 of do Carmo)  
 Green's first identity for chainlet boundaries of codimension one.

\begin{theorem} Let $f,g:A \to \R$ be smooth functions defined on a Riemannian $n$-manifold
$M$.   Suppose
$A$ is an $n$-chainlet in $M$.  Then $$\int_{*\p A} fdg = \int_A d(*f dg).$$ 
\end{theorem}
  Green's first identity follows.
$$\int_A <df,dg>\nu +  \int_A f \Delta^2g  = \int_{\p A} f<dg,N>\s.$$

Green's first identity for abritrary codimension.
\begin{theorem} Let $M$ be a Riemannian $n$-manifold.  Let $\o$ be an $n-p$-form, $f$ a
real valued function defined on $A$, and $A$ a
$p$-chainlet.  (The degree of differentiability of $f, \o$ depend on the fractal dimension of the 
chainlet. ) Then 
$$\int_{\p A} *(f d\o) = \int_A df \wedge + *d\o + f \wedge d*d\o.$$
\end{theorem}

Green's second identity for chainlets.

\begin{theorem} $M$ Riemannian $n$-manifold, $A$ an $n$-chainlet, $f$ and $g$ are forms
such that $dim \b = n-p$  Then  
$$\int_{*\p A} f \wedge dg - g \wedge df = \int_A f \wedge d *dg - g \wedge d*df.$$
\end{theorem}

  If $\o$ is a form with $|\o|^{\natural_{j}} < \i$, then we may find explicit expressions for the jet of $\o$ at a point $p$, similar to the Taylor's expansion.     (This will  shortly be expanded, with examples, such as $\o = P(x,y) dx.$ It will be shown that the discrete forms determined by a finite set $K$ and $\o$ limit to $\o$ as $|K| \to 0.$)

 \subsection*{Poincar\'e Lemma for chainlets}  Suppose $\p P = 0$ and the ambient
space is starlike.  Then there is a chainlet $C$ s.t. $P = \p C.$ (rought sketch)
The idea is to show that the join of a point $p$ with $P$ is a chainlet $C$.  This cone over $P$ will be the $C$ we
are looking for.   Suppose $P = \sum D^i + \p B.$  Then $P^* \sum D^{i*} + (\p B)^*$ the cones of each, are well defined.  Observe that $ (\p B)^* = \p (B^*) - B.$  Let $K$ be the diameter of the set containing $supp (P)$ and $p$.    Prove that  $$\|D^{i*}\|_i \le (K+1) \|D^i\|_i$$ and $$ |(\p B)^*|^{\natural_{r}} \le (K+1)|B|^{\natural_{r}}.$$ The first estimate can be found by approximating the cone over a difference cell by staircase difference chains.

\subsection*{Wavelets over chainlets}   The techniques of wavelets (and Fourier series) fit beautifully with chainlets.  Wavelets provide a way to write a form as an infinite sum of simpler forms:
$$\o = \sum_{j=1}^{\i} b_j \phi_j.$$  Chainlets provide a way to write a domain  as an infinite sum of simpler domains, such as cells:
$$J = \sum_{i=1}^{\i} a_i \s_i.$$   Then $$\int_J \o = \sum_{i=1}^{\i} \sum_{j=1}^{\i} \int_{\s_i}\phi_j.$$  In this way, integration of forms over chainlets reduces to integration of simple forms over simple domains.  
\begin{verse}
{\em And it is a noteworthy fact that ignorant men have long been in advance of the learned about vectors.  Ignorant people, like Faraday, naturally think in vectors.  They may know nothing of their formal manipulation, but if they think about vectors, they think of them as vectors, that is, directed magnitudes.  No ignorant man could or would think about the three components of a vector separately, and disconnected from one another.  That is a device of learned mathematicians, to enable them to evade vectors.  The device is often useful, especially for calculating purposed, but for general purpose of reasoning the manipulation of the scalar components instead of the vector itself is entirely wrong.}

\end{verse}  
\begin{flushright} -- Oliver Heavyside \end{flushright}
\newpage
\begin{table}[htdp]
\caption{notation}
\begin{center}
\begin{tabular}{|c|c|}
$T_vX$ &translation of a set $X \subset \R^n$ through $v \in \R^n$  \\
$L$ & linear subspace of $\R^n$\\
$F \subset E$ & affine subspaces $F \subset E = T_vL$\\
$\s$ & $k$-cell\\
$\t$ & facet of a $k$-cell $\s$\\
$\p \s$ & boundary of $\s$\\
$X \times Y$ & Cartesian product of sets $X$ and $Y$\\
$S$ & algebraic $k$-chain\\
$P$ & polyhedral $k$-chain\\
$supp(P)$ & support of a polyhedral chain $P$\\
$M(P)$ & mass of a polyhedral chain $P$\\
$\cal{P}_k$ & vector space of polyhedral $k$-chains\\
$\cal{N}_k^0$ & completion of $\cal{P}_k$ with the mass norm\\
$|v|$ & norm of $v \in \R^n$\\
$U^j$ & list of vectors $U^j = (u_1, \dots, u_j), j \ge 1$\\
$V^k$ & list of linearly independent vectors $V^k = (v_1, \dots, v_k), 0 \le k \le n$\\
$\s(V^k)$& parallelepiped determined by the list $V^k$\\

$\Delta_{U^j} \s(V^k)$& $j$-difference $k$-cell determined by translations of a $k$-cell through a list $U^j$.\\
$D^j$ & chain of $j$-difference $k$-cells\\
$\cal{D}_k^j$ & vector space of chains of $j$-difference $k$-cells\\
$\|D^j\|_j$ & $j$-difference norm\\
$|P|^{\natural_{r}}$& $r$-natural norm of $P$\\
$\cal{N}_k^r$ & completion of $\cal{P}_k$ with the $r$-natural norm\\
$J$ & a chainlet, a member of $\cal{N}_k^r$\\
$Q$ & a cube w.r.t. the inner product\\
$T_v J$ & translation of a chainlet $J$ through a vector $v$\\
$\a = \a(V^k)$& $k$-element determined by a list $V^k$\\
$Vec(\s)$& $k$-element of a $k$-cell $\s$\\
$\Delta_{U^j}\a(V^k)$& $j$-difference $k$-element\\
$\nabla_{U^j}\a(V^k)$& $j$-differential $k$-element\\

\end{tabular}
\end{center}
\label{notation}
\end{table}%

 \newpage
\begin{table}[htdp]
\caption{notation}
\begin{center}
\begin{tabular}{|c|c|}
$A^j$ & chain of $j$-differential $k$-elements\\
$\cal{A}_k^j$& vector space generated by $j$-differential $k$-elements\\
$\cal{A}_k$ & direct sum $\oplus_j \cal{A}_k^j$\\
$\cal{A}$ & direct sum $\oplus_k \cal{A}_k$\\
$\cal{C}_k^j$& dual space of coelements $(\cal{A}_k^j)'$  \\
$\cal{C}_k$ & direct sum $\oplus_j \cal{C}_k^j$\\
$\cal{C}$ & direct sum $\oplus_k \cal{C}_k$\\
$\g^j$& a member of $\cal{C}_k^j$\\
$\o$ & a $\cal{C}$-valued function on a set $U \subset \R^n$\\
$\o = \sum \g^j$&  the jet of $\o$\\
$\p J$ & boundary of a chainlet\\
$d$ & exterior derivative $d\o(\a) = \o(\p \a)$\\
$\perp J$ &\quad  \\
$\star$& Hodge star $\star \o(\a) = \o(\perp \a)$\\
$\diamond$& $\diamond = \perp d \perp$\\
$\Box$& $\Box = \p \diamond + \diamond \p$\\
$\cal{L}_v J$ & Lie derivative of a chainlet\\
$\nabla_v J$ & directional derivative of a chainlet\\
$e_{\b} J$ & exterior product of a chainlet\\
$i_{\b} \o$ & interior product of a form\\
 $\cal{I}_J \o$ & preintegral of a form over a chainlet\\ 
 $\a \wedge \b$& wedge product of elements\\
 $Vec(J)$& $k$-element of a chainlet $J$\\
 $Ch(\o)$& chainlet of a form $\o$\\

\end{tabular}
\end{center}
\label{notation}
\end{table}%

   \end{document}